\documentclass[11pt]{article}

\usepackage{eucal}
\usepackage{color}

\usepackage{tikz}
\usepackage{amsmath,bbm}
\usepackage{amssymb}
\usepackage{epic}

\usepackage{graphicx}
\usepackage{pict2e}

\title{Tilings of non-convex Polygons, skew-Young Tableaux and determinantal Processes}

\author{Mark Adler\thanks{2000
{\em Mathematics Subject Classification}. Primary:
60G60, 60G65, 35Q53; secondary: 60G10, 35Q58. {\em Key
words and Phrases}:Lozenge tilings, non-convex polygons, kernels. \newline
 Department of Mathematics, Brandeis University,
Waltham, Mass 02453, USA. E-mail: adler@brandeis.edu.
The support of a Simons Foundation Grant %National Science Foundation grant 
 \# 278931 is
 gratefully acknowledged. M.A. thanks the Simons Center for Geometry and Physics for its hospitality.}~~~~~~Kurt Johansson\thanks{Department of Mathematics,
KTH Royal Institute of Technology, Stockholm, Sweden. E-mail: kurtj@kth.se. The support of the Swedish Research Council (VR) and grant KAW 2010.0063 of the Knut and Alice Wallenberg Foundation are gratefully acknowledged.} ~~~~~ Pierre
van Moerbeke\thanks{ Department of Mathematics,
Universit\'e de Louvain, 1348 Louvain-la-Neuve, Belgium
and Brandeis University, Waltham, Mass 02453, USA. E-mail: pierre.vanmoerbeke@uclouvain.be . The support of a Simons Foundation Grant %National Science Foundation grant 
 \# 280945 is
gratefully acknowledged. PvM thanks the Simons Center for Geometry and Physics, Stony Brook, and the Kavli Institute of Physics, Santa Barbara, for their hospitality.\newline
 }
}

\date{}

\newcommand{\MAT}[1]{\left(\begin{array}{*#1c}}
\newcommand{\mat}{\end{array}\right)}
\newcommand{\qed}{\leavevmode\unskip\nobreak\penalty200\hskip2pt\null
\nobreak\hfill\rule{1.1ex}{1.1ex}%\parfillskip=0pt
\medbreak }

\newcommand{\x}{ \textcolor[rgb]{0.00,0.00,1.00}{\Large \! /\!\!  /\!\! /\!\!   /\!\! /\!\! /\! }
}

\newcommand{\y}{ \textcolor[rgb]{0.00,0.00,1.00}{\Large \backslash \! \backslash\!\!  \backslash\!\! \backslash\!\!   \backslash\!\! \backslash\!\! \backslash\! }
}

\newcommand{\I}{{\rm i}}

\newcommand{\CR}{{\cal C}}

\newcommand{\GR}{{\cal G}}

\newcommand{\LR}{{\cal L}}
\newcommand{\RR}{{\cal R}}

\newcommand{\PR}{{\cal P}}

\newcommand{\BC}{{\mathbb C}}

\newcommand{\BH}{{\mathbb H}}

\newcommand{\BP}{{\mathbb P}}

\newcommand{\BZ}{{\mathbb Z}}

\newcommand{\pl}{\partial}
\newcommand{\al}{\alpha}

 \newcommand{\Om}{\Omega}

\newcommand{\tom}{\widetilde \omega}
\newcommand{\haom}{\widehat \omega}
\newcommand{\tze}{\widetilde \zeta}
\newcommand{\haze}{\widehat \zeta}
\newcommand{\tU}{\widetilde U}
\newcommand{\haU}{\widehat U}

\newcommand{\Id}{\mathbbm{1}}

 \newcommand{\bl}{\begin{aligned}}
  \newcommand{\el}{\end{aligned}}

\newenvironment{proof}{\medskip\noindent{\it Proof:\/} }{\qed}
\newenvironment
        {remark}{\medskip\noindent\underline{\it Remark:\/} }{\medbreak}
\newenvironment
        {example}{\medskip\noindent\underline{\it Example:\/} }{\medbreak}

\newcommand{\om}{\omega}

\newcommand{\ga}{\gamma}
\newcommand{\Ga}{\Gamma}
\newcommand{\dt}{\delta}
\newcommand{\Dt}{\Delta}
 \newcommand{\vr}{\varepsilon}
\newcommand{\sg}{\sigma}
\newcommand{\ze}{\zeta}
\newcommand{\BR}{{\mathbb R}}
\newcommand{\lb}{\lambda}

\newcommand{\dis}{\displaystyle}

\newcommand{\diag}{\operatorname{diag}}
\newcommand{\Res}{\operatorname{Res}}

\def\be#1\ee{\begin{equation}#1\end{equation}}
\def\bea#1\eea{\begin{eqnarray}#1\end{eqnarray}}
\def\bean#1\eean{\begin{eqnarray*}#1\end{eqnarray*}}

 \newtheorem{definition}{Definition}[section]
 \newtheorem{theorem}[definition]{Theorem}
 \newtheorem{lemma}[definition]{Lemma}
 \newtheorem{corollary}[definition]{Corollary}
 \newtheorem{proposition}[definition]{Proposition}

\catcode `!=11

\newdimen\squaresize
\newdimen\thickness
\newdimen\Thickness
\newdimen\ll! \newdimen \uu! \newdimen\dd! \newdimen \rr! \newdimen
\temp!

\def\sq!#1#2#3#4#5{%
\ll!=#1 \uu!=#2 \dd!=#3 \rr!=#4
\setbox0=\hbox{%
 \temp!=\squaresize\advance\temp! by .5\uu!
 \rlap{\kern -.5\ll!
 \vbox{\hrule height \temp! width#1 depth .5\dd!}}%
 \temp!=\squaresize\advance\temp! by -.5\uu!
 \rlap{\raise\temp!
 \vbox{\hrule height #2 width \squaresize}}%
 \rlap{\raise -.5\dd!
 \vbox{\hrule height #3 width \squaresize}}%
 \temp!=\squaresize\advance\temp! by .5\uu!
 \rlap{\kern \squaresize \kern-.5\rr!
 \vbox{\hrule height \temp! width#4 depth .5\dd!}}%
 \rlap{\kern .5\squaresize\raise .5\squaresize
 \vbox to 0pt{\vss\hbox to 0pt{\hss $#5$\hss}\vss}}%
}%end of \hbox
 \ht0=0pt \dp0=0pt \box0
}%end of \sq!

\def\vsq!#1#2#3#4#5\endvsq!{\vbox to \squaresize{\hrule
width\squaresize height 0pt%
\vss\sq!{#1}{#2}{#3}{#4}{#5}}}

\newdimen \LL! \newdimen \UU! \newdimen \DD! \newdimen \RR!

\def\vvsq!{\futurelet\next\vvvsq!}
\def\vvvsq!{\relax
  \ifx     \next l\LL!=\Thickness \let\continue=\skipnexttoken!
  \else\ifx\next u\UU!=\Thickness \let\continue=\skipnexttoken!
  \else\ifx\next d\DD!=\Thickness \let\continue=\skipnexttoken!
  \else\ifx\next r\RR!=\Thickness \let\continue=\skipnexttoken!
  \else\def\continue{\vsq!\LL!\UU!\DD!\RR!}%
  \fi\fi\fi\fi
  \continue}

\def\skipnexttoken!#1{\vvsq!}

\def\place#1#2#3{\vbox to 0pt{\vss
\rlap{\kern#1\squaresize
  \raise#2\squaresize\hbox{$#3$}}
\vss}}

\def\Young#1{\LL!=\thickness \UU!=\thickness \DD! = \thickness \RR! =
\thickness \vbox{\smallskip\offinterlineskip
\halign{&\vvsq! ##
\endvsq!\cr #1}}}

\def\blank{\omit\hskip\squaresize}
\catcode `!=12

\begin{document}

\sloppy
%\vspace*{-1.7cm}
\maketitle
 \vspace*{-.7cm}
 \tableofcontents
 
 \newpage

 \vspace*{-3cm}

\begin{abstract}
This paper studies random lozenge tilings of general non-convex polygonal regions. We show that the pairwise interaction of the non-convexities leads asymptotically to new kernels and thus to new statistics for the tiling fluctuations. 
 The precise geometrical figure here consists of a hexagon with cuts along opposite edges. For this model we take limits when the size of the hexagon and the cuts tend to infinity, while keeping certain geometric data fixed in order to guarantee interaction beyond the limit. We show in this paper that the kernel for the finite tiling model can be expressed as a multiple integral, where the number of integrations is related to the fixed geometric data above. The limiting kernel is believed to be a universal master kernel. 
  
 %This type of limiting kernel arises in tiling models having pairwise interaction of non-convexities and, since many other statistics should be obtainable by proper scaling limit, w
 
  %the so as such that the cuts keep interacting
 
% . The asymptotics can be performed due to the fact that the kernel   manageableTaking a limit when the size of the hexagon and the cuts tend to infinity
 
 %The kernel for this model is also the inverse Kasteleyn matrix for the corresponding dimer problem.

\end{abstract}

 \section{Introduction and main results}

The purpose of this paper is to study random lozenge tilings of non-convex polygonal regions.
Non-convex figures are particularly interesting due to the appearance of new statistics for the tiling fluctuations, caused by the non-convexities themselves or by the interaction of these non-convexities. The final goal will be to study the asymptotics of the tiling statistical fluctuations in the neighborhood of these non-convexities, when the polygons tend to an appropriate scaling limit.

The tiling problems of hexagons by lozenges goes back to the celebrated  1911-formula on the enumeration of lozenge tilings of hexagons of sides $a,b,c,a,b,c$ by the Scottish mathematician MacMahon \cite{McMa}. This result has been extended in the combinatorics community to many different shapes, including non-convex domains, in particular to shapes with cuts and holes; see e.g., Ciucu, Fischer and Krattenthaler \cite{Ciucu,Krat2}.

Tiling problems have been linked to Gelfand-Zetlin cones by Cohn, Larsen, Propp \cite{Cohn}, and to non-intersecting paths, determinantal processes, kernels and random matrices by Johansson \cite{Jo01b,Jo02b,Jo03b}. In \cite{Jo05b}, Johansson showed that the statistics of the lozenge tilings of hexagons was governed by a kernel consisting of discrete Hahn polynomials; see also Gorin\cite{Gorin}. In \cite{Jo05c} and \cite{JN}, it was shown that, in appropriate limits, the tiles near the boundary between the frozen and stochastic region (arctic circle) fluctuate according to the Airy process and near the points of tangency of the arctic circle with the edge as the GUE-minor process.  

Tiling of non-convex domains were investigated by Okounkov-Reshetikhin \cite{OR} and Kenyon-Okounkov \cite{KO} from a macroscopic point of view. Further important phenomena for nonconvex domains appear in the work of Borodin, Gorin and Rains \cite{BGR}, Defosseux \cite{Defos}, Metcalfe \cite{Metc}, Petrov \cite{Petrov1,Petrov}, Gorin \cite{Gorin1}, Novak \cite{Nov}, Bufetov and Knizel \cite{BuK}, Duse and Metcalfe \cite{Duse,Duse1}, and Duse, Johansson and Metcalfe \cite{DJM}; see also the recent paper by Betea, Bouttier, Nejjar and Vuletic \cite{BBNV}.

The present study consisting of two papers leads to the so-called {\bf Discrete Tacnode Kernel} (\ref{Final0}), which we believe to be a master kernel, from which many other kernels can be deduced (see Fig.1); namely, 
\newline (1) the GUE-tacnode kernel for overlapping Aztec diamonds \cite{AJvM,ACJvM,AvM} (also a non-convex geometry), when the size of the overlap remains small compared to the size of the diamonds. See also coupled GUE-matrices \cite{AvM}.% It is not at all clear how this kernel relates to the master kernel
\newline (2) the Tacnode kernel in the context of  colliding Brownian motions and double Aztec diamonds. 
\newline (3) The cusp-Airy kernel  (\cite{DJM}) should also be a scaling limit of the Discrete Tacnode Kernel (\ref{Final0}), etc...

%Our present study is motivated by a new statistics which came up  It also appears in the context of GUE-matrices with a certain coupling. 

%In this paper, we address the question of their asymptotics when the size of the polygon tends to infinity together with two cuts along opposite sides. It is our belief that this is universal statistics, more general than the tacnode-GUE statistics for overlapping Aztec diamonds and that a number of other statistics can be obtained from this one.

\vspace*{2cm}
\setlength{\unitlength}{0.015in}\begin{picture}(100, 0) 
\put(  150,  35){\makebox(0,0) { $\fbox{$  \begin{array}{ccccc}
& \mbox{discrete-tacnode kernel ${\mathbb L}^{\mbox{\tiny dTac}}$} 
\\& \mbox{for non-convex hexagons} 
\end{array}
 $ }$}}
 \put(100,   10){\vector(-0.5, -1){20}}
 \put(160,   10){\vector( 0.7, -1){40}}
  \put(220,   10){\vector( 1.2, -1){60}}
   \put(140,   10){\vector( .2, -1){20}}
 
 \put(  60,  -65){\makebox(0,0) {
  $\fbox{$  \begin{array}{ccccc}
& \mbox{GUE-tacnode for} 
\\& \mbox{double Aztec-diamonds} 
\end{array}
 $ }$
 }}

\put(  230,  -65){\makebox(0,0) {$\fbox{Cusp-Airy kernel}$
}}

\put(  330,  -65){\makebox(0,0) {$\fbox{tacnode kernel}$
}}

\put(  180,  -115){\makebox(0,0) {$\fbox{Pearcey kernel}$
}}

  \end{picture}
  %$$.\label{Fig1}$$
 
 %\ref{Fig1}
 \vspace*{5cm}
  Fig.1. Is the statistics associated with the discrete-tacnode kernel for non-convex hexagons universal ?  Does it imply in some appropriate limit all these known statistics? Is it a master-kernel?
  
  \vspace*{.7cm}

   %@@@@@@

This led us to investigate determinantal processes for lozenge tilings of fairly general non-convex polygons, with non-convexities facing each other. %The present work is motivated by the question whether tuning the sizes of the polygons and the non-convexities might lead to some new statistics. 
 This is going much beyond Petrov's work \cite{Petrov} on the subject, and yet inspired by some of his techniques. 
 Consider a hexagon with several cuts as in Fig.2, and a tiling with lozenges of the shape as in Fig.3, colored blue, red and green; introducing a cut amounts to covering a region with red tiles. Note there is an affine transformation from our tiles to the usual ones in the literature; see e.g. the simulation of Fig.5.  The right-leaning blue tiles turn into our blue ones, the up-right red ones into our red ones and the left-leaning green tiles (30${}^o$) to our green tiles (45${}^o$), all as in Fig.2. 
 
 Two different determinantal pocesses, a ${\mathbb K}$-process and an ${\mathbb L}$-process, will be considered, depending on the angle at which one looks at the polygons; south to north for the ${\mathbb K}$-process or south-west to north-east for  the ${\mathbb L}$-process. In this series of two papers, the first one will focus on the 
${\mathbb K}$-process and its kernel, and the second one \cite{AJvM2} on the ${\mathbb L}$-process, its kernel and its asymptotic limit in between the non-convexities.  Nevertheless both processes will be introduced in this paper.

 \newpage
 
   \setlength{\unitlength}{0.017in}\begin{picture}(0,60)
\put(115,-70){\makebox(0,0) {\rotatebox{0}{\includegraphics[width=170mm,height=225mm]
 {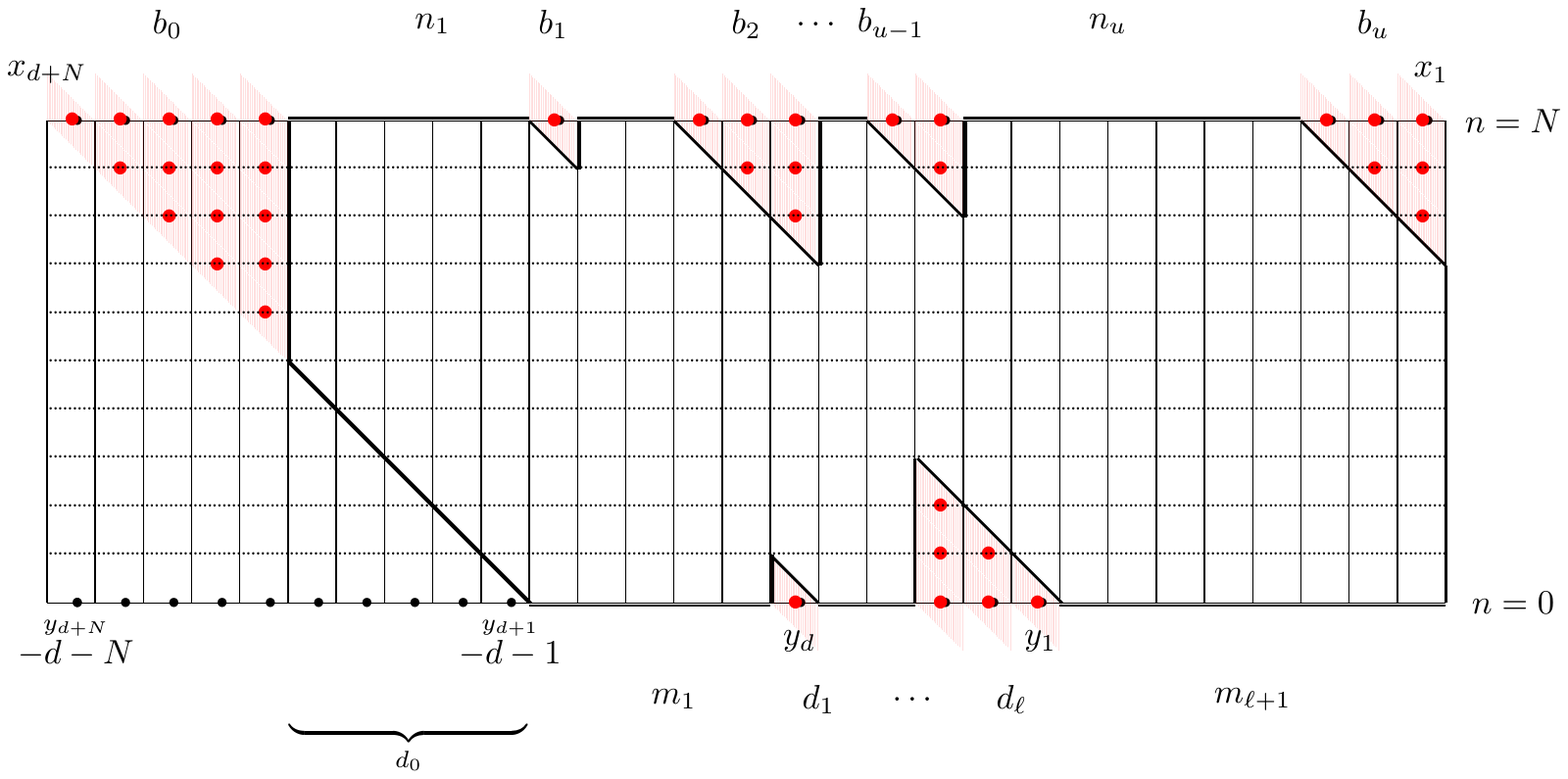} }}}
 \end{picture}
 
 \vspace{.9cm}
 
 \noindent Fig. 2\label{Fig1}: A non-convex polygon $\bf P$  (hexagon with $u+\ell-1$ cuts), and the hexagon $\widetilde{\bf P}={\bf P}\cup \{u+\ell+1 \mbox{  red triangles}\}$. ({\em multicut model})
 
 \vspace{.7cm}

% We will be dealing with three different models:
%\newline $\bullet$  {\bf Multi-cut model}: many cuts along the upper- and the lower-edge.
%\newline $\bullet$  {\bf Lower one-cut model}: many cuts on the upper-edge and one cut on the lower-edge.
%\newline $\bullet$  {\bf Two-cut model}: one cut on the upper-edge and one on the lower-edge.
 
 %\medbreak
 \noindent    A good part of the work will consist of reducing the number of integrations in the ${\mathbb K}$-kernel to $r+3$, where $r$, an integer defined in (\ref{rs}), relates to the geometry of the polygon and of the ${\mathbb L}$-process. In the second paper, the ${\mathbb L}$-kernel will require many more transformations in order to be in the right shape to perform asymptotics. As a preview of the second paper \cite{AJvM2} we merely state here the form of the ${\mathbb L}$-kernel and the asymptotics without proof. Incidentally, the ${\mathbb K}$-kernel should also lead to interesting open questions related to the Gaussian Free Field (Petrov \cite{Petrov1}) and also to open questions related to Petrov's \cite{Petrov1} and Gorin's \cite{Gorin1} work; see comments after Theorem \ref{Kr}.

%The first paper shows the ${\mathbb K}$-process is determinantal and we give its kernel and in the second paper we give the kernel of the ${\mathbb L}$-process and its asymptotic limit

%We believe the  ${\mathbb K}$-kernel is a {\em master-kernel}, from which many new and old kernels can be deduced in appropriate asymptotic limits.

 %\newpage

  To be precise, and as shown in Figure 2, we consider a {\bf general non-convex polygonal region $\bf P$ (multicut model)} consisting of taking a hexagon where two opposite edges have cuts, $ u-1$ cuts $b_1,~b_2,\dots,b_{ u-1}$ cut out of the upper-part and $\ell$ cuts $d_1,~d_2,\dots,d_{\ell}$ cut out of the lower-part\footnote{The $b_i$ and $d_i$'s also denote the size of the cuts.}; let $d:=\sum_{1}^\ell  d_i  $. Let $b_0$ and $b_{  u}$ be the 
  \noindent ``cuts" corresponding to the two triangles added to the left and the right of {\bf P} and let $d_0$ be the size of the lower-oblique side. Then $N:= b_0+d_0$ is the distance between the lower and upper edges. The intervals separating the upper-cuts (resp. lower-cuts) are denoted by $n_i$ (resp. $m_i$) and we require them to satisfy $\sum_1^{\ell+1} m_i=\sum_1^{  u} n_i$, which is equivalent to $\sum_0^{  u }b_i= d +N$.  Define $\widetilde{\bf P}$ to be the quadrilateral (with two parallel sides) obtained by adding the red triangles to ${\bf P}$, as in Fig.1.

Introduce the {\bf coordinates} $(n,x)\in \BZ^2$, where $n=0$ and $n=N$ refer to the lower and upper sides of the polygon, with $x$ being the running variable along the lines $n$. The vertices of $\bf P$ and $\widetilde{\bf P}$ all belong to the vertical lines $x=\{\mbox{half-integers}\}$ of the grid (in Fig.2). The $d=\sum_1^\ell d_i$ integer points in $\{\widetilde{\bf P}\backslash {\bf P}\}\cap \{n=0\}$ are labeled by $y_1>\ldots>y_d$; they are the integers in the cuts. We complete that set with the integer points to the left of $\widetilde{\bf P}$ along $\{n=0\}$; they are labeled by $y_{d+1} >\dots>y_{d+N} $ and we set $y_{d+1}=-d-1$ and $y_{d+N}=-d-N$. This fixes the origin of the $x$-coordinate. Similarly, the integer points $\{\widetilde{\bf P}\backslash {\bf P}\}\cap \{n=N\}$ are labeled by $x_1>\dots>x_{d+N}=-d-N$. We assume that $x_i\geq y_i$ for all $1\leq i\leq d+N$,  and that $y_d\notin \{\mbox{$x$-coordinates of an upper-cut}\}$. 

    Besides the $(n,x)$-coordinates, another set of coordinates $(\eta,\xi)$ will also be convenient (see Fig. 3):
\be\eta=n+x+\tfrac 12,~~~\xi=n-x-\tfrac 12 ~~~~~ \Leftrightarrow ~~~n =\tfrac 12(\eta+\xi),~~x=\tfrac 12 (\eta-\xi-1) .\label{Lcoord}\ee
%
%
%Define $ c:=\max \{i ~ |~x_i> y_d\}$, $b:=N-c$ and the set $\RR=\{x_1,\dots,x_c\}$; assume that both $y_1-x_c,~y_1-y_d\leq N-1$. It follows that $x_{c+1}<y_d< x_c$. 
%
Assuming $y_1-y_d\leq N-1$, we define polynomials\footnote{\label{ft1} For any integers $k\in \BZ$ and $N\geq  0$ we have $k_{0}=1$ and $k_{ N }=k(k+1)\dots(k+N-1)$.}:
\be \label{P} \mbox{$P(z) :=(z-y_d+1)_{N-(y_1-y_d+1)}~~\mbox{and}  ~~
Q(z) :=\prod_1^{d+N}(z-x_i)$}. \ee
 The roots $x_i$ of $Q(z)$, compared to the roots $y_1-N+1<\dots<y_d-1$ of $P(z)$, can be subdivided into three sets, the $\LR$(eft), the $\RR$(ight), the $ {\mathcal C}$(enter), and a set $\GR$(ap) not containing any $x_i$  :
  \be\bl
  \LR&:=\{x_i~,~\mbox{such that}~x_i<y_1-N+1\}\\
  \RR&:=\{x_i~,~\mbox{such that}~x_i\geq y_d \}\\
  {\mathcal C}&:=\{x_1,\dots,x_{d+N}\}\backslash (\LR\cup\RR)
  \\
\GR&:=\{y_1-N+1,\dots,y_d-1\}\backslash {\mathcal C},\label{LRB}  \el
  \ee
  ensuing the decompositions in polynomials
  \be P(z)=Q_{\mathcal C}(z)P_{\GR}(z)
  \mbox{,    and   ~~} Q(z)=Q_{\LR}(z)Q_{\mathcal C}(z)Q_{\RR}(z)    .\label{PQdecomp}\ee
  The number $r$, assumed positive, will play an important role:
  \be r:=|\LR|-d.  %~~~\mbox{and}~~~s:=|\RR|-d.
  \label{rs}
  \ee 
  Referring to contour integration in this paper, the notation $\Ga(\mbox{set of points})$ will denote a contour encompassing the points in question and no other poles of the integrands; e.g., contours like  
  \be
  \Ga(\RR), ~\Ga({\LR}), ~\Ga(x+\mathbb N),\dots.
  \label{GaRL}\ee
  
 %\be\mbox{``the contours $\Ga_\RR$ and $\Ga_\LR$ in $\BC$, encompassing the points of $\RR$ and $\LR$ only",}
%\label{GaRL}\ee
%and no other points of the integrands. The same holds for the notation $\Ga(\mbox{set of points})$.
  
     {\bf The ${\mathbb K}$ and ${\mathbb L}$-processes}. Given a covering of this polygonal shape with tiles of three shapes, colored in red, blue and green tiles, as in Fig.2, put a {\bf red and blue dot} in the middle of the {\bf red and blue tiles}. The {\bf red dots} belong to the intersections of the vertical lines $x=$ {\em integers} and the horizontal lines $n=0, \ldots , N$; they define a point process $(n,x)$, which we call the {\bf ${\mathbb K^{\mbox{\tiny red}}}$-process}. The initial condition at the bottom $n=0$ is given by the $d$ {\em fixed red dots} at integer locations in the lower-cuts, whereas the final condition at the top $n=N$ is given by the  $d+N$ {\em fixed red dots} in the upper-cuts, including the red dots to the left and to the right of the figure, all at integer locations. Notice that the process of red dots on $\widetilde {\bf P}$ form an interlacing set of integers starting from $d$ fixed dots (contiguous for the two-cut and non-contiguous for the multi-cut model) and growing linearly to end up with a set of $d+N$ (non-contiguous) fixed dots. This can be viewed as a ``truncated" Gel'fand-Zetlin cone!

 The {\bf blue dots} belong to the intersection $\in {\bf P}$ of the parallel oblique lines $n+x =k-\tfrac 12$ %for integer $-d+1\leq k\leq m_1+m_2+b-1$ 
  with the horizontal lines $n=\ell-\tfrac 12$ for $k,\ell \in \BZ$
 % $1\leq \ell\leq N$
 ; in terms of the coordinates (\ref{Lcoord}), the blue dots are parametrized by $(\eta,\xi )=(k,2\ell-k-1)\in \BZ^2$, with $( k,\ell)$   as above. It follows that the $(\eta,\xi)$-coordinates of the blue dots satisfy $\eta+\xi =1,3,\dots,2N-1$.
 This point process defines the {\bf ${\mathbb L}^{\mbox{\tiny blue}}$-process}. The blue dots on the oblique lines also interlace, going from left to right, but their numbers go up, down, up and down again. 
 The number of blue dots per oblique line $n+x =k-\tfrac 12$ is given by the difference between the heights computed at the points $n=0$ and $n=N$ along that line ; see Fig.7. A special feature appears in the two-cut model.
 
 One could also consider an {\bf ${\mathbb L}^{\mbox{\tiny green}}$-process}, by putting a green dot on the green tiles. It would lead to a a determinantal process of green particles on the vertical lines $x=$half integers, where the number $r$ in (\ref{rs}) would be replaced by $s= |\RR|-d$ ; see Fig.4. In the end, this kernel would be similar to the  ${\mathbb L}^{\mbox{\tiny blue}}$-kernel.%: their numbers along the oblique lines $x+n=-\tfrac 12+k$ (i.e., $\eta=k$) in the strip $\{\rho\}$ (defined just below) will exactly equal $r:=b-d$, a local minimum along those oblique lines; this will be shown later.
 
%Similarly the {\bf green dots} belong to the intersection $\in {\bf P}$ of the parallel vertical lines $x =k-\tfrac 12$ %for integer $-d+1\leq k\leq m_1+m_2+b-1$ 
%  with the horizontal lines $n=\ell-\tfrac 12$ for $k,\ell \in \BZ$. This point process will define the {\bf ${\mathbb L}^{\mbox{\tiny green}}$-process}.  ??????

As it turns out, the two kernels ${\mathbb K}^{\mbox{\tiny red}}$ and ${\mathbb L}^{\mbox{\tiny blue}}$ are highly related, as will be shown in \cite{AJvM2}, where also the asymptotics for the ${\mathbb L}$-kernel will be carried out.  Indeed, 
both point processes can be described by dimers on the points of the associated bipartite graph dual to ${\bf P}$; the two kernels are related by the inverse Kasteleyn matrix of the bipartite graph.  The aim of this first paper is to find a kernel for the ${\mathbb K}^{\mbox{\tiny red}}$-process, which involves no more than $r+3$ integrations which is a {\em conditio sine qua non} for taking asymptotics for ${\mathbb L}^{\mbox{\tiny blue}}$ while keeping $r$ fixed. As mentioned, the  ${\mathbb L}^{\mbox{\tiny blue}}$-kernel will require many more contour changes to do so in \cite{AJvM2}. This kernel could also be used for showing the Gaussian Free Field result in the bulk, again keeping $r$ fixed.

% I would explain very briefly explain the two-cut model

  The following rational function,  
  \be h(u):=\frac{  Q _{ \RR }(u)}{  P_\GR(u  ) Q_\LR(u )},\label{h}\ee
  %
 % \newpage
 %\vspace*{-.3cm}  
%\noindent
appears crucially in the following $k$-fold contour integral\footnote{Set $\Om^+_0(v,z)=1$. A shorthand notation for the Vandermonde is $\Dt_k(u):=\Dt_k(u_1,\dots,u_k)=\prod_{1\leq i<j\leq k}(u_i-u_j)$.} for $k\geq 0$, (see (\ref{GaRL}))
 \be\begin{aligned} 
 \Om^\pm_k (v,z)&:=\left(\prod_{\al=1}^k  \oint_{\Gamma(\LR)}
\frac{du_\al  h(u_\al) (z-u_\al)^{\pm 1} }{2\pi \I ~(v-u_\al)  }
    \right)
  \Dt^2_k (u) \left\{\bl &\widetilde E^{({\bf y}_{\mbox{\tiny cut}})}_g ) (u_1,\dots,u_k)\\
                              &\widetilde E^{({\bf y}_{\mbox{\tiny cut}})}_g  (z;u_1,\dots,u_{k} )\el\right. .
 \end{aligned}\label{Omr}\ee
where $\widetilde E^{({\bf y}_{\mbox{\tiny cut}})}_g$ is a symmetric function of the variables $u_1,\dots,u_k$, which depend on the integer points $y_d<\dots<y_1$ in the lower-cuts, with $g=y_1-y_d-d+1=$ the number of gaps in that sequence. The precise symmetric function will be given later in (\ref{detV}) and (\ref{symcompl}). When that sequence is contiguous, we have $g=0$ and the symmetric function equals $1$; e.g., this is so when ${\bf P}$ has one lower-cut.

   \begin{theorem} \label{Kr}%Under the same condition as in previous theorem, 
 For $(m,x)$ and $(n,y)\in {\bf P}$, the determinantal process of red dots is given by the kernel, involving at most $r+2$-fold integrals, with $r$ as in (\ref{rs}). %${\mathbb K}$-kernel is given by: %(remember $y_1=m_1-1$)
  \be \begin{aligned}
  & {\mathbb K}^{\mbox{\tiny red}} (m,x;n,y) 
 \\&= -\frac{(y-x+1)_{n-m-1}}{(n-m-1)!}\Id_{n>m}\Id_{y\geq x}
%\\
%&
 +  \frac{(N - n  )!}{(N\!-\!m\!-\!1)!}
\oint_{ {\Gamma( {x +{\mathbb N}}) }}  \!\!\!\!
 \frac{dv~  (v\!-\!x\!+\!1)_{N-m-1}   }{2\pi \I Q _{ \RR }(v)Q _{ \CR }(v)
      }\times 
\\
&%\qquad 
 ~ 
\left( \oint_{\Gamma_{\infty}}  
  \frac{dz  ~ Q _{ \RR }(z)Q _{ \CR }(z)}{2\pi \I(z\!-\!v)(z\!-\!y)_{N\!-\!n\!+\!1}}   
 \frac{  \Om^+_r (v,z)}{  \Om^+_r (0,0)} 
 +\!\tfrac{1}{r\!+\!1}\!\oint_{ \Ga_{ \tau%y_1\geq y+n  
  }}  
  \frac{dz ~ P(z)Q _{ \LR }(z)}{2\pi \I (z\!-\!y)_{N-n+1}}   
 \frac{ \Om^-_{r+1} (v,z)}{  \Om^+_r (0,0)} 
 \right),
  % \mbox{    for } y+n\geq m_1 ,
%  \end{aligned}\label{Kernlim4}\ee
 % 
%   \be \begin{aligned}
%   {\mathbb K} (m,x;n,y) 
  \end{aligned}\label{Kernlim4}\ee
  where
\be\begin{aligned}
\Gamma(x+{\mathbb N})&:=\mbox{contour containing the set}~  x+{\mathbb N}=\{x,x+1,\ldots\}
\\
\Gamma_{\infty}&:=\mbox{very large contour containing all the poles of the $z$-integrand}
\\
\Ga_{\tau }&:=\Ga(y+n-N,\dots,\min(y_1 -N,y) )\Id_{ \tau<0},\mbox{    with  }\tau:=(y+n)-(y_1+1).
%\\&=\Ga(x_{c+d}-\rho+\tau,\dots,\min(x_{c+d}-\rho-1,y))\Id_{\tau<0}.%\mbox{  with $y_1=$ }\left\{\begin{array}{lll}\mbox{most right point}
%\\
%\mbox{in the lower-cut}
%\end{array}\right.
\end{aligned}
\label{cont0}\ee
%\noindent The kernel ${\mathbb K}$ for the multi-cut case will be given in Proposition \ref{Kernlim} of section 7 
\end{theorem}

It is an interesting open problem to investigate the correlation function $\det ({\mathbb K}^{\mbox{\tiny red}} (m_i,x_i;n_j,y_j) )_{1\leq i,j\leq k}$ in the bulk (liquid region) and its limit when the size of the figure tends uniformly to $\infty$. For a configuration with cuts on one side only, Petrov \cite{Petrov1} has shown that the limit of the correlation function is given by the correlation of the incomplete beta-kernel for a ``slope" satisfying the Burger's equation (translation invariant Gibbs measure); these were introduced in \cite{OR1}. Gorin \cite{Gorin1} goes beyond by showing that the result remains valid if one allows the location of the cuts on one side to be random, yielding a measure in the limit. For a figure with two-sided cuts (above and below), what is the analogue of the incomplete beta-kernel and the slope? The present project actually deals with a very different limit, as will be explained below.

The ${\mathbb L}  ^{\mbox{\tiny blue}}$-kernel will be given in Theorem \ref{Theo:L-kernel} below, but the proof will appear in another paper \cite{AJvM2}.   
It is not clear how to obtain the ${\mathbb L}^{\mbox{\tiny blue}}$-kernel from scratch, due to the intricacy of the interlacing pattern, mentioned earlier. Therefore we must first compute the ${\mathbb K}^{\mbox{\tiny red}}$-kernel and then hope to compute the ${\mathbb L}^{\mbox{\tiny blue}}$-kernel by an alternative method. Indeed, we check that the inverse Kasteleyn matrix of the dimers on the associated bipartite graph dual to ${\bf P}$ coincides with the ${\mathbb K}$-kernel. This leads us to the following statement to appear in \cite{AJvM2}:

\begin{theorem}\label{Theo:L-kernel}
The ${\mathbb L} ^{\mbox{\tiny blue}}$-process of blue dots and the $ {\mathbb K}^{\mbox{\tiny red}}$-process of red dots have kernels related as follows:
%
%in the coordinates $(\xi, \eta)$ is given in terms of 
%as follows:
%
% is given by the kernel in terms of the coordinates $(\xi, \eta)$, as in (\ref{Lcoord})  :
\be\begin{aligned} {\mathbb L}^{\mbox{\tiny blue}}(\eta &,\xi ;\eta',\xi') 
 =-{\mathbb K}^{\mbox{\tiny red}}\left(m -\tfrac 12,x ;
m' +\tfrac 12,x' \right),
% \\&=-{\mathbb K}\left(\tfrac12 (\eta_1+\xi_1-1),\tfrac12 (\eta_1-\xi_1-1);
% \tfrac12 (\eta_2+\xi_2+1),\tfrac12 (\eta_2-\xi_2-1)\right)
%\\&=-{\mathbb K}\left(m,x;n,y\right),
\end{aligned}\label{L-kernel'}\ee
%$~~~~~~~~~~~~~~~~~~~~~~m~~~~~~~~~~~~~~~~~~~~x~~~~~~~~~~~~~~~~~n~~~~~~~~~~~~~~~y$
%\newline%
where $(m,x)$ and $(m',x')$ are the same geometric points as $(\eta,\xi)$ and $(\eta',\xi')$, expressed in the new coordinates (\ref{Lcoord}); see Fig. 6.
% \be
%(m,x;n,y)= \left(\tfrac12 (\eta_1+\xi_1-1),\tfrac12 (\eta_1-\xi_1-1);
%\tfrac12 (\eta_2+\xi_2+1),\tfrac12 (\eta_2-\xi_2-1)\right)
%\label{xy-to-xi eta}\ee
%where
%$\Gamma_\tau$ is as before, with $\tau=\eta_2-m_1$. %
%
\end{theorem}

 We state the asymptotic result (without proof), suggested by the simulation of Fig.5,  for the {\bf two-cut model} (one cut below and one above) and for even $N\to \infty$.  We do not expect the multi-cut case to lead to a fundamentally new universality class in the limit, as the upper- and lower-cuts will only interact pairwise locally. For the two-cut model, we concentrate on the polygonal shape ${\bf P}$, as in Figs.2\&4, with ${\ell}=1, ~{ u}=2$ and with equal opposite parallel sides, i.e., $b_{{ u} }=d_0$ and $N-b_u =b_0$; it has two cuts of same size $d :=d_1=b_1$, one on top and one at the bottom. 
 
    For that model, the {\bf oblique strip $\{\rho\}$} extending the oblique segments of the upper- and lower-cuts within ${\bf P}$ will play an important role (see Fig.4):
%\medbreak
that is the region containing the parallel lines $x+n=-\tfrac 12 +k$ (or, what is the same, the lines $\eta=k$) for $m_1\leq k\leq n_1+b-d$. Thus the strip $\{\rho\}$ has the following width: %(we use the same notation for the {\em name and the width} of the strip!)
\be \rho:=n_1-m_1+b-d=m_2-n_2+b-d , \mbox{    setting $b:=b_0$},\label{rho}\ee
and we assume $\rho\geq 0$.  
%
% \noindent(ii) a {\bf vertical strip $\{\sg\}$} extending the vertical segments of the upper- and lower-cuts; that is the region between the lines $x = n_1-c-\tfrac 12$ and $x=m_1-d-\tfrac 12. $
%The strip $\{\sg\}$ has width (again same notation for the {\em name and the width} of the strip!)
%\be \sg:=m_1-n_1+c-d=n_2-m_2+c-d.;\label{sg}\ee
%We assume $\rho,\sg\geq 0$, which amounts to the inequalities:
%$$d-b\leq n_1-m_1=m_2-n_2\leq c-d.$$
%
It will be shown that the $\rho+1$ parallel oblique lines $\eta=k,~k\in \BZ$ within the strip $\{\rho\}$ each carry the same number $r$ of blue dots, with $r$ defined in (\ref{rs}). In the simulation of Fig.5, $\{\rho\}$ is the ``finite"  oblique strip separating the two ``large" hexagons; see also Fig.4.

  %\be \label{r}r:=|\LR|-d\mbox{    number of blue dots.}\ee 
  
   The discrete-tacnode kernel $ {\mathbb L}^{\mbox{\tiny dTac}}  ( \tau_1, \theta_1 ;\tau_2, \theta_2)$ in the variables $(\tau_i,\theta_i)\in \BZ\times \BR$ is defined by  the following expression, where the integrations are taken along  upwards oriented vertical lines $\uparrow L _{0+}$ to the right of a (counterclock) contour $\Gamma_0$ about the origin and with integer $r\geq 0$ :
  \be \label{Final0}\begin{aligned}
%(-1)^{\tau_2-\tau_1}\frac{C_t^{(2)}}{C_t^{(1)}}&{\mathbb L}(\eta_1,\xi_1;\eta_2,\xi_2,) \frac12 \Dt \xi_2
 {\mathbb L}^{\mbox{\tiny dTac}}  (&\tau_1, \theta_1 ;\tau_2, \theta_2)  
  :=  - 
{\mathbb H}^{\tau_1-\tau_2}(  \theta_2-  \theta_1) 
\\& +\oint_{\Ga_0}\frac{dV} {(2\pi\I)^2}\oint_{\uparrow L_{0+}}  \frac{ dZ}{Z-V}\frac{V^{\rho-\tau_1}}{Z^{\rho-\tau_2}}
\frac{e^{-V^2  -  \theta_1    V  }}
{e^{-Z^2 - \theta_2   Z  }}
      \frac{ \Theta_r( V, Z )} { \Theta_r(0,0)}  
\\& +\oint_{\Ga_0}\frac{dV} {(2\pi\I)^2}\oint_{\uparrow L_{0+}}\frac{ dZ}{Z-V} \frac{V^{  \tau_ 2}}{Z^{  \tau_1}}
\frac{e^{ -V^2 +  (\theta_2- \beta  )V  }}{e^{- Z^2  + (   \theta_1- \beta    )Z  }}
     \frac{ \Theta_r( V  , Z )} { \Theta_r(0,0)} 
\\& +r\oint_{\uparrow L_{0+}  }    \frac{dV} {(2\pi\I)^2} \oint_{\uparrow L_{0+}}dZ \frac{V^{ -\tau_1}}{Z^{\rho-\tau_2}}
\frac{e^{ V^2 -    (\theta_1- \beta     )V  }}
{e^{-Z^2 - \theta_2    Z  }}
    \frac{ \Theta^+_{r-1}(  V, Z )} {  \Theta_r(0,0)} 
 \\
 & -\tfrac1{r+1}\oint_{\Ga_0}\frac{dV} {(2\pi\I)^2} \oint_{\Ga_0}dZ 
 \frac{V^{ \rho-\tau_1}}{Z^{ -\tau_2}}
\frac{e^{-V^2 - \theta_1 V  }}
{e^{ Z^2 - (\theta_2- \beta    )Z  }}
    \frac{ \Theta^-_{r+1}( V, Z  )} {  \Theta_r(0,0)} =:
    \sum_0^4 {\mathbb L}^{\mbox{\tiny dTac}}_i  ,
%\\
%=&   \sum_{k=0}^4  {\mathbb L}_k^{\mbox{\tiny dTac}}
%(\tau_1, \theta_1;\tau_2, \theta_2),
\end{aligned}\ee

%The final aim of this work is to obtain the ${\mathbb L}^{\mbox{\tiny blue}}$-kernel and to show that in the asymptotic limit, a new  kernel appears, the discrete-tacnode kernel  ${\mathbb L}^{\mbox{\tiny dTac}}$, which has universal properties. 

\newpage
 
 \vspace*{-2.4cm}

\setlength{\unitlength}{0.017in}\begin{picture}(0,60)
\put(145,-70){\makebox(0,0) {\rotatebox{0}{\includegraphics[width=160mm,height=225mm]
 {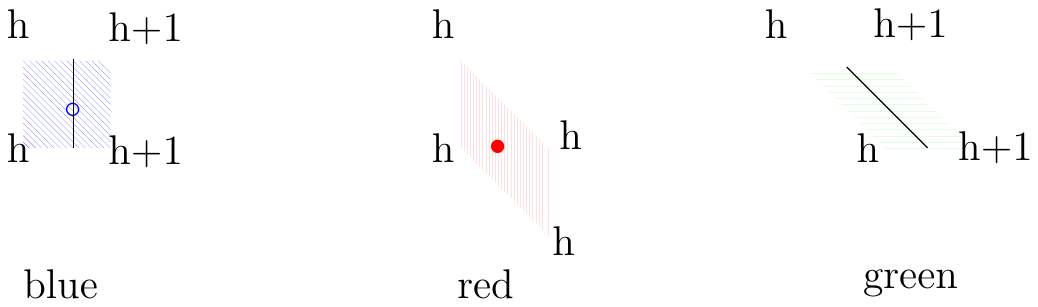} }}}

 \end{picture}
 
 \vspace*{.5cm}
Fig. 3. Three types of tiles, with the height function and with corresponding level line. The red tiles (blue tiles) have a red dot (blue dot) in the middle.

\vspace*{-4.4cm}

%\vspace*{-7cm}

\setlength{\unitlength}{0.017in}\begin{picture}(0,170)
\put(175,-70){\makebox(0,0) {\rotatebox{0}{\includegraphics[width=160mm,height=225mm]
 {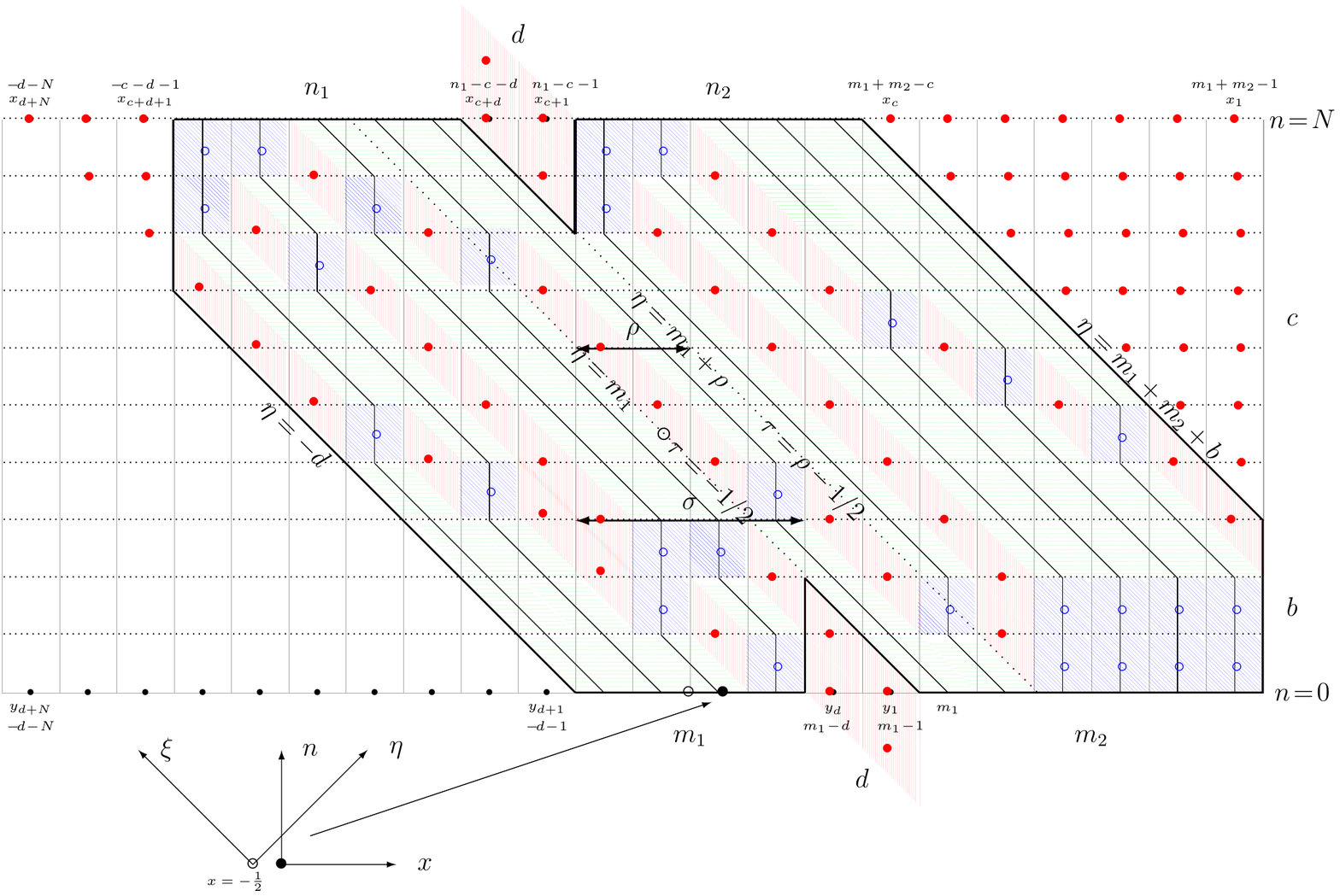} }}}
 %\put(75,-43){\makebox(0,0) {\line
 
%         \put(115,-333){\makebox(0,0) {   $
 %        \begin{aligned}&\mbox{Figure 1:  lozenge tiling of a Hexagon with red dots (${\mathbb K}$-process)}\\
%        &\mbox{and blue dots (${\mathbb K}$-process), with non-intersecting paths, with $n_1=n_2=5,~m_1=4,~m_2=6,~ b=3,~c=7,~d=2,$ and thus $~ r=1,~ \rho=2,~\sigma=4.$}
 %       \end{aligned}$   }$
        % }}

\end{picture}

\vspace*{9.8cm}

Fig. 4 \label{Fig3}: Tiling of a hexagon with two opposite cuts of equal size ({\em Two-cut model}), with red, blue and green tiles. Here $d=2$, $n_1=n_2=5,~m_1=4,~m_2=6,~ b=3,~c=7$, and thus $~ r=1,~ \rho=2,~\sigma=4.$ The $(x,n)$-coordinates have their origin at the black dot and the $(\xi, \eta)$-coordinates at the circle given by $(x,n)=(-\tfrac 12,0)$.
Red tiles carry red dots on horizontal lines $n=k$  for $0\leq k\leq N$ (${\mathbb K}$-process) and blue tiles blue dots on oblique lines $\eta=k$ for $-d+1\leq k\leq m_1+m_2+b-1$ (${\mathbb L}$-process). The left and right boundaries of the strip $\{\rho\}$ are given by the dotted oblique lines $\eta=m_1$ and $\eta=m_1+\rho$. Finally, the tilings define $m_1+m_2$ non-intersecting paths.  %Note the origin, given by a circle, has coordinates $(x_0,y_0)=(m_1-\frac N2,\frac N2-\frac 12)$ or $(\xi_0,\eta_0)=(N-m_1-1,m_1)$. The new rescaled variables $(\sg,\tau)$ will have their origin on the black circle, belonging to the dotted line $\eta=m_1$.

 \newpage

  \newpage
  
   \vspace*{-7.5cm}
  
 \setlength{\unitlength}{0.017in}\begin{picture}(0,60)
\put(145,-70){\makebox(0,0) {\rotatebox{-90}{\includegraphics[width=120mm,height=160mm]
 {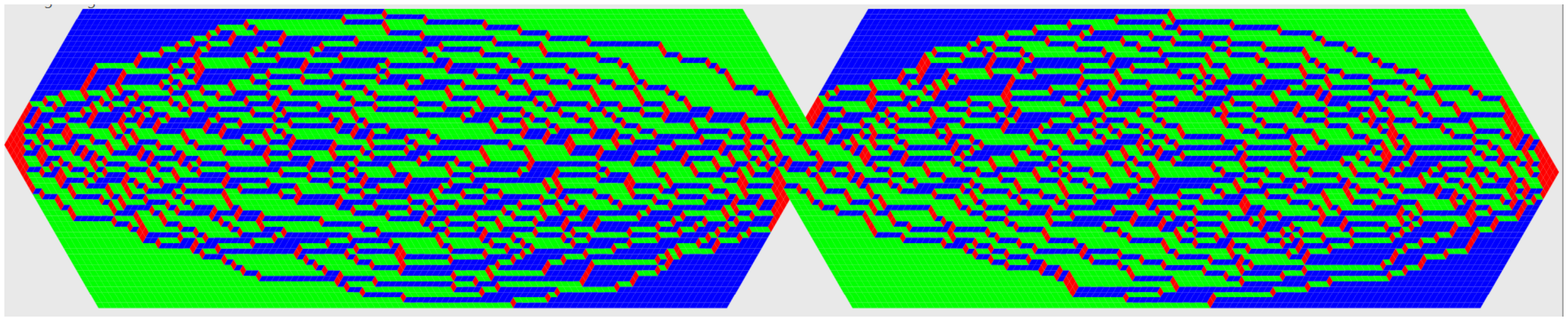}}}}
   \put( 170,-226){\makebox(0,0) {  \mbox{\footnotesize $\rho$} }}    
  \put(142, -226){\vector(1, 0){56}}
  \put(142, -226){\vector(-1, 0){.1}}
  \end{picture}
 
\vspace*{1.6cm}

  \setlength{\unitlength}{0.017in}\begin{picture}(0,60)
\put(145,-70){\makebox(0,0) {\rotatebox{-90}{\includegraphics[width=63mm,height=90mm]
 {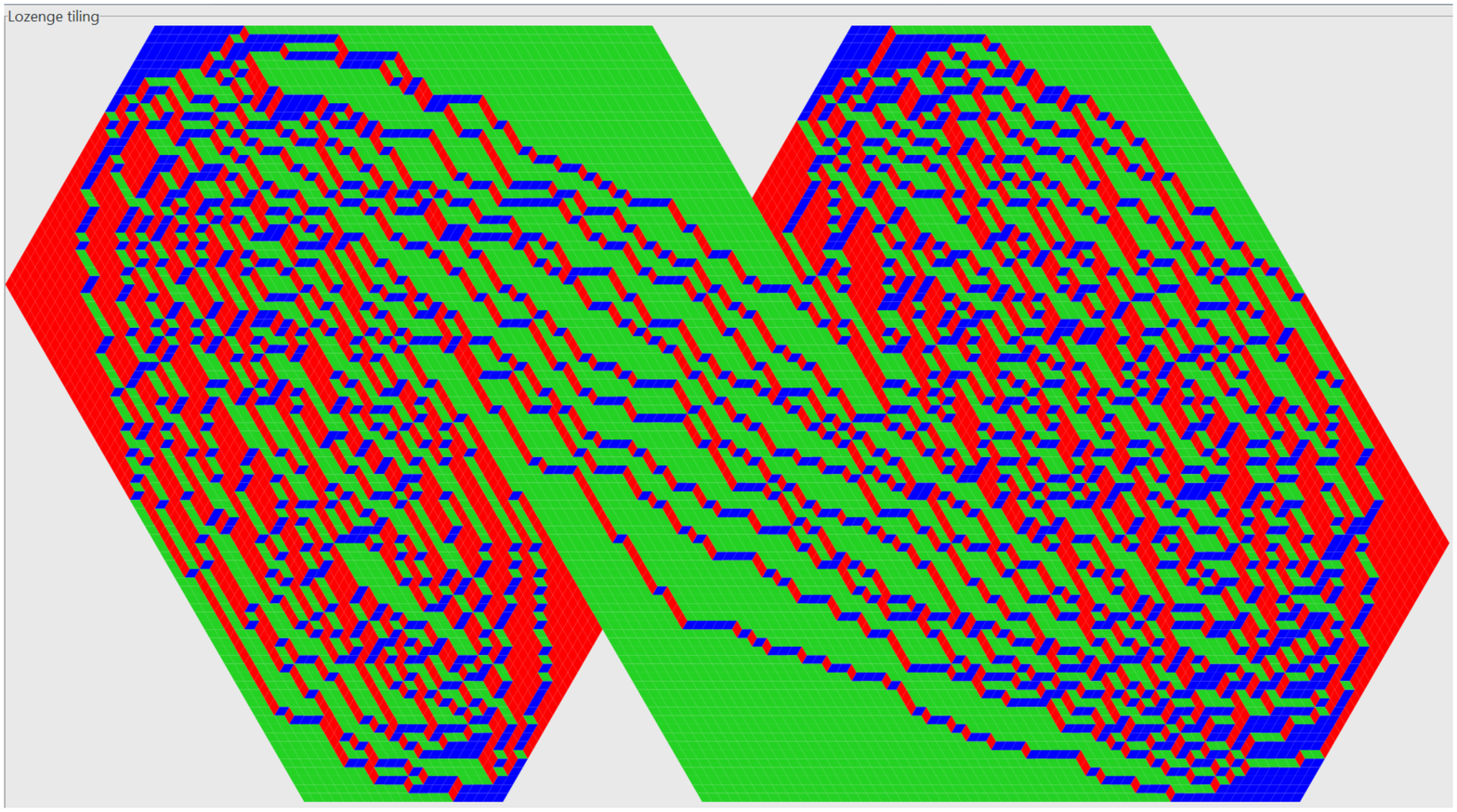}}}} \end{picture}

 %Antoine_graph_100_30_60_50_20_20-1
 \vspace*{6cm}
  \noindent Fig. 5. Computer simulation for $n_1 = 105, ~n_2=95,~m_1=m_2=100,~b=25,~c=30,~d=20$
 %n2 = 30    d = 20  b = 30  c = 70   m1 = 40 m2 = 40
 %
 and $n_1 = 50, n_2 = 30, m_1 = 20, m_2 = 60, b = 30, c = 60, d = 20$. Courtesy of Antoine Doeraene.

 % \vspace*{.7cm}
  
 %\vspace*{-3.5cm}

% \noindent

 %\vspace{-.5cm}
%\newpage

\vspace*{-.1cm}

 %\begin{figure}
  \setlength{\unitlength}{0.017in}\begin{picture}(0,170)
\put(175,0){\makebox(0,0) {\rotatebox{0}{\includegraphics[width=160mm,height=225mm]
 {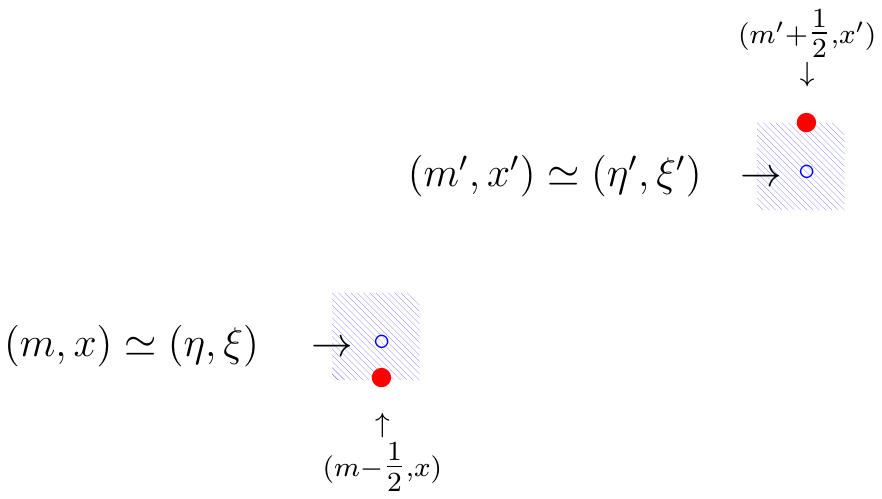} }}}

 \put(140, 50){\makebox(0,0) {$\begin{aligned}\mbox{Fig. 6.}& ~\mbox{The ${\mathbb L}^{\mbox{\tiny blue}}$-kernel of blue dots expressed in terms of the ${\mathbb K}^{\mbox{\tiny red}}$-kernel of}\\
 &\mbox{neighbouring red dots, using the $(\eta,\xi)$-variables in (\ref{Lcoord}).}
 \end{aligned} $  }}
  \end{picture}
% \end{figure}
 
 \vspace{-1.0cm}

 %
 
 %\newpage
 %
 %@@@@@@@@@@@@@@@
 %
\noindent where
  \be   \begin{aligned}
    \BH^{m}(z)&:=\frac{z^{m-1}}{(m-1)!}\Id _{z\geq 0}\Id_{m\geq 1},~~~~(\mbox{Heaviside function})
    \\
 \Theta_r( V, Z)&:=\left[     
\prod_1^r\oint_{\uparrow L_{0+}}\frac{e^{    2W_\al^2+\beta   W_\al}}{    W_\al  ^{\rho }}
 ~\left(\frac{Z\!-\!W_\al}{V\!-\!W_\al}\right) \frac{dW_\al}{2\pi \I}\right]\Dt_r^2(W_1,\dots,W_r)
  %\end{aligned}$$
%
%$$  \begin{aligned}
 \\  \Theta^{\pm}_{r\mp1}(  V, Z)     &:=\left[     
\prod_1^{r\mp 1}\oint_{\uparrow L_{0+}}\frac{e^{    2W_\al^2+\beta  W_\al}}{    W_\al  ^{\rho }}
 ~\left( ({Z\!-\!W_\al} )\ ({V\!-\!W_\al})\right)^{\pm 1} \frac{dW_\al}{2\pi \I}\right]
 \\
 &~~ ~~\qquad~ ~\qquad \qquad~~~~~~~~~~~~~~~~~~~~~\Dt_{r\mp1}^2(W_1,\dots,W_{r\mp1}).
 \label{Theta}
\end{aligned}\ee
We now state the main Theorem of this project, which will appear in \cite{AJvM2}:

% We assume $N$ even.
\begin{theorem}\label{Theorem2}
Given the polygon {\bf P} with cuts of equal size $d$, above and below, keeping $ \rho,r\geq 0$ fixed, as defined in (\ref{rho}),(\ref{rs}), let the polygon $\bf P$ and the size $d$ of the two cuts go to $ \infty$, according to the following scaling of the geometric variables $b,c,m_i,n_i>0 $, in terms of $d\to \infty$ and new parameters $1<\gamma<3$,   $a:=2\sqrt{\frac{\ga}{\ga-1}}$, $  \beta_1<0$, ~$  \beta_2, ~ \gamma_1,~ \gamma_2\in \BR$, %Define 
%$a:=2\sqrt{\frac{\ga}{\ga-1}}$.
\be\begin{array}{lllll}
b=d+r&& c=\ga d
\\
n_i = m_i-(-1)^i(\rho-r) &&  m_i=\tfrac{\ga+1}{\ga-1}
 ( d+\tfrac a2  \beta_i\sqrt{d}+ \ga_i)\mbox{  for  }i=1,2.
%\\
%n_2 = m_2-(\rho-r) &&  m_2=\tfrac{\ga+1}{\ga-1}
 %( d+\tfrac a2 \bar\beta_2\sqrt{d}+\bar\ga_2).   
\end{array} \label{geomscaling}\ee
%with 
%$$
%\alpha_i: = \frac{\gamma+1}{\gamma-1}
%~~~\mbox{and}~~\beta_i:=\tfrac a2 \frac{\ga+1}{\ga-1}\bar \beta_i \qquad \mbox{and}~~~a:=2\sqrt{\frac{\ga}{\ga-1}}
%$$
%
%
The variables $(\eta,\xi)\in \BZ^2$ with $\xi-\eta\in 2\BZ+1$ get rescaled into new variables $(\tau,\theta)\in \BZ\times\BR$, having their origin at the halfway point $(\eta_0,\xi_0)$ along the left boundary of the strip $\{\rho\} $, shifted by $(-\tfrac 12, \tfrac 12)$  (see little circle along the line $\eta=m_1$ in Fig.4): %(assume $N$ even)
\be\begin{aligned}
 (\eta_i,\xi_i) &=(\eta_0,\xi_0)+(\tau_i,~\tfrac {\ga+1}a (\theta_i +  \beta_2)\sqrt{d} )\mbox{    with   }
 %\left\{\begin{aligned}&
  (\eta_0,\xi_0)  =(m_1,N-m_1 -1).%\\  %&   \tilde \sg_i=\tfrac {\ga+1}a \sg_i \end{aligned} \right.
% \\
%  &=
\end{aligned}
\ee
With this scaling and after a conjugation, the kernel (\ref{L-kernel'}) of the ${\mathbb L}$-process tends to the  new kernel ${\mathbb L}^{\mbox{\tiny dTac}} $, as in (\ref{Final0}), depending only on the width $\rho$ of the strip $\{\rho\}$, the number $r=b-d$ of blue dots on the oblique lines in the strip $\{\rho\}$ and the parameter 
$ \beta :=- \beta_1- \beta_2$, to be precise,
%= \lim_{d\to \infty} \bigl({2d(d+c) +(m_1+m_2)(d-c)} \bigr)\tfrac{\sqrt{d^{-1}-c^{-1}}}{d+c};%@\sqrt{\frac{c-d}{cd}}
%\el\label{beta}\ee to be precise,%\ref{beta}
%
\be\begin{aligned}
 \lim_{d\to \infty} (-1)^{\tfrac 12 
   (\eta_1+\xi_1-\eta_2-\xi_2)}
%which is an integer, because \xi -\eta=odd. Also the second exponent should be \eta_2-\eta_1. } 
% (-1)^{\frac 12((\tau_1\!-\!\tau_2)+\sqrt{d}\frac{\ga+1}a(\eta_2-\eta_1) )}
&\left(\sqrt{d}\frac{\ga\!+\!1}{2a}\right)^{\eta_2-\eta_1%@@\tau_2-\tau_1
 }
{\mathbb L}^{\mbox{\tiny blue}}( \eta_1 ,\xi_1;\eta_2,\xi_2)\frac 12 \Dt\xi_2
\\
&={\mathbb L}^{\mbox{\tiny dTac}} (\tau_1, \theta_1;\tau_2, \theta_2)d\theta_2.\end{aligned}
 \label{limit}\ee 
The kernel satisfies the following involution:
$$
{\mathbb L}^{\mbox{\tiny dTac}} (\tau_1, \theta_1;\tau_2, \theta_2)
=
{\mathbb L}^{\mbox{\tiny dTac}} (\rho-\tau_2, \beta-\theta_2;\rho-\tau_1, \beta-\theta_1).
$$
This involution exchanges ${\mathbb L}_1^{\mbox{\tiny dTac}}\leftrightarrow
{\mathbb L}_2^{\mbox{\tiny dTac}}$, with  ${\mathbb L}_k^{\mbox{\tiny dTac}}$ being self-involutive for $k=3,4$. Also ${\mathbb L}_1^{\mbox{\tiny dTac}}$ has support on $\{\tau_1>\rho\}$, ${\mathbb L}_2^{\mbox{\tiny dTac}}$ has support on $\{\tau_2<0\}$ and ${\mathbb L}_4^{\mbox{\tiny dTac}}$ on $\{\tau_1>\rho\}\cap \{\tau_2<0\}$.
\end{theorem}

%\newpage
 %\newpage

% \vspace*{-.7cm}

These formulas of Theorems \ref{Kr} and \ref{Theo:L-kernel} can be specialized to known situations: to hexagons with no cuts (Johansson\cite{Jo05b}), to non-convex polygons $\bf P$ with cuts at the top only (Petrov \cite{Petrov}), and to the case where the strip $\rho$ reduces to a line (i.e., $\rho=0$), in which the polygon ${\bf P}$ (as in Fig.3) can be viewed as two hexagons glued together along one side (Duse-Metcalfe \cite{Duse}).  %Two special cases of non-convex polygons $\bf P$ have been considered in the recent literature. On the one hand, Leonid  considered an hexagon, with cuts at the top only and obtained a $\mathbb K$-kernel, which is considerably simpler than the one in Theorem \ref{Kr}. On the other hand, when the strip $\rho$ reduces to a line (i.e., $\rho=0$), then the polygon ${\bf P}$ (as in Fig.3) can be viewed as two hexagons glued together along one side. For this situation, see Duse-Metcalfe \cite{Duse}. 

\medbreak

\noindent{\bf Outline}. Many steps are necessary to prove Theorem \ref{Kr}.  

\noindent
Section 2:  Instead of putting the uniform distribution on the red dot-configuration on $\bf P$, it will be more convenient to first consider a non-uniform distribution depending on a parameter $0<q\leq 1$; this distribution will tend to the uniform one when $q\to 1$.  The red dot-configuration will be shown to be equivalent to the set of semi-standard skew-Young Tableaux of a given shape; the latter can be read off the geometry of ${\bf P}$. The $q$-probability on this set will have a Karlin-McGregor type of formula, which is not surprising due an equivalent formulation in terms of non-intersecting paths. The use of $q$-deformations have been initiated by Okounkov-Reshetikhin \cite{OR} and Kenyon-Okounkov \cite{KO}. Also it has been used effectively in the work of Petrov \cite{Petrov} for lozenge tilings of hexagons with cuts on the upper-side only. Section 2 also contains a brief description of the two-cut model.
\newline\noindent Section 3 deals with some determinantal identities, but also with a useful, but unusual, integral representation of the elementary symmetric function $h_{y-y_j}(1^n)$.
\newline\noindent Section 4: Adapting Eynard-Mehta techniques, further refined in \cite{BR} and \cite{BFP}, will lead to the construction a kernel ${\mathbb K}_q$, involving the inverse of a matrix $M$. 
\newline\noindent Section 5: The matrix $M$ can be transformed so that the inverse is readily computable. 
\newline\noindent Section 6: The transformed kernel ${\mathbb K}_q$ has a multiple integral representation, using integral representations of the different ingredients.
\newline\noindent Section 7: Taking the limit when $q\to 1$ yields a kernel ${\mathbb K}^{\mbox{\tiny red}}$, involving contour integrals about $\RR$, with at the worst $d+2$-fold integrations, where $d$ is the size of the cuts. 
\newline\noindent Section 8: The kernel ${\mathbb K}^{\mbox{\tiny red}}$ will then further be reduced  to a sum of contour integrations, mostly about $\LR$, with at the worst  $r+3$-fold integrals. 
%\newline\noindent Section 9: The polygon $\bf P$ has a natural involution, which leads to an alternative form of the kernel. Sections 8 and 9 establish Theorems \ref{Kr} and \ref{Kri}.
 %\newline\noindent Section 10: This section explains what modifications need to be done for the {\bf multi-cut model}. It gives another form of the kernel than the one found in Section 7.

%END of introduction

%\newpage

%  As shown in Figure 1, we consider a {\bf general non-convex polygonal region $\bf P$ (multi-cut model)} consisting of taking a hexagon where two opposite edges have cuts, $ u-1$ cuts $b_1,~b_2,\dots,b_{ u-1}$ cut out of the upper-part and $\ell$ cuts $d_1,~d_2,\dots,d_{\ell}$ cut out of the lower-part\footnote{The $b_i$ and $d_i$'s also denote the size of the cuts.}; let $d:=\sum_{1}^\ell  d_i  $. Let $b_0$ and $b_{  u}$ be the 
 %

  %\newpage
  
  \section{Interlacing and measures on skew-Young Tableaux}
  
 % {\bf Mark:  In this section, one has to combine the items in the old section 2 with the two-cut model}

  {\bf The two-cut model}, already mentioned before,  deserves some further discussion. Remember for the two-cut model we have $b:=b_0=|\LR|$ and $c:=b_{{ u}} $; see Fig. 2\&4. 
 In other terms, this is now a hexagon with edges of size $m_1+m_2+d,~ b,~c\sqrt{2},~n_1+n_2+d, ~b,~c\sqrt{2}$ with two cuts, one below and one above, both of same size $d$, satisfying $m_1+m_2=n_1+n_2$ and with  $N=b+c$ (see Fig.4). The quadrilateral $\widetilde {\bf P}$ associated with $ {\bf P}$ is depicted in Fig.7.  Note that the most-right point $y_1=m_1-1$ in the lower-cut plays an important role!

In {\bf the two-cut model}, two strips within ${\bf P}$ will play a special role: (Fig.4)
%\medbreak
\noindent(i) the {\bf oblique strip $\{\rho\}$}, already introduced in (\ref{rho}). %extending the oblique segments of the upper- and lower-cuts; that is the region between the lines $x+n=-\tfrac 12 +k$ or what is the same $\eta=k$  for $m_1\leq k\leq n_1+b-d$. The strip $\{\rho\}$ has width %(we use the same notation for the {\em name and the width} of the strip!)
%\be \rho:=n_1-m_1+b-d=m_2-n_2+b-d,\label{rho}\ee
%and we assume that $\rho\geq 0$. 

 \noindent(ii) a {\bf vertical strip $\{\sg\}$} extending the vertical segments of the upper- and lower-cuts; that is the region between the lines $x = n_1-c-\tfrac 12$ and $x=m_1-d-\tfrac 12. $
The strip $\{\sg\}$ has width (again same notation for the {\em name and the width} of the strip!)
\be \sg:=m_1-n_1+c-d=n_2-m_2+c-d\geq 0.\label{sg}\ee

\newpage

\vspace*{-3cm}

 \setlength{\unitlength}{0.017in}\begin{picture}(0,60)
\put(145,-70){\makebox(0,0) {\rotatebox{0}{\includegraphics[width=115mm,height=160mm]
 {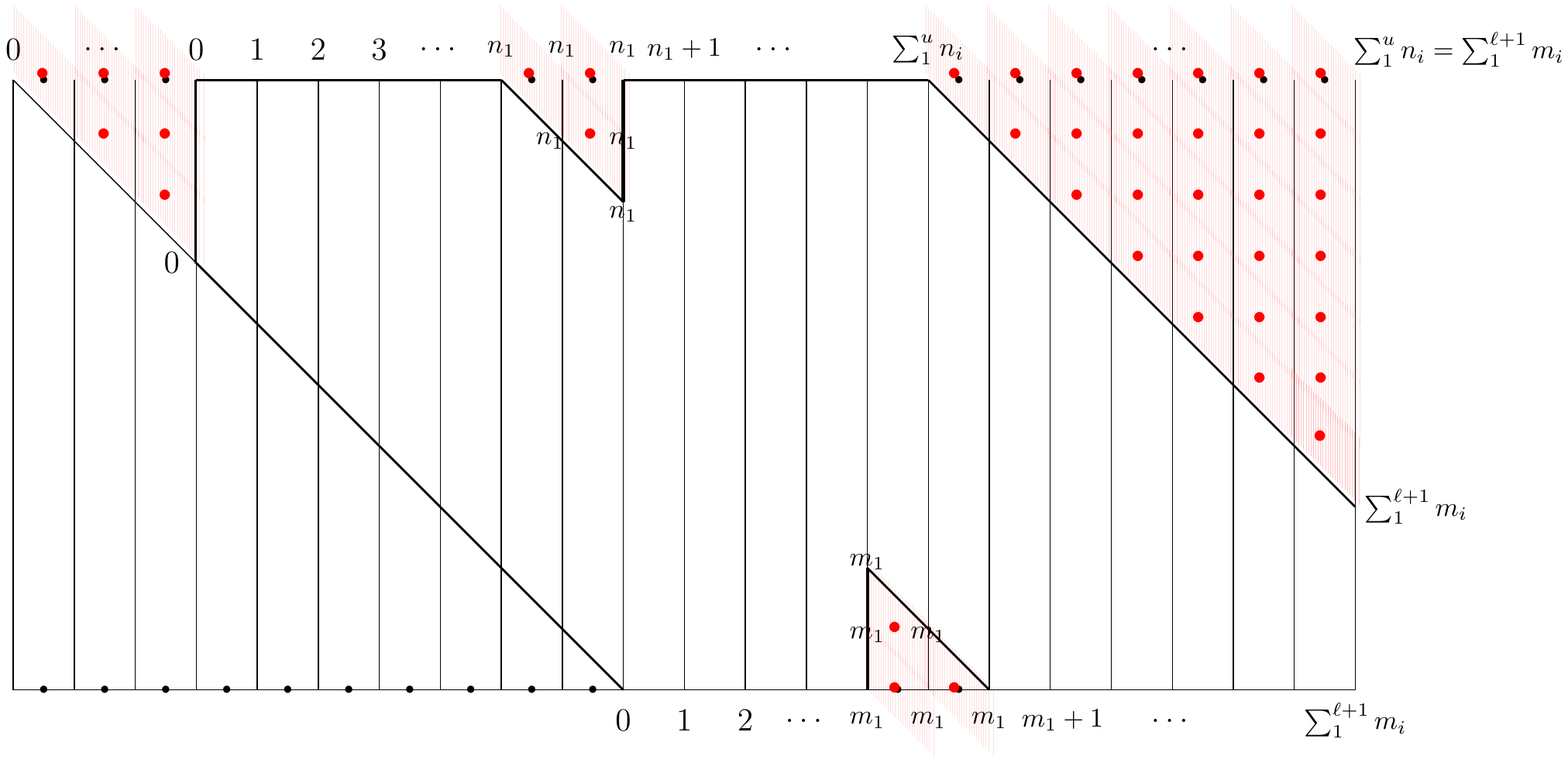}}}}

 \end{picture}
 
   \vspace{4.8cm}
  
  Fig. 7. The polygon $\bf P$, with $u-1$ upper-cuts and $\ell$ lower-cuts and the parallelogram $\widetilde {\bf P}={\bf P}\cup \{4\mbox{ red triangles}\}$. The heights along the boundary of ${\bf P}$ are given by the numbers next to the figure.

 \vspace*{.55cm}

\noindent That $\rho,\sg\geq 0$ amounts to the inequalities:
$$d-b\leq n_1-m_1=m_2-n_2\leq c-d.$$

It is natural to assume that the strips $\{\rho\}$ (respectively $\{\sg\}$) have no point in common with the vertical parts (respectively oblique parts) of the boundary $\pl {\bf P}$. This condition for $\{\rho\}$ 
implies $\{x+n=-\tfrac 12+m_1\}\cap \{n=N\}> -c-d-\tfrac 12$ and 
 $\{x+n=-\tfrac 12+n_1+b-d\}\cap \{n=0\}< m_1+m_2-\tfrac 12$, implying $d-b> \max(-m_1,-n_2)$. The condition for the strip $\{\sg\}$ implies $n_1-c-\tfrac 12> -d-\tfrac 12$ and $m_1-d-\tfrac 12<m_1+m_2-c-\tfrac 12$, implying $c-d< \min(n_1,m_2)$. These inequalities combined with condition $\rho, \sg\geq 0$ above 
 imply 
 %
 %
%For future use in {\bf the two-cut model}, two strips within ${\bf P}$ will play a role: (see Fig. 3)
%\medbreak
%\noindent(i) an {\bf oblique strip $\{\rho\}$} extending the oblique segments of the upper- and lower-cuts; that is the region between the lines $x+n=-\tfrac 12 +k$ or what is the same $\eta=k$  for $m_1\leq k\leq n_1+b-d$. The strip $\{\rho\}$ has width %(we use the same notation for the {\em name and the width} of the strip!)
%\be \rho:=n_1-m_1+b-d=m_2-n_2+b-d,\label{rho}\ee
%and we assume that $\rho\geq 0$. 
%
 \be
\max (-n_2,-m_1)<d-b \leq m_2-n_2=n_1-m_1\leq c-d <\min(m_2,n_1).
\label{ineq}
\ee
The four regions $\subset \mbox{line}\{n=N\}$ mentioned in (\ref{LRB}) can now be written as %consist of three separate sets of integers~  $  \backslash  \!\!\!\!\!\!\!\subset \mbox{polygon}~{\bf P}$, a  $\LR$(eft) region, an upper-${\mathcal C}$(ut)-region and the $\RR$(ight) region, containing respectively $b,~d$ and $c$ integers; in total $N+d$ integers; to wit:%Consider the three sets of integers, not belonging to the polygon ${\bf P} $, the $L$(eft), the upper-$C$(ut) and the $R$(ight)-sets :
\be\begin{aligned}\LR& =\{ x_{d+c+b},\dots , x_{d+c+1}\}
,~~
\RR  =\{ x_{c},\dots , x_{ 1}\},~~
%\\
\CR  =\{ x_{c+d},\dots , x_{ c+1}\}\\
{\mathcal G}& =\{\rho\}\cup \{\sg\},
\end{aligned}\label{LCR}\ee
and so the polynomials $P$ and $Q$, as in (\ref{PQdecomp}), spelled out,  are given by: 
\be\begin{aligned}
P(z)&  %=(z-y_d+1)_{N-d} 
 =:P_{ \rho }(z)Q_{ \CR }(z)P_{ \sg }(z)
  %\\&
  =(z-x_{d+c}+1)_\rho (z-x_{c+1})_d(z-y_d+1)_\sg
 \\
Q(z) &:=  Q_{ \LR }(z)Q_{ \CR }(z)Q_{ \RR }(z)  
%\\&
 =(z-x_{d+c+1})_b(z-x_{c+1})_d(z-x_1)_c,
\end{aligned}\label{PQ}\ee
where $P_{ \rho },~ P_{ \sg }, ~Q_{ \LR },~ Q_{ \CR },~ Q_{ \RR }$ are monic polynomials whose roots are given by the sets $\{\bar\rho\},~\{\bar\sg\}, ~ \LR,  \CR$ and $  \RR$, respectively, where $\{\bar\rho\} :=\{\rho\}\cap\{n=N\}$ and $ \{\bar\sg\}  :=\{\sg\}\cap\{n=N\}$.

%Still for the two-cut model, we define two other sets of $\sg$ and $\rho$ contiguous integers on the line $y=N$:
%$$\begin{aligned}
% \{\bar\sg\}  :=\{\sg\}\cap\{n=N\}&=\{x_{c+1}+1,\dots, y_d-1\}=\{n_1-c,\dots,m_1-d-1\}
%\\
 %  \{\bar\rho\} :=\{\rho\}\cap\{n=N\}&=  \{ x_{c+d}-\rho,\dots, x_{c+d}-1\}
 %  \\
 %  &=\{ m_1-b-c,\dots,n_1-c-d-1\}.   
%\end{aligned}$$
%
The inequalities (\ref{ineq}) imply that each of the sets $\CR\cup \{\bar\sg\}$ and $\{\bar\rho\}\cup\CR$ form a contiguous set of integers, such that each of the three sets are completely separated: $\CR<\{\bar\sg\}<\RR$ and $\LR<\{\bar\rho\}<\CR$.
\medbreak

 {\bf Nonintersecting paths, level lines and the ${\mathbb K}^{\mbox{\tiny red}}$- and ${\mathbb L}^{\mbox{\tiny blue}}$-processes}. 
 The height function on the tiles, given in Fig.3, imply that the heights along the boundary 
 %
 % \newpage
  %
 % \vspace*{-2cm}
%
 %\noindent 
  of the polygon $\bf P$ are independent of the tiling; for the heights along $\pl {\bf P}$, see Fig. 7.  The level lines of heights $\tfrac 12,~\tfrac 32 ,\dots,~ m_1-\tfrac 12,~ m_1+\tfrac 12,~\dots, \sum_1^{\ell+1} m_i -\tfrac 12
$ pass obliquely through green tiles, vertically through blue tiles, and avoid the red tiles; the level lines are the nonintersecting paths going from top to bottom in Fig.4. 
It follows that the intersection (within $\bf P$) of the level lines with the oblique lines $\{\eta=\mbox{integer}\}$ %for integer $-d\leq k\leq n_1+n_2+b$ 
 determine the blue dots and with the lines $\{n=\mbox{integer}\}$ the red dot at the integers $x\notin $ level lines. 
 In other terms drawing a horizontal line from left to right through the middle of the blue and green tiles (say, in Fig. 4) increases the height by $1$ and remains flat along the red tiles. Thus the tilings of the hexagon ${\bf P}$ are equivalent to 
 $\sum_{1}^{\ell+1}m_i =\sum_{1}^{u}n_i$ non-intersecting level-lines. %, which in the multi-cut case gets replaced by $\sum m_i=\sum n_i$ non-intersecting level-lines. 
  It follows that  
$$\begin{aligned}\#&\{\mbox{red dots of the $\mathbb K$-process on the horizontal line $n=k$ within 
 $\widetilde {\bf P}$}\}
 \\=&\#\{\mbox{flat segments of the heigth function along the line $n=k$}\}
\\=&\left[\mbox{length of the (line $n=k$)$\cap \widetilde {\bf P}$}\right]-\left[\begin{array}{lll}\mbox{increment in height between the }
\\ \mbox{two points (line $n=k$)$\cap \pl \widetilde {\bf P}$}
\end{array}\right]
\\=& (\sum_1^{\ell+1} m_i+d+k)-(\sum_1^{\ell+1} m_i)=d+k.
\end{aligned}$$
When an oblique line $\eta=k$ traverses a tile, as in Fig.2, the height increases by $1$ for a blue tile and stays flat for a red and green tile. Therefore, we have 
$$\begin{aligned}\#&\{\mbox{blue dots of the $\mathbb L$-process on the oblique line $\eta=k$ within 
 $\widetilde {\bf P}$}\}
 %\\=&\#\{\mbox{flats on that segment}\}
 \\&
=\left[\mbox{increment in height along (line $\eta=k$)$\cap \pl \widetilde {\bf P}$}\right],
%\\=&= d+k
\end{aligned}$$
leading to the following pattern for the blue dots for the {\bf two-cut model}:  
$$\mbox{\footnotesize$\begin{array}{ccccccccccccccccccccccc}
 k&&& \# \{\mbox{blue dots on $\eta=k$}\}
\\ \hline
\\
n_1+n_2+b&&&0\\
\vdots
\\
n_1+n_2&&& b\\  
\vdots
\\
n_1+b  &&&b
\\ \vdots \\
m_1+\rho   &&&b-d=r
\\ \vdots \\
m_1   &&&b-d=r
 \\
 \vdots \\
 m_1-d &&& b
 \\
 \vdots
 \\
 b-d&&&b
 \\
 \vdots
 \\
 -d&&&0
\end{array}$}
$$
This implies that the number of blue dots are $=r$ (local minimum) along each of the oblique lines $\eta=k$ within the strip $\{\rho\}$, including its boundary. Outside,  that number starts growing linearly in steps of $1$ up to $b$, stays fixed for a while and then goes down to $0$ linearly in steps of $1$.
 
 \bigbreak
 
\noindent {\bf The red dot-configurations and skew-Young tableaux}. The red dots on the horizontal lines $\{n=k\}\cap \widetilde {\bf P}$ for $0\leq k\leq N$ are parametrized by  $x^{(k)}\in \BZ^{d+k}$
 $$
 x^{(k)}_{N}<\ldots<x^{(k)} _1,
 $$
subjected to the interlacing pattern below, in short $x^{(k-1)} \prec x^{(k)}$, 
\be
x_{i+1}^{(k)}<
x_{i}^{(k-1)}\leq x_{i}^{(k)}.\label{interlac}\ee
This interlacing follows from an argument similar to \cite{Petrov,ACJvM}. 
So, we have a 
truncated Gelfand-Tsetlin cone with prescribed top and bottom and for completion, set $x_k^{(0)}=y_k$ and $x_k^{(N)}=x_k$: %(set $N=b+c$)

%\newpage
%\vspace*{-2cm}
 
$$\hspace*{-.5cm}\begin{array}{ccccccccccccccccccccc}
  % =\!-\!d\!-\!N    \!\!\!  \!\!\!  \!\!\!            &               &=n_1-d-c          &             & \!\!\!=n_1\!-\!1\!-\!c                     &&\!\!\!  \!\!\!  \!\!\!  \!\! =n_1\!+\!n_2\!-\!1
  x^{ }_{d+N} &< \ldots <&x^{ }_{d+c} &<\ldots <&x^{ }_{1+c} <&\ldots & \!\!\!  \!\!\!< x^{}_{1 }
\\ \\ 
      & x^{(N-1)}_{d+N-1}<&  &\ldots&&      < x_1^{(N-1)}
%\\
\\&&&  \vdots&&
%\\&&&  \vdots&&
\\
  & &y_{d} <&  \dots <y_{2} <   &\dots <y_{1}.  &
  \\
 &&%=m_1-d  & = m_1-2 &  =m_1-1
 \end{array}
$$
The interlacing pattern above, with $x^{(k-1)} \prec x^{(k)}$, is equivalent to a red dot configuration of a polygon $\bf P$.

As an example, for the two-cut model, the top consists of the three regions $\LR$,~$\CR$ and $\RR$ of contiguous integers; and the bottom of one contiguous region, given by the lower-cut. 

 %For ease of notation, we set $x_k:=x_k^{(N)}$, thus omitting the superscript and , which we continue below ${\bf P}$ by setting $y_{d+i}:=-d-i$ for $1\leq i\leq N$. See Fig. 1. 
%The inequalities (\ref{interlac}) imply that, given (contiguous) red dots all along $n=N$ in $\widetilde{\bf P}\backslash {\bf P}$, one automatically has the regular pattern of red dots in $\widetilde{\bf P}\backslash {\bf P}$, as illustrated in Figure 1; same remark for $n=0$.   

Moreover, setting for $0\leq k\leq N$,
\footnote{\label{ft2} Define $|x^{(k)}|:= \sum_{i=1}^{d+k} x_i^{(k)}$ for $0\leq k\leq N$; in particular for $k=0$ and $k=N$, we have $|y|=\sum_1^{d}y_i$ and $|x|=\sum_1^{d+N}x_i$. That means, the sum is always taken within the parallelogram $\widetilde {\bf P}$. Also define $|\nu^{(k)}|:=\sum_i \nu^{(k)}_i$}
$$\nu_i^{(k)} =x_i^{(k)}+i ,
$$
leads to a sequence of partitions
\begin{equation}
 \nu^{(0)} \subset \nu^{(1)} \ldots \subset \nu^{(N )}  ,
\label{1}
\end{equation}
with prescribed {\bf initial and final condition},
$$ \mu:=  \nu^{(0)}   ~~~~~  \mbox{    and    }   ~~~~~ \lb:=  \nu^{(N)}  
$$
and 
such that each skew diagram $\nu^{(i)} \backslash \nu^{(i-1)}$ is a horizontal strip\footnote{\label{ft3}An horizontal strip is a Young diagram where each row has at most one box in each column. }. 
Note the condition $x_i\geq y_i$ for all $1\leq i\leq d+N$ on the cuts, mentioned just before (\ref{P}), guarantees that $\mu\subset \lb$. In fact the consecutive diagrams $\nu^{(i)} \backslash \nu^{(i-1)}$ are horizontal strips, if and only if the precise inequalities $x_{i+1}^{(m)}<
x_{i}^{(m-1)}\leq x_{i}^{(m)}$ hold.

 Putting the integer $i$'s in each skew-diagram $\nu^{(i)} \backslash \nu^{(i-1)}$, for $1\leq i\leq N$, is equivalent to a skew-Young tableau filled with $N$ numbers $1,\dots,N$ and exactly $N$. This leads to the following {\bf semi-standard skew-Young tableau} in the two-cut case, as in Fig.8. 
 An argument, using the height function, which is similar to the one in \cite{ACJvM}, shows that all configurations are equally likely. So we have {\bf uniform distribution} on the set of configurations.
\noindent To conclude, we have:
   $$\begin{aligned}
   &\{\mbox{All configuration of red dots}\}   
    %\\& \Longleftrightarrow  \{\mbox{Non-intersecting paths}\}
   \\& \Longleftrightarrow  \{\mbox{skew-Young Tableaux of shape $\lb\backslash \mu$ filled with numbers $1$ to $N $ }\}        
 %\end{aligned}
 %$$ 
 \\%$$
  &\Longleftrightarrow  \left\{\begin{aligned}
%&%\mbox{the set of $\nu^{(1)} \subset \nu^{(2)} \ldots \subset 
%\nu^{(N)}$ such that}
%\\
&%$\lb\backslash \mu$}\\&
  \mbox{$\mu  = \nu^{(0)} \subset \nu^{(1)} \ldots \subset \nu^{(N )} = \lambda  $, with $\mu\subset \lb$ fixed,}
  \\ 
   &\mbox{with $\nu^{(m)}\backslash \nu^{(m-1)}=\mbox{horizontal strip} 
 %\left\{ 
 %\begin{array}{l} \mbox{at most one box}\\
  %\mbox{ in each column}
 %\end{array}
% \right\}
 $},
\end{aligned}\right\}
\end{aligned}
$$
with equal probability for each configuration. 

  \vspace*{2.7cm}
 
% \newpage
 
  \setlength{\unitlength}{0.017in}\begin{picture}(0,60)
\put(75,-70){\makebox(0,0) {\rotatebox{0}{\includegraphics[width=170mm,height=240mm]
 {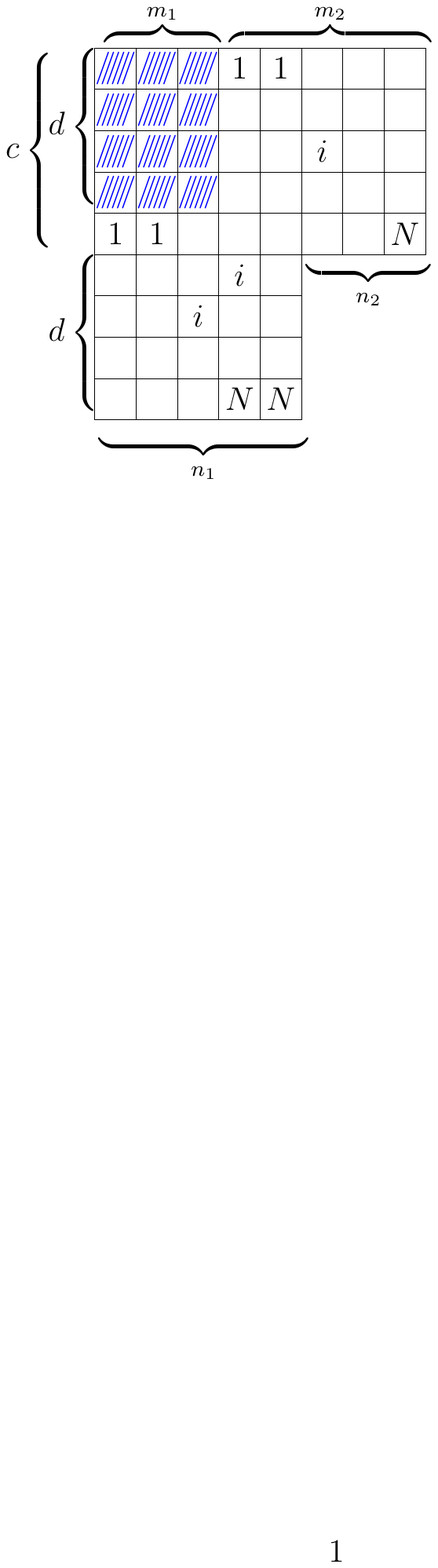}}}} \end{picture}
  \setlength{\unitlength}{0.017in}\begin{picture}(0,60)
\put(205,-120){\makebox(0,0) {\rotatebox{0}{\includegraphics[width=170mm,height=240mm]
 {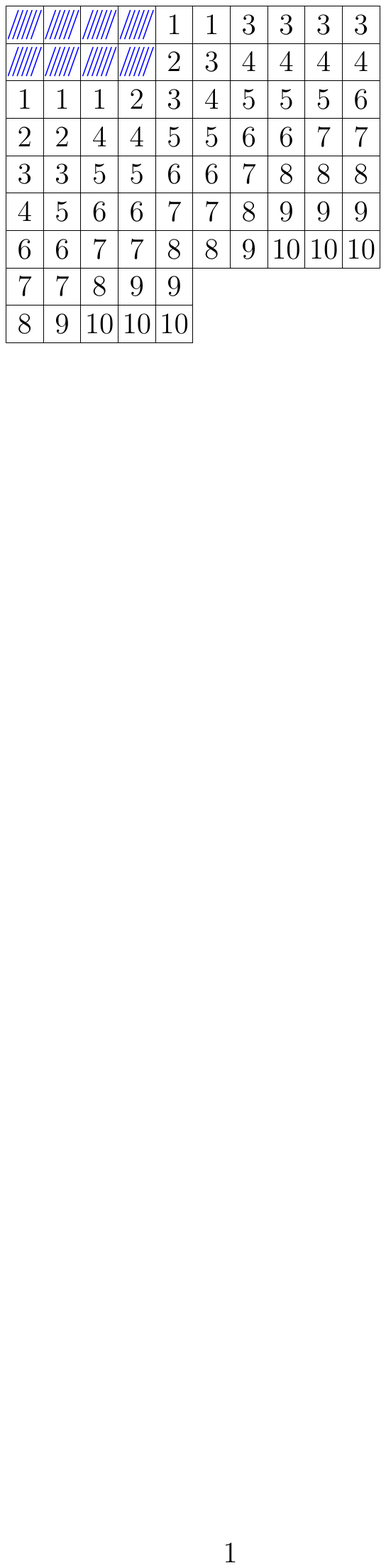}}}} \end{picture}

% \newpage

 \vspace*{.7cm}

Fig. 8: Example for the two-cut model. Semi-standard skew-Young Tableau $\lb  \backslash  \mu$ filled with numbers $1,\cdots, N$ in one-to one correspondence with the red dot-process. On the right is the precise semi-standard skew-Young Tableau associated with Fig. 4.

\vspace{.5cm}

%It is readily seen that there is a one-to-one correspondence between semi-standard skew-Young Tableau $\lb /\ \!\mu$ filled with numbers $1,\cdots, N$ and the red dot-process described before. 

% \bigbreak

 \noindent{\bf Uniform and $q$-probability on the set of red dot-configurations}. The last statement above implies that 
 \be\begin{aligned}
   &\#\{\mbox{configuration of red dots}\}   
    %\\& \Longleftrightarrow  \{\mbox{Non-intersecting paths}\}
    = s_{\lb\backslash \mu} (\underbrace{1,\dots,1}_N,0,0\ldots)  ,   
 \end{aligned}
 \label{number}\ee
and thus we have {\bf uniform probability measure $\BP $} on the space of all $N+1$-uples of partitions $\{ \nu^{(0)},\ldots,  \nu^{(N )}\}$
\be\begin{aligned}
{\mathbb P}&(\mbox{given configuration of red dots})
\\&=\BP( \nu^{(0)}, \nu^{(2)} , \ldots \, \nu^{(N )},~~\mbox{such that}~~\mu=\nu^{(0)}\subset \nu^{(1)} \subset \ldots \subset \nu^{(N )}=\lb)
\\
&=\frac{1}{s_{\lb\backslash \mu} (\underbrace{1,\dots,1}_N,0,0\ldots)}\Id_{\nu^{(0)}\subset \nu^{(1)} \subset \ldots \subset \nu^{(N )}}
\Id_{\nu^{(0)}=\mu}  \Id_{\nu^{(N )}=\lb}.\label{ProbUnif}
\end{aligned}\ee 

 As will be seen, we define a {\bf $q$-dependent probability measure $\BP_q$}, for $0<q\leq 1$, on the same space of all $N+1$-uples of partitions, as follows 
\be\begin{aligned}
\BP_q(&\nu^{(0)}, \nu^{(2)} , \ldots \, \nu^{(N )},~~\mbox{such that}~~\mu=\nu^{(0)}\subset \nu^{(1)} \subset \ldots \subset \nu^{(N )}=\lb)
\\
&=
\frac{(q^{-1})^{ \sum^{N-1}_{i=1}|\nu^{(i )}|- (N-1)|\nu^{(0)}|} }
{s_{\lambda\backslash \mu} (q^{1-N}, q^{2-N},\ldots, q^{-1}, q^0) }
\Id_{\nu^{(0)}\subset \nu^{(1)} \subset \ldots \subset \nu^{(N )}}
\Id_{\nu^{(0)}=\mu}  \Id_{\nu^{(N )}=\lb},
\end{aligned}\label{qprob}
\ee
which for $q\to 1$ leads to the uniform probability (\ref{ProbUnif}). This formula will be explained below.

%@@@The expression (\ref{skewSchur1}) for $s_{\lambda\backslash \mu} (q^{1-N}, q^{2-N},\ldots, q^{-1}, q^0)$ leads naturally 

MacDonald \cite{McD} is an excellent reference for the following discussion. Recall skew-Schur polynomials are defined as: for any $\mu\subset \lb$,% have on the one hand a determinantal expression and on the other hand an expression in terms of (\ref{1}):
\begin{equation}
s_{\lambda\backslash \mu} (x) := \det (h_{\lambda_i -i -\mu_j  +j} (x))_{1\leq i, j \leq n} =\det(h_{x_i -y_j   } (x))_{1\leq i, j \leq n} ,~~\ \textrm{for} \ n \geq \ell(\lambda),
\label{2}
\end{equation}
where $h_r(x)$ is defined for $r\geq 0$ as
\be\label{seriesh}
\prod_{i\geq 1} (1-x_i z)^{-1} = \sum_{r\geq 0} h_r (x) z^r,
\ee 
and $h_r(x)=0$ for $r < 0$. Skew Schur polyomials also satisfy the properties 
\be
s_{\lb\backslash\mu}(x,y)=\sum_{\nu} 
s_{\lb\backslash\nu}(x)s_{\nu\backslash\mu}( y)
,\label{Schursum}\ee
and for $x=(x_1,\ldots,x_N)$,
\begin{equation}
s_{\lambda\backslash \mu} (x_1,\ldots,x_{N }) = \sum_{\mu=\nu^{(0)}\subset\ldots\subset\nu^{(N )}=\lambda}  \prod^{N-1}_{i=0} x_{i+1}^{|\nu^{(i+1)}|-|\nu^{(i )}|}.
\label{3}
\end{equation}
In particular, for $0< q \leq 1$, from (\ref{3})
\be\begin{aligned} \label{skewSchur1}
s_{\lambda\backslash \mu} (q^{1-N}, q^{2-N},\ldots, q^{-1}, q^0)  &= 
\!  \sum_{\mu=\nu^{(0)} \subset\nu^{(1)} \subset\ldots\subset \nu^{(N )} = \lambda} (q^{-1})^{%\displaystyle 
   \sum^N_{i=1}(N-i)(|\nu^{(i )}|-|\nu^{(i-1)}|)} 
\\
&= \! \sum_{\mu=\nu^{(0)}\subset\nu^{(1)}  \subset\ldots\subset \nu^{(N )} = \lambda} (q^{-1})^{%\displaystyle 
 \sum^{N-1}_{i=1}|\nu^{(i )}|- (N-1)|\nu^{(0)}|} .
\end{aligned}\ee
The expression  $\sum^N_{i=1}(N-i)(|\nu^{(i )}|-|\nu^{(i-1 )}|)$ in  formula (\ref{skewSchur1}) is the volume of the three-dimensional figure obtained by putting $N-i$ cubes of size $1\times 1\times 1$ on top of the squares containing $i$ in the skew-Young diagram, for $1\leq i\leq N-1$. This shows why (\ref{qprob}) defines a probability measure.

%Instead of putting the uniform probability on the set of red dot configurations, we consider a non-uniform probability depending on a parameter $0<q\leq 1$, for which we shall compute the kernel ${\mathbb K}_q$ and then let $q\to 1$. 

\medspace

\noindent {\bf Some useful formulas}:
Notice that
 \be\begin{aligned}
    &h_r(\al^\gamma,\ldots,\al^{\gamma-N+1})=\al^{r(\gamma-N+1)} {h_{r}  (\al^0, \ldots,\al^{ N-1}) }
 \end{aligned}\ee
 and thus for $\al=q^{-1}$ and $\gamma=-d$, we have \be\label{hr(q)}\begin{aligned}
    h_{r }  (q^0,q^{-1},\ldots,q^{1-N}) 
 &=q^{r(1-N-d)}  h_{r  } (q^d,\ldots,q^{d+N-1}).
 \end{aligned}\ee
 Defining for $y,z\in\BR$  (in terms of the $q$-Pochhammer symbol $(a;q)_k=\prod_{j=0}^{k-1} (1-aq^j)$ for $k>0$ and $=1$ for $k=0$),
\be\PR_{n }(z):=\prod_{i=1}^{n }\frac { 1-zq ^{ i}  }{ 1-q ^{i } }=:\frac{(zq;q)_{n }}{( q;q)_{n }}=:\sum_{i=0}^{n } e_{i } z^{i } \mbox{   and  }
\tilde \PR_{n}^{y}(z):=q^{-dy  }\PR_{n}(zq^{ -y }),
\label{calP}\ee
 one checks, using the generating series (\ref{seriesh}) and the definition of ${\cal P}_n$, that for $n \geq 0$, 
 $$\begin{aligned} 
 h_r(q^0,\dots,q^{n}) &=\frac {1}{2\pi \I}\oint_{\Gamma_0}\frac{du}{u^{r+1}}\prod_{\ell=0}^{ n }\frac 1 {1-uq^\ell}
%= \left.\prod_{i=1}^{n }\frac { 1-zq ^{ i}  }{ 1-q ^{i } }\right|_{z=q^r} 
 \\&={\cal P}_n(q^r) \mbox{     for  } r\geq 0.
  \end{aligned}$$
 This is valid for general $r\in \BZ$ and $n,d \geq 0$, upon inserting an indicator function: 
 \be\begin{aligned} 
 h_r(q^d,\ldots,q^{d+n }) =q^{rd}h_r(q^0,\ldots,q^{ n })
 &= \frac {1}{2\pi \I}\oint_{\Gamma_0}\frac{du}{u^{r+1}}\prod_{\ell=d}^{ d+n }\frac 1 {1-uq^\ell}
 \\
 &=q^{rd}  {\cal P}_n(q^r) (\Id_{ r\geq -n} \mbox{ or }\Id_{ r\geq 0}).
  %\\.
 \end{aligned}\label{hP}\ee
Indeed the left hand side equals $0$ for $r<0$, whereas ${\cal P}_n(q^r)=0$ for $-n\leq r<0$, but $\neq 0$ for $r<-n$. The integral equals $ 0$ for $r<0$. So, the identity is valid for each of the two indicator functions; both of them will be used. Also by (\ref{seriesh}), 
\be\begin{aligned}
   s_{\lambda\backslash \mu}  (q^0,q^{-1},\ldots,q^{1-N})
&=  \det (h_{x_i-y_j  } (q^0,q^{-1},\ldots,q^{1-N}))_{i\leq i,j \leq N+d}
\\ \\
&= 
q^{(  d+N-1)(\sum_1^{d+N} y_i-\sum_1^{d+N}x_i)} \det (h_{x_i-y_j} (q^d,\ldots,q^{d+N-1}))_{1\leq i,j \leq N+d}
\\ \\
&=q^{- N(d+N-1)(d+(N+1)/2) } q^{(  d+N-1)(|y|-|x| )} \det(H),
\label{skewSchur2}\end{aligned}\ee
where $H$ is a square matrix of size $d+N$ :
\be\begin{aligned}
H = (H_{ij})_{1\leq i,j \leq N+d}&:=(h_{x_i-y_j} (q^d,q^{d+1},\ldots,q^{d+N-1}))_{1\leq i,j \leq N+d}
\\
&=    ( q^{(x_i-y_j)d}h_{x_i-y_j} (q^0, \ldots,q^{ N-1}))_{1\leq i,j \leq N+d},
\end{aligned}\label{Hdef}\ee
where formula  (\ref{hr(q)}) is used in the second line.

\bigbreak

\noindent {\bf q-Calculus}. The polynomial $Q(z)$ has a natural $q$-extension, namely  (we often write $q_i:=q^{x_i}$)
\be\label{Qq}
Q_q(z):=\prod _1^{d+N} (z-q^{x_i}) =\prod _1^{d+N} (z-q _i ),
%,~~~Q(z)=\prod _1^{d+N} (z-x _i )
\ee
%Also recalling the definition (\ref{calP}) of $\PR_{n}(z)$, we define 
%an expression, depending on the points $y_i$ in the lower-cut,%$$
%\PR_{n}(z)=\prod_{i=0}^{n-1}\frac{1-zqq^i}{1-qq^i} =
%   \frac{(zq;q)_n}{(q,q)_n}
%  $$ 
%  \be\label{calPb} \tilde \PR_{n}^{y}(z):=q^{-dy  }\PR_{n}(zq^{ -y }).\ee
  and from (\ref{calP}) we have
  $$\begin{aligned}
\PR_{N-m}(vq^{-x})=\prod_{i=1}^{N-m}\frac{1-zq^i}{1-q^i}\Bigr|_{z=vq^{-x}}=
  \prod_{i=1}^{N-m}\frac{1-vq^{i-x}}{1-q^i} =\frac{(vq^{-x+1};q)_{N-m}}{(q ;q)_{N-m}}
.  \end{aligned} $$
Following Petrov \cite{Petrov}, we define the $q$-hypergeometric function ${}_{2}\phi_1$ by means of
\be\begin{aligned} \label{Phiq}
z\sum_{k=1}^{n }z^{-k}q^{ (k-1) y}
\!\prod_{ r=n+1 }^{N  }\!(1\!-\!q^{r-k})&=: (q^{N-1};q^{-1})_{N\!-\!n\! } ~{}_{2}\phi_1
(q^{-1},q^{n-1};q^{N-1}\Bigr|q^{-1},\frac{q^y}z)
\\&=:\Phi_q(\frac {q^y }z).
  \end{aligned} \ee
$\Dt_k$ stands for the Vandermonde determinant of size $k$.

\begin{lemma}\label{qlim}
For any $a\in \BZ$, remembering the definitions (\ref{Qq}) and (\ref{calP}) of $Q(z)$ and $\tilde {\cal P}_n^{\beta}(z) $, the following limits hold
$$
\lim_{q\to 1}\PR_{n}(q^x)=\prod_{k=1}^n\frac{k+x }{k}=\frac {(x+1)_n}{n!}
,~~ \lim_{q\to 1}\tilde \PR_{n}^y(q^x)= 
%\prod_{\ell=1}^{N-1}
 \frac{{(x  -y +1)}_{n}}{n!},$$
\be\label{limPQ} \lim_{q\to 1}\frac{Q_q'(q^{x_i  })}{(q-1)^{d+N-1}}=%\prod^{d+N}_{{j=1}\atop{j\neq i}} %~~\lim_{q\to 1}\frac{(q^a;q}{}
% (x_i-x_j)=
 Q'(x_i)
 ,~~~~\lim_{q\to 1}\frac{\Dt_d(q^{x_1},\dots,q^{x_d})  }{(q-1)^{d(d-1)/2}}=
\Dt_d( {x_1},\dots, {x_d}) ,
\ee
and (due to Petrov\cite{Petrov})
 \be\label{Petrov}\begin{aligned}%(q^{N-1};q^{-1})_{N-n+1}
 \lim_{q\to 1} \frac1{(q-1)^{N-1}}&\oint_{\Gamma_\infty } \frac {dz}{2\pi \I z}~\Phi_q(z^{-1})%{}_{2}\phi_1(q^{-1},q^{n-2};q^{N-1}\Bigr|q^{-1},z^{-1})
\prod_1^{N-1}(z-q^{\rho_r-y})
\\&=
 (N-n+1)!\oint_{\Gamma_\infty}\frac{dz}{2\pi \I (z-y)_{N-n+2}}\prod_1^{N-1}  {(z-\rho_r)} .
 \end{aligned}\ee
 %
%We have the following identity (for some constant $c_{N,d}>0$)\footnote{For any positive integers $k$ and $N-1$ we have $k_{(N-1)}=k(k+1)\dots(k+N-2)$.}
%\be \label{detV}
%\det\left( \frac{(x_\al+\beta)_{ N-1 }}{(N-1)!} \right)_{1\leq \al,\beta\leq d}=
%c_{N,d}\Dt_d(x_1,\dots,x_d)\prod_{\al=1}^{d}  (x_\al+d)_{ N-d }\ee
%%with
%$$c_{N,d}:= \prod_{j=1}^d\frac{1}{(N-j)!}.$$
%For $1\leq n\leq N$, we have the following identity, with the second term vanishing, when $y+\beta+n \geq 1$, to wit,
%\be\begin{aligned}
%\left({N-1}\atop{n-1}\right) (y+\beta)_{n-1} =
 % \oint_{\Gamma_0}&\frac{du}{2\pi \I u^{y+\beta+1}(1-u)^n}
 %  =\frac{(N-n)!}{(N-1)!} \left(\oint_{\Gamma(y,y-1,\dots,y-N+n)} \frac{dz(z +\beta+1 )_{N-1} }{2\pi \I(z-y)_{N-n+1}} \right.   % \frac{(z +\beta )_{N-1}}{(N-1)!}
%\\&\left.+(-1)^n\Id_{y+\beta+n \leq   0}
%\oint_{\Gamma(-y,-y-1,\dots,-y-N+n)}\frac{dz(z  -n-\beta+1)_{N-1} }{2\pi \I(z+y)_{N-n+1}}\right).\end{aligned}\ee
%from which one deduces the identity,
%$$
%  \sum_{\al=0}^{c-m} (-1)^\al \left(    { c-m} \atop{\al} \right)
% \left(    { c+k-\al-1} \atop{c} \right)
  %
%=    
%   \left(    {k+m-1} \atop{m} \right)
 % . $$
\end{lemma}
 \proof  The limits are straightforward, except for Petrov's limit formula.\qed% its proof can be found in Petrov\cite{Petrov}. \qed   %The integrand has no pole at $\infty$, when  $ N+k \geq 1$ and so we have 
%\be
%\oint_{\Gamma_0}\frac{dw}{2\pi \I w^{1+k}(1-w)^N}=\frac{(k+1)_{N-1}}{(N-1)!}.
%\ee
%

 %\newpage

\section{An alternative integral representation for $h_{y-y_j}(1^n)$ and some determinantal identities}

\begin{proposition} \label{detV'}   Given $d\geq 1$ and integers ${\bf y}=\{y_d<\dots<y_1\}$ satisfying $N-1+y_d-y_1\geq 0$, and given the sum of the gaps $g:=y_1-y_d-d+1$ between these integers, the following holds, with the  constant \footnote{The explicit form of the constant will never be used.}
 $C_{N,d}=C'_{N,d}C''_{N,d}$ and with $C'_{N,d}= \prod_{j=1}^d\frac{1}{(N-j)!} $ and $C''_{N,d}= \frac{\Dt_d(y_1,\dots,y_d)}{\prod_{j=1}^d(d-j)!}$ :
  \be \label{detV}%\label{detV}
  \begin{aligned}
\det&\left( \frac{(x_\al-y_\beta+1)_{ N-1 }}{(N-1)!} \right)_{1\leq \al,\beta\leq d}
\\&=
C _{N,d} \Dt_d(x_1,\dots,x_d){  E}^{({\bf y})}_g(x_1,\dots,x_d)\prod_{\al=1}^{d}  (x_\al-y_d+1)_{ N-(y_1-y_d +1)}.
 \end{aligned}\ee
where ${ E}^{({\bf y})}_g$ is a symmetric function of $x_1,\dots,x_d$ with coefficients depending on the $y_i$'s, 
\be {  E}^{({\bf y})}_g(x_1,\dots,x_d)= (x_1\dots x_d)^{g}+\mbox{lower order terms},
\label{Esymm}\ee
where the ``lower order terms" refer to terms of total degree $<gd$ , with degree in any $x_i \leq g$. 
%and where
%$$C_{N,d}= \frac{1}{\prod_{j=1}^d(N-j)!},~~~C'_{N,d}= \frac{\Dt_d(c_1,\dots,c_d)}{\prod_{j=1}^d(d-j)!} .$$ 
 %
 For contiguous $(y_1,\dots,y_d)=(d,\dots,1)$, we have $g=0$, and so the symmetric function $=1$ and $C''_{N,d}=1$.
\end{proposition}

 %For future use we set
 % $${ F}^{({\bf y})} (t_1,\dots,t_d):={ E}^{({\bf y})} (x_1,\dots,x_d) , \mbox{   where   } 
 % t_\al(\LR)=\sum_{i=1}^d x^\al_i+\sum_{i=1}^r x'^\al_i$$
 Given an arbitrary  choice of $k+\ell$ points $(x_1,\dots,x_k,x'_1,\dots,x'_\ell)$ in the set $\LR$ (having $k+\ell$ points), and the associated power sum symmetric polynomials\footnote{ Given variables $u_1,\dots,u_k$, %$v_1,\dots,v_\ell$,
   any symmetric polynomial $E(u)$ in these variables can be written as a polynomial in power sum symmetric polynomials $t_\al(u_1,\dots,u_k)=\sum_{i=1}^k u_i^\al$ for integer $\al\geq 0$ with rational coefficients in the coefficients of the symmetric polynomial. } $t_\al(\LR):=\sum_{i=1}^k x^\al_i+\sum_{i=1}^\ell x'^\al_i$,  we have that any symmetric function $S(  x_1,\dots,x_k)$ (resp. $S(z; x_1,\dots,x_k)$ with some  additional variable $z$) can be expressed in terms of a symmetric function in the complementary variables $x'_1,\dots,x'_\ell$
    (resp. $z,x'_1,\dots,x'_\ell$). To be precise,% the operation from $E^{\bf y}(x_1,\dots,x_k)$ to $\widetilde E^{\bf y}(x'_1,\dots,x'_\ell)$ consists of doing the following:
 \be\bl
 S(z, x_1,\dots,x_k)&=
  F \left( z+\sum_1^\al  x ^{}_i,\dots,  z^\al+\sum_1^\al x ^{\al}_i \right)
  \\
  &=
    F^{ }\left(z+t_1(\LR)-\sum_1^\ell x'^{}_i,\dots, z^\al+t_\al(\LR)-\sum_1^\ell x'^{\al}_i \right)
 \\
 &=: \widetilde S(z,x'_1,\dots,x'_\ell),
 \el\label{symcompl}
 \ee 
 the latter being a new symmetric function, with coefficients depending polynomially on the variables $t_\al(\LR)$. Formula (\ref{symcompl}) is also valid in the absence of the $z$-variable.
 
 As an example, with one gap ($g=1$), upon defining $\sg_i(x)$ by $\prod_1^d(u+x_i)=\sum_{i=0}^d \sg_i u^{d-i}$, 
 $$  \mbox{for}\bl\mbox{${\bf y}=(2<4<5<6)$:~~~~}&
 E^{\bf y}_1(x)=\sg_4-\sg_3-\sg_2+11\sg_1-49 
\\  ~~~~~\mbox{${\bf y}=(2<3<5<6<7)$:~~~~}&
 E^{\bf y}_1(x)=\sg_5-\tfrac 45 (\sg_4+\sg_3)+4(\sg_2+\sg_1)-\tfrac{604}5.
 \el $$
 %
%Given integers $d,r\geq 0$ such that $d+r=b$ and variables $x_1,\dots,x_b$. 
%
%\begin{proposition} \label{detV'}The following determinantal identity holds, with the  constant\footnote{The explicit form of the constant will never be used.}
% $C_{N,d}= \prod_{j=1}^d\frac{1}{(N-j)!} $, \be 
%\det\left( \frac{(x_\al+\beta)_{ N-1 }}{(N-1)!} \right)_{1\leq \al,\beta\leq d}=
%C_{N,d}\Dt_d(x_1,\dots,x_d)\prod_{\al=1}^{d}  (x_\al+d)_{ N-d }.\ee
%\end{proposition}
 
 \proof We give the proof for $(y_d,\dots,y_1)=(1,\dots,d)$. The determinant above has the general form $\det((x_\al+\beta)_{N-1})_{1\leq \al,\beta\leq d}$. Each column of the matrix in the determinant $\det((x_\al+\beta)_{N-1})_{1\leq \al,\beta\leq d}$ has the common factor $(x_\al+d)_{N-d}$, which, taken out, leaves us a determinant of the type below; by Lemma 3 in  Krattenthaler \cite{Krat}, this determinant  equals:
 $$\begin{aligned}\det_{1\leq i,j\leq n} &\left((x_i+A_n) (x_i+A_{n-1})\dots(x_i+A_{j+1})(x_i+B_j)(x_i+B_{j-1})\dots(x_i+B_2)
 \right)
 \\&=\Dt_n(x)\prod_{2\leq i,j\leq n}(B_i-A_j).
 \end{aligned}$$
 In this expression $A_j=j-1,~B_j=N+j-2$ for $2\leq j\leq d$, from which it follows that $\prod_{2\leq i,j\leq n}(B_i-A_j)=((N-1)!)^d\prod_{k=1}^{d }((N-k)!)^{-1}$, proving identity (\ref{detV}).\qed
 
 \medbreak
 
%The left-most oblique line of (red) ${\mathbb K}$-dots within the strip $\rho$ is given by $\eta=m_1+\tfrac 12 $ (i.e., $y+n=m_1=y_1+1$). It plays a special role: it divides the polygon ${\bf P}$ into two regions $y+n\geq m_1=y_1+1$ and $y+n< m_1=y_1+1$. As a result, the distance $$\tau:= (y+n)-m_1=\eta-(m_1+\tfrac12)$$ between the oblique lines $\eta=(\mbox{\it integer}+\tfrac 12)$ and the oblique line $\eta=
%m_1+\tfrac12$, will figure prominently into the problem, with $\tau<0$ strictly to the left of the line $\eta=m_1+\tfrac12$ and $\geq 0$ to the right of that line. 

 Defining $B(y):=\max (y+n ,y_1+1  )-N$, we now define the contours 
\be\label{gammay}\begin{aligned}
\gamma_y &:= \Gamma(y,y-1,\dots,B(y)), \mbox{   if  }B(y)\leq y
\\ &~=\emptyset~~~~~~~~~~~~~~~~~~~~~~~~, \mbox{   if  }B(y)>y
,\end{aligned}
\ee
and the contour  $\Ga_\tau$, defined such that 
\be \begin{aligned}
\gamma_y&=\Ga(y,y-1,\dots,y+n-N)\backslash   \Ga_\tau
, \end{aligned}
 \label{gay}\ee
 given that a contour integral will be taken over a rational function with denominator equal to $(z-y)_{N-n+1}$. 
 Then it is easily seen that 
\be\label{gatau}\Ga_\tau:=%\Ga(y_1 -N  ,\dots,y_1 -N+( \tau+1) )=
 \left\{
\begin{aligned}& \Ga(\min(y_1 -N,y) ,\dots,y+n-N),
 \mbox{   if  $\tau< 0$ }
\\& \emptyset,
 \mbox{  if 
$\tau\geq 0$}
\end{aligned}\right.,
  \ee 
with $\Ga_\tau$ containing exactly $-\tau>0$ points, if $y_1-N\leq y$ and $N-n+1$ points, if  $y_1-N\geq y$, and none, when $\tau\geq 0$. This confirms definition (\ref{cont0}).

\medbreak

We now state the following general Lemma, which will play a crucial role in the expression of the ${\mathbb K}$-kernel.

\begin{proposition} \label{LemmaInt}Given $0\leq n\leq N-1$ and  integer points $(y_1>\dots >y_j>\dots)%=(m_1-1,\dots,m_1-d)$ in the lower-cut, 
$ satisfying $0 \leq y_1-y_j \leq N-1$,% and for  set $  B(y)=\max (y+n ,y_1+1  )-N$,
the following holds, with the contours $\ga_y$ and $\Ga_\tau$ as in (\ref{gammay}) and (\ref{gatau})
\be\label{IntId1}\begin{aligned}
 h_{y-y_j}(1^n)& =\oint_{\Gamma_0} \frac{du}{2\pi \I u^{y-y_j+1}(1-u)^n}%\\
\\& =%&=
\frac{(N-n)!}{(N-1)!}  \oint_{\gamma_y } \frac{dz(z -y_j+1 )_{N-1} }{2\pi \I(z-y)_{N-n+1}}
\\& =%&=
\frac{(N-n)!}{(N-1)!} \left( \oint_{\Gamma(y, y-1,\dots , y-N+n) } -\oint
_{\Gamma_\tau}\right)\frac{dz(z -y_j+1 )_{N-1} }{2\pi \I(z-y)_{N-n+1}}  .
\end{aligned}
\ee

\end{proposition} 
\proof At first, notice that by Cauchy's theorem, the integral below, for $n\geq 0$, equals minus the sum of the residues at $z=1$ and $  \infty$, with the residue at $  \infty$ vanishing when $ n+k-1 \geq 0$,
\be \label{Int2}\begin{aligned} %\label{Int2}
\oint_{\Gamma_0}\frac{dz}{2\pi \I z^{1+k}(1-z)^n} 
&=\prod_{\ell=1}^{n-1}\frac{k+\ell}{\ell}-\mbox{Res}_{z=\infty}
 =\frac{(k+1)_{n-1}}{(n-1)!}  -\mbox{Res}_{z=\infty}   \Id_{n+k\leq 0}.
%+(-1)^{n }\frac{(-n-k+1)_{n-1}}{(n-1)!}\Id_{n+k\leq 0}.%\mbox{Res}_{w=\infty}.
\end{aligned}\ee

Consider the contour $\ga_y$ as in (\ref{gammay}). If $B(y)>y$, then we have $\max(y+n,y_1+1)>y+N$, implying that $y_1-N> y-1$, since $0\leq n\leq N$, and thus $N \geq y_1-y_j+1>y-y_j+N $, and so $y-y_j+1 \leq 0$,
   implying that both sides of (\ref{IntId1}) vanish. In that case, since $\ga_y=\emptyset$, we have that, in view of (\ref{gay}), 
   $$\Ga_\tau=\Ga(y,y-1,\dots,y+n-N).$$
   So, we now consider $B(y)\leq y$. We have thus $\emptyset \neq \{y, y-1,\dots, B(y)\}\subseteq \{y, y-1,\dots , y-N+n\}$. Then, since $y-B(y)=N-n-\max (  y_1+1-y-n,0)$, we have for the right hand side of (\ref{IntId1}), upon evaluating by residues and upon using the change of variables $\ell=y-k$ and $i=y-j$ in $\stackrel{ \ast}{=}$,
$$\begin{aligned}
\hspace*{-4cm} \frac{(N-n)!}{(N-1)!}&\sum_{\ell=B(y)}^y\frac{(\ell-y_j+1)\dots(\ell-y_j+N-1)}{\displaystyle\prod^y_{{i=y-N+n}\atop{i\neq \ell}}( \ell -i)}
\\
&\stackrel{ \ast}{=}
\frac{(N-n)!}{(N-1)!}\sum_{k=0}^{y-B(y)}\frac{(y-y_j-k+1)\dots(y-y_j-k+N-1)}{\displaystyle\prod^{N-n}_{{j=0}\atop{j\neq k}}( j-k)}
\end{aligned}
$$
\be\begin{aligned}%\\&\stackrel{ \ast}{=}
%\frac{(N-n)!}{(N-1)!}\sum_{k=0}^{y-B(y)}\frac{(y-y_j-k+1)\dots(y-y_j-k+N-1)}{\displaystyle\prod^{N-n}_{{j=0}\atop{j\neq k}}( j-k)}
\\&=
\frac{(N-n)!}{(N-1)!}\sum_{k=0}^{y-B(y)}\frac{(-1)^k(y-y_j-k+1)_{N-1}}{k!(N-n-k)!}
\\&=\sum_{k=0}^{N-n-\max(0, y_1\!+\!1\!-y-n )}(-1)^k
\left({{N-n}\atop{k}}\right)\frac{ (y-y_j-k+1)_{N-1}}{(N-1)!}
\\&\stackrel{\ast\ast}{=}\sum_{k=0}^{N-n-\max (0, y_1\!+1\!-y-n )}(-1)^k
\left({{N-n}\atop{k}}\right) \oint_{\Gamma_0}\frac{dz}{z^{ y-y_j-k+1}(1-z)^N},%@@\frac{ (y-y_j-k+1)_{N-1}}{(N-1)!}
\label{IntId1'}\end{aligned}
\ee
provided $y-y_j-k+N-1\geq 0$ ($\ast$) holds for all $0\leq k\leq N-n-\max(0,y_1+1-y-n )$ in the expression on the right hand side of $\stackrel{\ast\ast}{=}$. It suffices to check this for the largest value $k=N-n-\max(0,y_1+1-y-n )$. Indeed, this is so:
\newline (i) \underline{If $y+n\geq y_1+1$}, then $k=N-n$ and  $y-y_j-k+N-1=y+n-y_j -1\geq y_1-y_j \geq 0$. %for all $j=1,\dots ,d$ (points in the lower-cut).
\newline (ii) \underline{If $y+n< y_1+1$}, then $k=N+y-y_1-1$ and so $y-y_j-k+N-1=y_1-y_j \geq 0$; this checks the inequality ($\ast$) above. For this same case, namely $y+n< y_1+1$, and for $N-n-(y_1+1-y-n)+1\leq k\leq N-n$, we also have that the integrand in the last expression (\ref{IntId1'}) has no residue at $z=0$, because the $z$-exponent in its denominator equals
$$
y-y_j-k+1\leq y-y_j-(N-n-(y_1+1-y-n)+1)+1=y_1-y_j-N+1\leq 0,
$$
%
%\newpage
%
%\vspace*{-2cm}
%\noindent 
 using the inequality in the statement.
 
\noindent Therefore, in the last expression in (\ref{IntId1'}) the sum can be extended to $0\leq k\leq N-n$ and so the final expression reads 
$$\begin{aligned}\sum_{k=0}^{N-n}(-1)^k \left({N-n}\atop{k}\right)
\oint_{\Gamma_0}\frac{~z^k~dz  }{2\pi \I z^{y -y_j+1}(1-z)^N}
 &=\oint_{\Gamma_0}\frac{dz ~(1-z)^{N-n}}{2\pi \I z^{y-y_j+1 }(1-z)^N}
\\& =\oint_{\Gamma_0} \frac{dz   }{2\pi \I z^{y-y_j+1  }(1-z)^n}.
 \end{aligned}
$$
The last formula of (\ref{IntId1}) follows from formula (\ref{gay}), thus ending the proof of Proposition \ref{LemmaInt}.\qed

 Both Propositions will be applied to certain determinants, which will come up later, and which depend on the geometry of $\bf P$. In the corollary below, the $u_1,\dots,u_d$ are arbitrary complex variables. Referring to the model $\bf P$, remember ${\bf y}_{\mbox{\tiny cut}}=\{y_1>\dots >y_d\}$ are the integer points in the $\{\mbox{lower-cuts}\}\cap \{n=0\}$ and $ P(z)$ is as in (\ref{P}). We define three expressions, which will play an important role in Section 7:
 \be \begin{aligned}
 \Dt_{ d}^{({\bf y}_{\mbox{\tiny cut}})}( u_1,\dots, u_d)&:=\det\!\left( \!\frac{(u_\al -y_\beta+1)_{N-1 }  }{(N-1)!}\!\!\right)_{\!\!1\leq \alpha,\beta \leq d} \\
 \widetilde \Dt_{ d}^{({\bf y}_{\mbox{\tiny cut}})}( w;u_2,\dots, u_d)&:= \det\left(\begin{array}{cccccccc}
w^{ y_1}\!\!\!\!\!\!&\dots\!\!\!\!\!\!&\!\!\!\!\!\!w^{y_d}\\ \\
 &   \left(  \frac{(u_\al -y_\beta+1)_{N-1 }  }{(N-1)!} \right)_{ {2\leq \alpha  \leq d}\atop{1\leq \beta\leq d}}
 \end{array}
\right)
 \\
   \widetilde\Dt_{ d,n}^{({\bf y}_{\mbox{\tiny cut}})}( y;u_2,\dots, u_d)%={\mathcal D}_2( y,u_2,\dots,u_d) 
  &:=  \oint_{\Gamma_0}  \frac {dw}{2\pi \I w^{  y  +1}  { (1-w )^n}} 
 \widetilde \Dt_{ d}^{({\bf y}_{\mbox{\tiny cut}})}( w;u_2,\dots, u_d)
\\&= \det\left(\begin{array}{cccccccc}
h_{y-y_1}(1^n)\!\!\!\!\!\!&\dots\!\!\!\!\!\!&\!\!\!\!\!\!h_{y-y_d}(1^n)\\ \\
 &   \left(  \frac{(u_\al -y_\beta+1)_{N-1 }  }{(N-1)!} \right)_{ {2\leq \alpha  \leq d}\atop{1\leq \beta\leq d}}
 \end{array}
\right).
 \end{aligned}
\label{Deltacut}\ee

 \noindent We now apply formula (\ref{detV}) (or the first formula (\ref{Deltacut}))  to the (integer) points $y_1>\dots>y_d$ in the lower cuts, thus leading to the symmetric function $E^{({\bf y}_{\mbox{\tiny cut}})}_g(u_1,\dots,u_d)$, as in (\ref{detV}).

 \begin{corollary}\label{Dt23}

 The following identities hold  for the determinantal expressions in (\ref{Deltacut}), with $C_{N,d} $ as in (\ref{detV}), $P(u)$ as in (\ref{P}), and the contour $\Ga_\tau$ as in (\ref{gatau}):% (\ref{detV}) and the fact that $(z-y_d+1)_{N-d}=P(z)$, as defined in (\ref{PQ}), it follows immediately that 
\be \begin{aligned}
 \Dt_{ d}^{({\bf y}_{\mbox{\tiny cut}})}( u_1,\dots, u_d)
=C_{N,d} E^{({\bf y}_{\mbox{\tiny cut}})}_g(u)\Dt_d(u)\prod_{\al=1}^dP(u_\al),
\end{aligned}\label{Dt2}\ee
and
 \be \begin{aligned}
%\widetilde \Dt^{({\bf y}_{\mbox{\tiny cut}})}_{d,n }&(y,u_2,\dots, u_d)
   \widetilde\Dt_{ d,n}^{({\bf y}_{\mbox{\tiny cut}})}( y;u_2,\dots, u_d)
 =&
C_{N,d}  \left( \oint_{\Gamma(y, y-1,\dots , y-N+n) } -\oint
_{\Gamma_\tau}\right)\frac{(N-n)!dz}{2\pi \I(z-y)_{N-n+1}}
\\
&  ~\times~E^{({\bf y}_{\mbox{\tiny cut}})}_g(z,u_2 ,\dots,u_d )\Dt_d(z,u_2 ,\dots,u_d )  
 P(z)\prod_{\al=2}^dP(u_\al ).
\end{aligned}\label{Dt3}\ee

%@@@
%\be \begin{aligned}
 %\Dt_{ d}^{({\bf y}_{\mbox{\tiny cut}})}( u_1,\dots, u_d)
%=C_{N,d}\Dt_d(u)\prod_{\al=1}^dP(u_\al),
%\end{aligned}\label{Dt2}\ee
%and
%
%
%
 %\be \begin{aligned}
%\widetilde \Dt^{({\bf y}_{\mbox{\tiny cut}})}_{d,n }&(y,u_2,\dots, u_d)
 %  \widetilde\Dt_{ d,n}^{({\bf y}_{\mbox{\tiny cut}})}( y;u_2,\dots, u_d)
 %=&
%C_{N,d} \left( \oint_{\Gamma(y, y-1,\dots , y-N+n) } -\oint_{\Gamma_\tau}\right)\frac{(N-n)!dz}{2\pi \I(z-y)_{N-n+1}}
%\\
%& ~\qquad~~~ ~\times~\Dt_d(z,u_2 ,\dots,u_d )  
% P(z)\prod_{\al=2}^dP(u_\al ).
%\end{aligned}\label{Dt3}\ee

 \end{corollary}
 
 \proof The first identity (\ref{Dt2}) follows at once from Proposition \ref{detV'}. The second identity (\ref{Dt3}) follows from moving the $w$-integral to the first row of $ \widetilde \Dt_{ d}^{({\bf y}_{\mbox{\tiny cut}})}( w,\dots, u_d)$, using Proposition \ref{LemmaInt} :
  $$ \begin{aligned}
&  \widetilde\Dt_{ d,n}^{({\bf y}_{\mbox{\tiny cut}})}( y;u_2,\dots, u_d)
%\\&=\det\left(  
% \right)
\\&=\oint_{\ga_y}\frac{(N-n)!dz}{2\pi \I(z-y)_{N-n+1}}
\det\!\left( \!\frac{(u_\al -y_\beta+1)_{N-1 }  }{(N-1)!}\!\!\right)_{\!\!1\leq \alpha,\beta \leq d} \Bigr|_{u_1=z}
\\&=\oint_{\ga_y}\frac{(N-n)!dz}{2\pi \I(z-y)_{N-n+1}}
 \Dt_{ d}^{({\bf y}_{\mbox{\tiny cut}})}( z,u_2,\dots, u_d)
%
%\\&=C_{N,d} \oint_{\ga_y}\frac{(N-n)!dz}{2\pi \I(z-y)_{N-n+1}}
%\Dt_d(z,u_2 ,\dots,u_d )
%P(z)\prod_{\al=2}^dP(u_\al ).
.\end{aligned}
$$
Finally using the first identity (\ref{Dt2}), applied to $\Dt_{ d}^{({\bf y}_{\mbox{\tiny cut}})}(z,u_2,\dots,u_d)$ ends the proof of Corollary \ref{Dt23}.\qed

 %\remark {\bf In the multi-cut case}, the same identities hold for the polynomial $P(z)$, as defined in (\ref{P}), except the right hand side of (\ref{Dt2}) and the integrand of (\ref{Dt3}) will involve a symmetric function, as will be discussed in Section 10.% of $u_i$ parametrized by the $y_j$, with monic leading coefficient.
 %\newpage
%Using (\ref{Int2}) in the fourth line, 
%we check for $1\leq n\leq N$ and $y+\beta+n \geq 1$, 
%$$
%\begin{aligned}
%  \oint_{\Gamma_0} \frac{dz   }{2\pi \I z^{y+\beta+1  }(1-z)^n}
% &=\oint_{\Gamma_0}\frac{dz ~(1-z)^{N-n}}{2\pi \I z^{y+\beta+1 }(1-z)^N}
%\\
%&=\sum_{k=0}^{N-n}\left({N-n}\atop{k}\right)(-1)^k 
%\oint_{\Gamma_0}\frac{dz  ~z^k}{2\pi \I z^{y +\beta +1}(1-z)^N}
%\\
%&=(N-n)!\sum_{k=0}^{N-n}\frac{(-1)^k }{k!(N-n-k)!}\frac{(y +\beta+1 -k)_{N-1}}{(N-1)!}
%\\
%&=
% \oint_{\Gamma(y,y-1,\dots,y-N+n)}\frac{(N-n)!dz}{2\pi \I(z-y)_{N-n+1}}    \frac{(z +\beta+1 )_{N-1}}{(N-1)!} ,
%\end{aligned}
%$$
%where we  use for $0\leq k\leq N-n$,
%$$\frac{(z-y)_{N-n+1}}{z-y+k}\Bigr|_{z=y-k}=(-1)^k k! (N-n-k)!.$$
%When the second term in (\ref{Int2}) is present, one applies the same argument as above to that second term with the flip $y\to -y$; this leads the the second term in (\ref{IntId1}). This ends the proof of Lemma \ref{qlim}.\qed

\medbreak

  The next lemma deals with identities involving Vandermonde's with certain rows removed. For ${\bf x}\in \BR^m$ and ${\bf y}\in \BR^n$, define the Vandermonde
$$\begin{aligned}
\Dt_m({\bf x})&=\det\left(\begin{array}{cccccccc}
x_1^{m-1 }&\dots&x_1^0\\
\vdots &\dots&\vdots\\
x_m^{m-1}&\dots&x_m^0\\
\end{array}\right)=\prod _{1\leq i<j\leq m}(x_i-x_j)
\\
\Dt_m^{\hat k}({\bf x})&=
\det\left(\begin{array}{cccccccc}
x_1^{m }&\dots&x_1^{k+1}& x_1^{k-1}&\dots&x_1^0\\
\vdots & &\vdots&\vdots &&\vdots\\
x_{m } ^{m }&\dots&x_{m }^{k+1}& x_{m }^{k-1}&\dots&x_{m }^0\\
\end{array}\right)=\Dt_m({\bf x})e_{m-k}({\bf x}),
\end{aligned}$$
where ${} {\hat k}$ refers to removing the column containing the $k$th power. Also
$$
e_k({\bf y})=\sum_{i_1<i_2<\dots<i_k} y_{i_1} \dots y_{i_k} .
$$

\noindent We will need the following Lemma:
\begin{lemma}\label{Vanderm1} Then we have (here $Q(z):=\prod_1^m (z-x_i)\prod_1^n (z-y_i)$)
$$\begin{aligned}
\frac{\Dt_n^{\widehat {n-k}}({\bf y})}
{\Dt_{n+m}({\bf x},{\bf y})}
=(-1)^{m(m-1)/2}\frac{ \Dt_m({\bf x}) }{\prod_1^m Q'(x_\ell)}e_k({\bf y}),\end{aligned}$$
and so in particular, setting $k=0$, 
\be \begin{aligned}
\frac{\Dt_n^{}({\bf y})}
{\Dt_{n+m}({\bf x},{\bf y})}
=(-1)^{m(m-1)/2}\frac{ \Dt_m({\bf x}) }{\prod_1^m Q'(x_\ell)} .\end{aligned}\label{Vanderm2}
\ee
The following symmetric function, with variables removed, has the integral representation: (with $Q_q$ defined in (\ref{Qq}))
\be
\begin{aligned}
(-1)^{r-1}e_{r-1}&(q_1,\dots,\hat q_k,\hat q_{i_1},\dots,\hat q_{i_d},\dots ,q_{d+N})
\\
&=\oint_{\Gamma_\infty}\frac {dz}{2\pi \I z^{N-r+1}}\frac{Q_q(z)}{(z-q_k)\prod_{\ell=1}^d (z-q_{i_\ell})},
\end{aligned}
\label{symmf}\ee

\end{lemma}
\proof The first relation is shown by multiplication by $z^{n-k}(-1)^k$ and summing from $k=0$ to $n$, which leads to an obvious identity . The second is straightforward and the third is just the residue Theorem. \qed
%\section{Further form of the q-kernel}

%  \newpage

%\end{document}
%\newpage

\section{From Karlin-McGregor to the ${\mathbb K}_q$-kernel}

%The $q$-dependent probability measure $\BP_q$ on the space of all $N+1$-uples of partitions $\{ \nu^{(0)},\ldots,  \nu^{(N )}\}$, given by (\ref{qprob}) namely 
%which for $q\to 1$ leads to the uniform probability (\ref{ProbUnif}).
Since the $\mathbb K$-process of red dots has a nonintersecting paths description, one expects to have a Karlin-McGregor formula for the $q$-dependent probability measure $\BP_q$, as in  (\ref{qprob}) above. This will be shown in Proposition \ref{Karlin}. Then in Proposition \ref{Prop:Kern0}, we will obtain a kernel by adapting the arguments in Borodin-Ferrari-Pr\"ahofer \cite{BFP} and in Borodin-Rains \cite{BR} to these new circumstances. %This section refers to the {\bf multi-cut model}.

We first need some notation. Remember from (\ref{interlac}) the interlacing pattern $\Id_{x^{(m-1)} \prec x^{(m)}}$ for $1\leq m\leq N$.
%$$
%x^{(m )}_{d+m-1} \leq x^{(m-1)}_{d+m-2}   \leq x^{(m )}_{d+m-2} \ldots \leq \ldots   
% \leq x^{(m )}_2\leq x^{(m-1)}_1  \leq x^{(m )}_1
%$$
%\[
%x^{(m-1)}_m \leq x^{(m-1)}_{m-1} \leq \ldots \leq x^{(m-1)}_2 \leq x^{(m-1)}_1
%\]
To the left of each sequence $x^{(m-1 )}$, we add an extra point, a  so-called {\em virtual point}  $x^{(m-1 )}_{d+m} =$ virt; it can be thought of as a point at $-\infty$. 
Then we define for $x \in \{virt\} \cup \BZ$, $ z,k\in \BZ$ and $k\geq 0$, the following functions, where the matrix $H$ was defined in (\ref{defphipsi}):
\be\begin{aligned}
\chi_i(z)&:=\Id_{z=y_i},  \mbox{   for   } 1\leq i\leq  d\\
\varphi_k (x,z)& := q^{(k-1)(z-x)} \Id_{x\in\BZ} \Id_{x \leq z} + q^{(k-1)z} \Id_{x=\textrm{virt}} %\Id_{x\leq z}
 \\
&=\frac 1{2\pi \I}\oint_{\Ga_0}\frac {du}{1-uq^{k-1} }~\left(\frac {1}{u^{z-x+1}}\right)   +   q^{(k-1)z} \Id_{x=\textrm{virt}} 
\\&=h_{z-x}(q^{k-1})+q^{(k-1)z}\Id_{x=\textrm{virt}} 
\\
\psi_i (z) &:= \sum^{N+d}_{j=1} (H^{-1})_{ij} \Id_{z=x_j}
%= \sum^{N+d}_{j=1} (H^{-1})_{ij} \Id_{z=x_j } 
, \mbox{   for   } 1\leq i\leq d+N.
\end{aligned}\label{defphipsi}\ee
So, in particular we have $\varphi_k(\mbox{virt},z)=q^{(k-1)z}$.

\begin{proposition} \label{Karlin}The probability (\ref{qprob}) can be written as a product of $N+2$ determinants
\be\label{KMcG}\begin{aligned}
\BP_q(&\nu^{(0)}, \nu^{(1)} , \ldots \, \nu^{(N )},~~\mbox{such that}~~\mu=\nu^{(0)}\subset \nu^{(1)} \subset \ldots \subset \nu^{(N )}=\lb)
\\
&= C_{N,d,q} \det (\chi_i (x^{(0)}_j))_{1\leq i,j\leq  d}\\
& ~~~\times~
\prod^{N }_{m=1} \det (\varphi_{d+m }(x^{(m-1)}_i,x^{(m)}_j))_{1\leq i,j \leq d+m }
 \det (\psi_i (x^{(N )}_j))_{1\leq i,j\leq N+d},%
\end{aligned}\ee
with
 $$\begin{aligned}
 C_{N,d,q}&=q^{dN(d+N)+\frac 13 N(N^2-1)}.%\\
% &=q^{\sum_1^N(d+i-1)(d+i)/2} q^{  (N-1)d(d+1)/2}
\end{aligned} $$

\end{proposition}

\bigbreak

 Define the following $\ast$-operation between two real or complex functions $f(k)$ and $g(k)$ defined for $k\in \BZ$:
\be f\ast g:= \sum_{k\in \BZ}f(k)g(k)
\label{ast};\ee
e.g., for functions $\varphi_k$ as above, the operation $\ast$ means a convolution $(\varphi_k\ast\varphi_{k+1})(x,y)= \varphi_k(x,\mbox{\tiny$\circ$})\ast\varphi_{k+1} ( \mbox{\tiny$\circ$},y)= \sum_{z\in \BZ}\varphi_k(x,z)\varphi_{k+1}(z,y)$.
Subsequently, define 
\be \begin{aligned} \label{conv}
\varphi^{n ,m+1}(x,y) := (\varphi_{n  }\ast \ldots\ast\varphi_{m})(x,y)\dt_{n\leq m}.
\end{aligned}\ee 
The functions $\psi_{j}$, as in (\ref{defphipsi}), have a natural extension in terms of the geometry of $\bf P$:  namely, 
 for $0\leq m\leq N-1$ and $1\leq k\leq d+N$, define:
\begin{equation}\begin{aligned}
%\Psi^{d+k}_{d+k-j}(x):=
 \psi_k^{(m+1)} (x)&:= \varphi_{d+m+1 }{(x,\mbox{\tiny$\bullet$})}\ast\dots\ast \varphi_{d+N }(\mbox{\tiny$\bullet$},\mbox{\tiny$\circ$})\ast \psi_k (\mbox{\tiny$\circ$})
 %\\&=
 = \varphi^{d+m+1 ,d +N+1}{ }    (x,\mbox{\tiny$\circ$})\ast \psi_k (\mbox{\tiny$\circ$})
 \label{Psi}
\end{aligned}\end{equation}
and for $m=N$ we have 
  $
 \psi_j^{(N+1)}=\psi_j .
 $

\begin{proposition}\label{Prop:Kern0} The $  {\mathbb K}_q$-point-process is determinantal with kernel, for $0\leq m,n\leq N$, given by\footnote{The $\ell$-summation in the third term of (\ref{Kernel0}) goes up to $\ell=d+n$; that term contains $\varphi^{d+n+1,d+n+1} $, which we declare $=1$.
 }:%  kernel The general theory redevelopped for the case of nontrivial initial and final conditions leads to the following kernel for the dot-process: (The proof still needs to be written!)
\be\begin{aligned}
   {\mathbb K}_q (m&,x;n,y)\\
=&-  \varphi_{d+m+1}\ast\ldots\ast \varphi_{d+n } (x,y)\\
&+\sum_{{1\leq k\leq d+N}\atop{1\leq \ell\leq d}} \psi_{k}^{(m+1)}(x)(  M^{-1})_{k\ell}(\chi_{\ell}\ast
\varphi^{d+1,d+1+n})(y) 
%\varphi_{d+1 }\ast \ldots\ast\varphi_{d+n }
\\
&+\sum_{{1\leq k\leq d+N}\atop{d+1\leq \ell\leq d+n }} \psi_{k}^{(m+1)}(x)(  M^{-1})_{k\ell} (\varphi_\ell (\mbox{virt},\mbox{\tiny$\bullet$})\ast  \varphi^{\ell+1,d+1+n }   %@\varphi_{\ell+1}\ast\dots\ast \varphi_{d+n })
 (\mbox{\tiny$\bullet$},y)),
\end{aligned}\label{Kernel0}\ee
where for $1\leq i, j\leq N+d$:   
\be\begin{aligned}
M_{k\ell} :=\left\{\begin{array}{llll}
%&
 \chi_k(\mbox{\tiny$\circ$})\ast\psi^{(1)}_{ \ell}(\mbox{\tiny$\circ$})%=\chi_i(\mbox{\tiny$\bullet$})\ast (\varphi_{d+1}\ldots \varphi_{d+N})(\mbox{\tiny$\bullet$},\mbox{\tiny$\circ$})\ast \psi_{ j}(\mbox{\tiny$\circ$})%(z_{d+N+1})
,&1\leq k\leq d\\ \\
 %\varphi_{i }(\mbox{virt},\cdot)\ast\psi_j^{(i-d+1)}(\cdot)=
%&
 \varphi_{k }(\mbox{virt},\mbox{\tiny$\circ$})  \ast\psi^{(k+1-d )}_{ \ell}(\mbox{\tiny$\circ$})
 %=(\varphi_{i +1}\ast\ldots\ast \varphi_{d+N})(\mbox{\tiny$\bullet$},\mbox{\tiny$\circ$}) \ast \psi_{ j}(\mbox{\tiny$\circ$})
  ,& d+1\leq k\leq d+N.
%\\ \\
%&
 %\varphi_{i }(\mbox{virt},\mbox{\tiny$\circ$})\ast \psi_{ j}(\mbox{\tiny$\circ$}),&i=N+d
  \end{array}
 \right.
\end{aligned}\label{M}\ee

\end{proposition}

\noindent Before giving the proof of Propositions \ref{Karlin} and \ref{Prop:Kern0}, the following Lemma will be needed:

\begin{lemma}\label{lemma:det}The following determinantal identities hold: for $1\leq m\leq N$,  
\be\begin{aligned}
&\det (\chi_i (x^{(0)}_j))_{1\leq i,j\leq  d} =\prod_1^d \Id_{x^{(0)}_j=y_j}
%\\
 \\& \det ( \varphi_k (x^{(m-1)}_i, x^{(m)}_j))_{1\leq i,j\leq d+m } 
 %
 %=  
% \sum^{d+m }_{i=1} x^{(m)}_i - \sum^{d+m-1}_{i=1} x^{(m-1)}_{i })} \det (\Id_{x^{(m-1)}_i \leq x^{(m)}_j})_{1\leq i,j\leq d+m}
%\\   
%&
  =  q^{(k-1)(|x^{(m)}|-|x^{(m-1)}|)}    \Id_{x^{(m-1)} \prec x^{(m)}}
\\
 &\det (\psi_i (x^{(N )}_j))_{1\leq i,j\leq N+d}    =  \det (h_{x_i-y_j} (q^d,\ldots,q^{d+N-1}))^{-1}_{1\leq i, j \leq N+d} ~ \prod^{d+N}_{j=1} \Id_{x^{(N )}_j=x_j}.
\end{aligned}\label{det}\ee
Remembering the $\ast$-operation (\ref{ast}), we have the following convolution properties for $x,y\in \BZ$, with the last identity valid for $1\leq k\leq d+N,~1\leq m\leq N$:
$$\begin{aligned}
\varphi^{n+1,m+1}(x,y) = (\varphi_{n+1 }\ast \ldots\ast\varphi_{m})(x,y)
 &=h_{y-x}(q^{n},\ldots,q^{m-1})%\Id_{x\leq y}
 \Id_{n <m}\\
 &%\hspace*{5cm}
 =\frac {1}{2\pi \I}\oint_{\Gamma_0}\frac{dz}{z^{y-x+1}}\prod_{\ell=n}^{m-1}\frac {1} {1\!-\!uq^\ell} \Id_{n <m}
%\end{aligned}$$
%$$
 \\
 \chi_{\ell}\ast %\varphi^{n+1,m-1 }%
  (\varphi_{n+1 }\ast \ldots\ast\varphi_{m})
 ( y)
&=h_{y-y_\ell }(q^{n},\ldots,q^{m-1}) \Id_{n <m}%\Id_{x \leq y_\ell}
%
%
%$$
%\varphi_{n_1+d}\ast \ldots\ast\varphi_{n_2+d-1}(x,y)
%=h_{y-x}(q^{n_1+d-1},\ldots,q^{n_2+d-2})\Id_{x\leq y}\Id_{n_1<n_2}
%$$
\\
\varphi_n (\mbox{virt},\cdot)\ast  (\varphi_{n+1}\ast\dots\ast \varphi_{m})
 %\varphi^{n+1,m-1}
  (\cdot,y)&=
\Id_{n<m}\frac{q^{(n-1)y}}{(1-q)\ldots(1-q^{m-n })}
\\
%$$\begin{aligned}%d+N-
 \varphi_{n } (\mbox{virt},\cdot)\ast  (\varphi_{n +1} \ast\ldots\ast \varphi_{m})(\cdot,\mbox{\tiny$\circ$}) \ast \psi_{j}(\mbox{\tiny$\circ$}) 
&=\\%
  \frac{1}{(1-q)\ldots(1-q^{m-n })} &\sum_{\ell=1}^{
N+d} 
(H^{-1})_{ j\ell}~ q^{(n -1)x_\ell} ,
\end{aligned}$$
 
\be\begin{aligned}
\varphi^{n+1,m+1}%(\varphi_{n +1}\ast \ldots\ast\varphi_{m })
 (x,\mbox{\tiny$\circ$}) \ast \psi_j(\mbox{\tiny$\circ$})
%&=\sum_{z\in \BZ} h_{z-x}(q^{ n  },\ldots,q^{m-1})\sum_{\ell=1}^{N+d}(H^{-1})_{j\ell}\Id_{z=x_\ell}\\
&= \sum_{\ell=1}^{d+N}(H^{-1})_{j\ell}h_{x_\ell-x}(q^n,\dots,q^{m-1})
 %\Id_{x\leq y}\Id_{n_1<n_2}
\\
\chi_i(\cdot)\ast\varphi^{n+1,m+1}%(\varphi_{n +1}\ast \ldots\ast\varphi_{m })
 (\cdot,\mbox{\tiny$\circ$}) \ast \psi_j(\mbox{\tiny$\circ$})
 &= \sum_{\ell=1}^{d+N}(H^{-1})_{j\ell}h_{x_\ell-y_i}(q^n,\dots,q^{m-1}).\end{aligned}\label{phipsi}\ee
\be \label{psikm}\begin{aligned}
\psi_k^{(m)}(x)%&:=(\varphi_{d+m}\ast \ldots\ast\varphi_{d+N })(x,\cdot) \ast \psi_k(\cdot)
%\\&
  =\sum_{\ell=1}^{d+N}(  H^{-1})_{k\ell} h_{x_\ell-x}(q^{d+m-1 },\ldots,q^{d+N-1}).%\Id_{x\leq y}\Id_{n_1<n_2}
\end{aligned}\ee

\end{lemma}          

\proof The proof of the first three determinantal identities is straightforward; for instance,
$$\begin{aligned}
\det &( \varphi_k (x^{(m-1)}_i, x^{(m)}_j))_{1\leq i,j\leq d+m } 
 \\&=  q^{(k-1)(%\displaystyle 
 \sum^{d+m }_{i=1} x^{(m)}_i - \sum^{d+m-1}_{i=1} x^{(m-1)}_{i })} \det (\Id_{x^{(m-1)}_i \leq x^{(m)}_j})_{1\leq i,j\leq d+m}
\\
&= q^{(k-1)(|x^{(m)}|-|x^{(m-1)}|)}    \Id_{x^{(m-1)} \prec x^{(m)}},\end{aligned}$$
taking into account the fact that the last row of the matrix above
%$\varphi_k (x^{(m-1)}_i, x^{(m)}_j)$ 
 is given by $(q^{(k-1)x^{(m)}_1}   ,\dots,  q^{(k-1)x^{(m)}_{d+m}} )$.

The convolution properties (\ref{phipsi}) follow from the identity (\ref{Schursum}), applied to  the partition $\lb=(r,0,0,\ldots)$ and $\mu=(s,0,0,\ldots)$ with $r>s$, i.e., $\mu \subset \lb$. Indeed one has:
$$
s_{\lb}(x)=h_r(x), ~~s_{\lb\backslash\mu}(x)=h_{r-s}(x),
$$
and so, setting $\nu=(\al,0,0,\ldots)$, one has
\be 
h_{r-\alpha}(x,y)= \sum_{\al \in \BZ%s\leq \al\leq r
 } h_{r-\alpha}(x ) h_{ \alpha-s}( y)
.\label{convolution}\ee

\newpage

\vspace*{-2.5cm}

\noindent Another example of proof is:
$$\begin{aligned}%d+N-
\varphi_{n } (\mbox{virt},\cdot)\ast  (\varphi_{n +1}&\ast\ldots\ast \varphi_{m})(\cdot,\mbox{\tiny$\circ$}) \ast \psi_{j}(\mbox{\tiny$\circ$})\\
&=\sum_{z\in \BZ}\frac{q^{(n- 1)z} }{(1-q)\ldots(1-q^{m-n })}  \sum_{\ell=1}^{
N+d}
(H^{-1})_{j\ell}\Id_{z=x_\ell}
\\
&=\frac{1}{(1-q)\ldots(1-q^{m-n })} \sum_{\ell=1}^{
N+d} 
(H^{-1})_{ j\ell}~ q^{(n -1)x_\ell} ,
\end{aligned}$$
ending the proof of Lemma \ref{lemma:det}.\qed

\medbreak

\noindent{\em Proof of Proposition \ref{Karlin}}: 
 The denominator of the probability (\ref{qprob}) can be expressed as a determinant, using formula (\ref{skewSchur2}). Also from the second formula (\ref{det}), it follows that 
 $$\begin{aligned}
    \prod^{N }_{m=1} \det &(\varphi_{d+m }(x^{(m-1)}_i,x^{(m)}_j))_{1\leq i,j \leq d+m }
 \\
 &
  = q^{(N-1)|x^{(N )}|+d (|x^{(N )}|-|x^{(0)}|)}    q^{-\sum^{N-1}_1|x^{(i)}|}
  \Id_{x^{(0)}\prec  x^{(1)} \prec\ldots \prec  x^{(N )}}
  \\
 &
  = q^{(d+N-1)|x| -d |y|)}    q^{-\sum^{N-1}_1|x^{(i)}|}
  \Id_{x^{(0)}\prec  x^{(1)}\prec\ldots  \prec  x^{(N )}}
\end{aligned}$$
 and then using the two remaining formulas of (\ref{det}), one checks   \footnote{One uses $\sum_1^N (d+i-1)(d+i)/2 =  \tfrac 12( dN(d+N) +\tfrac 13 N(N^2-1))$} :
$$\begin{aligned}
\BP(&\nu^{(0)}, \nu^{(2)} , \ldots \, \nu^{(N )},~~\mbox{such that}~~\mu=\nu^{(0)}\subset \nu^{(1)} \subset \ldots \subset \nu^{(N )}=\lb)
\\
&=
\frac{(q^{-1})^{ \sum^{N}_{i=2}|\nu^{(i-1)}|- (N-1)|\nu^{(0)}|} }
{s_{\lambda\backslash \mu} (q^{1-N}, q^{2-N},\ldots, q^{-1}, q^0) }
\Id_{\nu^{(0)}\subset \nu^{(1)} \subset \ldots \subset \nu^{(N )}}
\Id_{\nu^{(0)}=\mu}  \Id_{\nu^{(N )}=\lb}\\
& =q^{ N(d+N-1)(d+(N+1)/2) }q^{-\sum_1^N (d+i-1)(d+i)/2+d(d+1)/2} q^{(N-1)d(d+1)/2}\\
&\frac{q^{(  d+N-1)(|x|-|y|)}(q^{-1})^{ \sum^{N-1}_{i=1}|x^{(i)}|- (N-1)|y |  } }
{ \det (h_{x_i-y_j} (q^d,\ldots,q^{d+N-1}))_{1\leq i,j \leq N+d}
}\Id_{\nu^{(0)}\subset \nu^{(1)} \subset \ldots \subset \nu^{(N )}}
\Id_{\nu^{(0)}=\mu}  \Id_{\nu^{(N )}=\lb}
\\
%&=\frac{ q^{(  N+d-1) |x|  -d|y|} (q^{-1})^{ \sum^{N}_{i=2}|\nu^{(i)}| } }
%{  \det (h_{x_i-y_j} (q^d,\ldots,q^{d+N-1}))_{1\leq i,j \leq N+d}
%}
%\Id_{\nu^{(1)}\subset \nu^{(2)} \subset \ldots \subset \nu^{(N+1)}}
%\Id_{\nu^{(1)}=\mu}  \Id_{\nu^{(N+1)}=\lb}
&=  C_{N,d,q} \frac{q^{(  d+N-1) |x|  -d|y|}   (q^{-1})^{ \sum^{N-1}_{i=1}|x^{(i)}|   }}
{  \det (h_{x_i-y_j} (q^d,\ldots,q^{d+N-1}))_{1\leq i,j \leq N+d}
}\Id_{\nu^{(0)}\subset \nu^{(1)} \subset \ldots \subset \nu^{(N )}}
\Id_{\nu^{(0)}=\mu}  \Id_{\nu^{(N )}=\lb}
\\
&= C_{N,d,q} \det (\chi_i (x^{(0)}_j))_{1\leq i,j\leq  d}\\
& ~~~\times~
\prod^{N }_{m=1} \det (\varphi_{d+m }(x^{(m-1)}_i,x^{(m)}_j))_{1\leq i,j \leq d+m }
 \det (\psi_i (x^{(N )}_j))_{1\leq i,j\leq N+d},%
\end{aligned}$$
  establishing the Karlin-MacGregor formula of Proposition  \ref{Karlin}.\qed
 %
%\newpage

 %\newpage

%\newpage

%\vspace*{-2cm}

\noindent {\em Proof of Proposition \ref{Prop:Kern0}}:  
The technology explained in \cite{BFP,BR} will now be adapted to these new circumstances. 
%To be consistent with their notation, we will replace the levels $0,\dots,N$  by $1,\dots,N+1$; so $x_i^{(m)}$ for $0\leq m\leq N$ becomes  $x_i^{(m)}$ for $1\leq m\leq N+1$. In the end we will obtain a kernel $\widetilde{\mathbb K}_q$, from which ${\mathbb K}_q$ can be deduced  upon flipping the roles of $m,n$,
%\be
%{\mathbb K}_q(m  ,x;n  ,y):=\widetilde{\mathbb K}_q(n+1,x;m+1,y)
%,~~~\mbox{for $0\leq m,n\leq N$.}\label{Kern0}\ee
%It also will be convenient to set $\tilde \varphi_k:= \varphi_{d+k}$ and in view of the notation (\ref{phipsi}), $\tilde \varphi^{n+1,m+1}:= \varphi^{d+n+1,d+m+1}$
%\ref{Kern0}
%
Given a set ${\frak X}$, let $L$ be a positive-definite $|{\frak X}|\times |{\frak X}|$-matrix with rows and columns parametrized by the points $x\in {\frak X}$, with $L_X$ a $X\times X$-submatrix with rows and columns parametrized by $x\in X$. A random point process ($L$-ensemble) on $ {\frak X}$ will be defined by the probability ${\mathbb P}(X)=\frac{\det(L_X)}{\det (\Id+L)}$ 

\newpage

\vspace*{-2cm}

\noindent for $X \in {\frak X}$. It is determinantal with kernel $K=L(\Id+L)^{-1}$; that means that the correlation $\rho(Y)={\mathbb P}(X\subset {\frak X}~|~Y\subset X)=\det K_Y$. Given a decomposition ${\frak X}= {\frak Y}\cup {\frak Y}^c$, with ${\frak Y}\neq \emptyset$ and with corresponding block matrix $\Id_{\frak Y}$, the conditional $L$-ensemble process on ${\frak Y}$ is defined by ${\mathbb P}(Y)=\frac{\det L_{Y\cup {\frak Y}^c}}{\det (\Id_{\frak Y}+L)}$ and is determinantal with correlation kernel $K=\Id_{\frak Y}-(\Id_{\frak Y}+L)_{{\frak Y}\times {\frak Y}}^{-1}$.

This is now applied to ${\frak X}= {\frak Y}\cup {\frak Y}^c=:
({\frak X}^{(0)}\cup\dots\cup {\frak X}^{(N)})\cup 
\{1,2,\dots,d, x_{d+1}^{(0)},\dots,x_{d+N}^{(N-1)}  \}
$, where ${\frak X}^{(k)}=\BZ$ are all the point configurations at level $0\leq k\leq N$ and the $x_{d+k+1}^{(k)}=\mbox{\em virt}$ are {\it virtual variables}, playing the role of point at $-\infty$ for each set ${\frak X}^{(k)}$ for $0\leq k\leq N-1 $;  .

%Removing the constant, the sum of the right hand side of the Karlin-McGregor formula (\ref{KMcG}) over all possible point configurations equals $\det M$, where \be \label{M}
%M=\Phi^{(d)}T^{0,1}\dots  T^{N-1,N}\Psi^{(d+N)}  \ee
%In view of the notation (\ref{phipsi}), it will be convenient to set $\tilde \varphi_k=\tilde \varphi^{k,k+1}:= \varphi_{d+k}=\varphi^{d+k,d+k+1}$ and $\tilde \varphi^{n+1,m+1}:= \varphi^{d+n+1,d+m+1}$. 

We now apply this set-up to the specific situation of the paper, although this is completely general.  Indeed, we now define (infinite) matrices in terms of the functions appearing in the Karlin-McGregor formula (\ref{Kernel0}), where $\Phi^{(d)} $ contains the initial condition (points not in $\bf P$ at level $y=0$, in particular the lower-cut) for the red dots, where the $T^{k-1,k}$ are the transition functions from one level to another and  where $\Psi^{(d+N)}$ contains the final condition for the red dots (points not in $\bf P$ at level $y=N$, in particular the upper-cuts):
\be\begin{aligned}
\bullet &~ \Phi^{(d)}  =\mbox{ $\{1,\dots,d\}\times {\frak X}^{(0)}$-matrix, with
$(\Phi^{(d)}) (i,x^{(0)})  = \chi_i (x^{(0)} )$}
\\
 \bullet &~T^{k-1,k}=\mbox{ $({\frak X}^{(k-1)}\cup \{x_{d+k}^{(k-1)}\}) \times {\frak X}^{(k)}$-matrix, for $1\leq k\leq N $}
 \\
 &~~~~~~~~~~~~~~~~~~\mbox{ with $T^{k-1,k}(x^{(k-1)} , x^{(k )} )  =  %\tilde\varphi^{d+k,d+k+1} (x^{(k-1)} , x^{(k )} )=
  \varphi_{d+k }(x^{(k-1)},x^{(k)})$}
\\
 \bullet &~\Psi^{(d+N)}=\mbox{ $ ({\frak X}^{(N )}  \times \{1,\dots,d+N\})$-matrix,} 
  \\ &~~~~~~~~~~~~~~~~~~~
 \mbox{with $\Psi^{(d+N)}(x^{(N )},j):=\psi_j(x^{N })   %=:\Psi^{(d+N)}_{d+N-j} (x^{N })
   $.}
\end{aligned}
\label{matrices}\ee
%By the Eynard-Mehta Theorem, the $(i,j)$th block of the correlation kernel is given by @@$K_{ij}=T^{[i,N+1)}\Psi^{(d+N)}M^{-1} \Phi^{(d)}T^{[1,j)}-T^{[i,j)}$ and 

 The conditional $L$-ensemble on ${\frak X}$ with ${\frak Y}$ as above is given by the block matrix, where each block has infinite size. The only nonzero determinants of the form $\det L_{Y\cup {\frak Y}^c} $ appearing in $\BP(Y)$ above are suitable finite minors of this matrix $L$ below. These lead exactly to the Karlin-McGregor formula (\ref{KMcG}), as is easily seen by picking appropriate rows and columns according to the coordinates on top and to the left of the matrix $L$ below :

%\newpage

   \vspace{.3cm}

% $L=$
 
% \newpage
 
  \setlength{\unitlength}{0.017in}\begin{picture}(0,60)
  
  \put(30,0){\makebox(0,0){$L$=}}
\put(155,-70){\makebox(0,0) {\rotatebox{0}{\includegraphics[width=190mm,height=170mm]
 {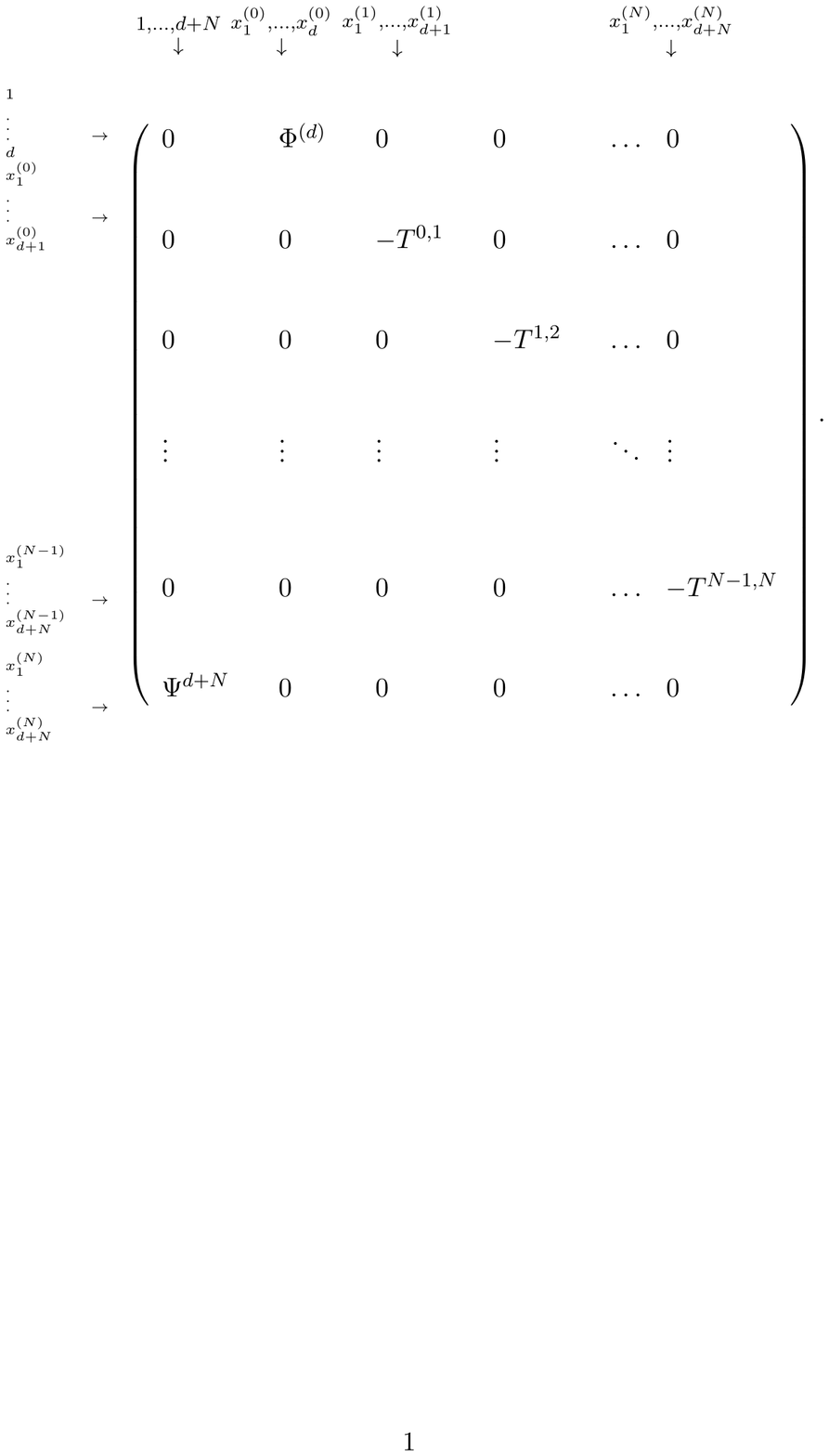}}}} \end{picture}
 
 \newpage

 %\newpage

%@@@

%

%\newpage
 
% \[\xymatrix{
%&  & 1  \cdots  d+N       & x_1^{(1)}  \cdots   x_d^{(1)}    & x^{(2)}_1 \cdots x^{(2)}_{d+1}    & \ldots  &  x^{(x)+1}_1\cdots x^{(x)+1)_{d+N}}\\
%& &   \ar@{<->}[l]       &  \ar@{<->}[l]                              &   \ar@{<->}[l]   &     &   \ar@{<->}[l] \\
%1 & \ar@{<->}[dd]       &   &  & &  \\
%\vdots & &  & &  & \\
%d & & &  &  & }\]

%\newpage

%\vspace*{-.9cm}

\noindent We now move rows and columns of $L$ in order to write the matrix as a block $({\frak Y}^c\otimes {\frak Y})$-matrix $L_{ {\frak Y}^c\otimes {\frak Y} }$. Indeed, move the rows corresponding to the virtual variables
$$
\begin{array}{ccc}
x^{(0)}_{d+1}
,~
x^{(1)}_{d+2}
,~
\dots
,
x^{(k-1)}_{d+k}
,~
\dots
,~
x^{(N-1)}_{d+N }
\end{array}
$$
respectively to the $d+1$, $d+2$, $\ldots,d+k,\ldots$, $d+N^{th}$ row, yielding the matrix below, where the ${\frak X}^{(k-1)}\times{\frak X}^{(k)}$-matrices $W_{k-1,k }$ are the matrices $T^{k-1,k }$ with the last row removed,

   \vspace{.3cm}

% $L=$
 
% \newpage
 
  \setlength{\unitlength}{0.017in}\begin{picture}(0,60)
  
%  \put(30,0){\makebox(0,0){$L$=}}
\put(115,-70){\makebox(0,0) {\rotatebox{0}{\includegraphics[width=190mm,height=170mm]
 {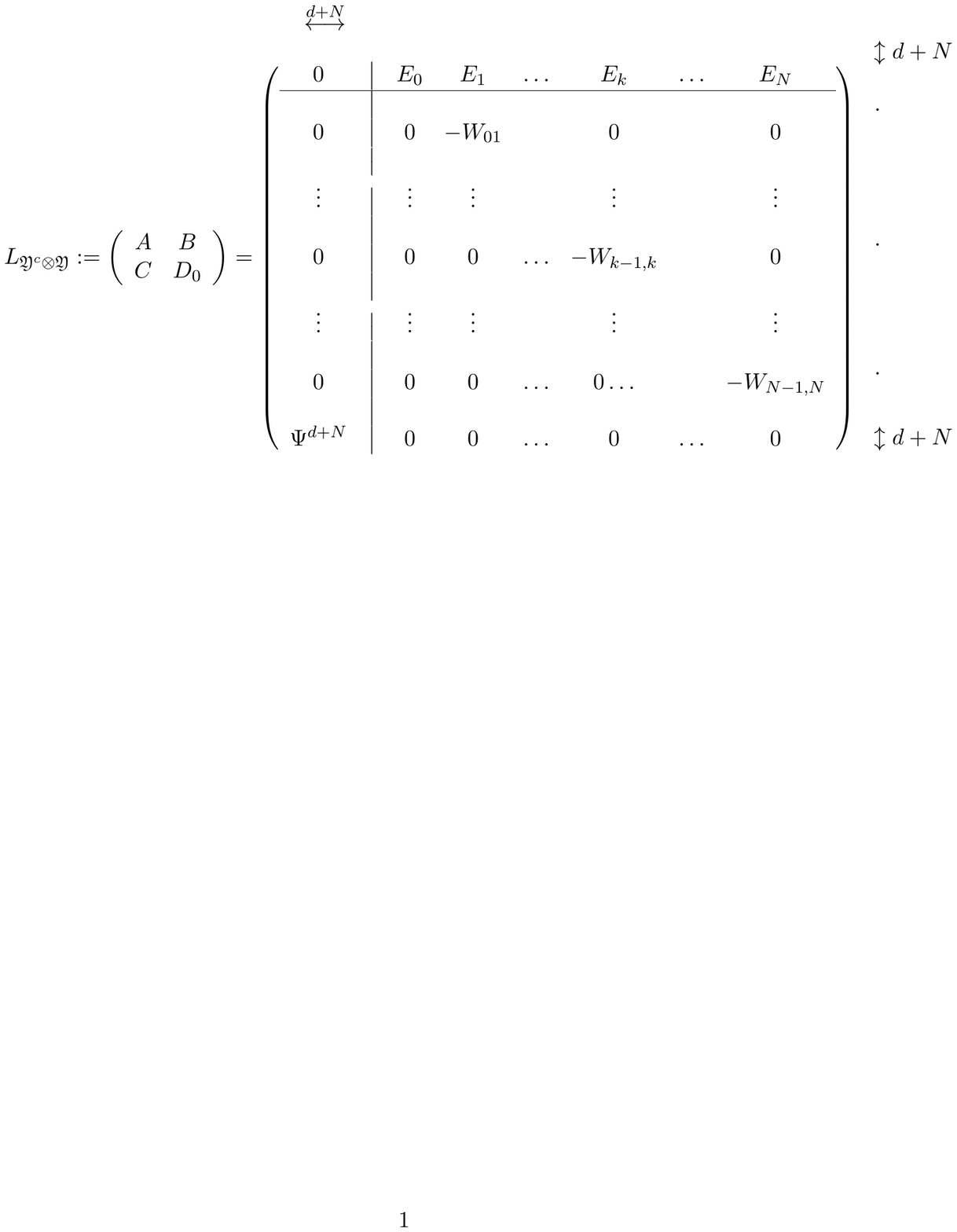}}}} \end{picture}

%  \newpage

 \vspace*{2.5cm}
 %\newpage

\noindent and
$$  \Id_{\frak Y} =\left(
\begin{array}{ccc}
0 & 0\\
0 & \Id
\end{array}
\right),
$$
with
$$
 \begin{aligned}
 A=0,~~&~~~ ~~~~~~~~B  := (E_0,\ldots,E_N)
 ,~~~~
 ~~~~~~~~~~~~~~C : = (0,\ldots,0,\Psi^{d+N})^T
 \\ \\
 D_0 &:= 
 \left(
\begin{array}{cccccccc}
0 & -W_{01} & 0 & \cdots & 0
\\
0 & 0 & -W_{12} & \cdots & 0
\\
\vdots & \vdots & \vdots & & \vdots
\\
0 & 0 & 0 &\cdots & -W_{N-1,N}
\\
0 & 0 & 0 &\cdots & 0
\end{array}
\right).%,~~~~~D:= \Id_{\frak Y} + D_0
 \end{aligned}
$$

%\vspace*{5mm}
 
 \noindent In the matrix $B$, the $E_k$ is a $(d+N)\times  {\frak X}^{(k )} $-matrix for $0\leq k \leq N$, defined by, for $k>0$, %($1\leq j\leq d+k$)
 $$
 E_k(i,x^{(k )}) =
 \left\{
 \begin{array}{llllll}
 0 & \mbox{for}  \   1\leq i\leq d+k-1
 \\
T^{k-1,k}(x^{(k-1)}_{d+k}, x^{(k )})=\varphi_{d+k}(x^{(k-1)}_{d+k}, x^{(k)} )%=@ \tilde \varphi^{k,k+1} (x^{(k)}_{d+k}, x^{(k+1)} ) 
   &\mbox{for}  \  i=d+k
 \\
  0 & d+k+1 \leq i \leq d+N  
 \end{array}
 \right. ,
 $$
% and $E_0$ a $(d+N)\times {\frak X}^{(0)}$-matrix
 and, for $k=0$,
% 
%\hspace*{62mm} $\stackrel{d}{\longleftrightarrow}$
 $$
 E_0 = \left(
\begin{array}{ccc}
\Phi^d\\
0\\
\vdots\\
0
\end{array}
\right).
 $$

\vspace*{-24mm}

\hspace*{78mm}
\begin{minipage}{60mm}
$\updownarrow d $

\vspace*{.5cm}

$\updownarrow N$
\end{minipage}

   \vspace*{15mm}

 \noindent Defining $D:= \Id_{\frak Y} + D_0$, the point measure on ${\frak Y}$ is determinantal with correlation kernel, given by
 \be\label{Ktilde}\begin{aligned}
 \widetilde {\mathbb K} :=\Id_{\frak Y}-(\Id_{\frak Y}+ L_{ {\frak Y}^c\otimes {\frak Y} })^{-1}\bigr|_{{\frak Y}\times {\frak Y}}, % \mbox{with}~~~M = BD^{-1} C-A,
=& \Id_{\frak Y}-\left(
\begin{array}{ccc}
A & B\\
C & D 
\end{array}
\right)^{-1}\\
=&\Id_{\frak Y}-\left(
\begin{array}{ccc}
\star & \star\\
\star & D ^{-1}-D^{-1}CM^{-1}BD^{-1}
\end{array}
\right)\bigr|_{{\frak Y}\times {\frak Y}}
\\=& \Id_{\frak Y} - D^{-1} + D^{-1} C M^{-1} B D^{-1} ,\end{aligned}\ee
with $M = BD^{-1} C-A $; it will be shown that this is precisely the matrix $M$ in (\ref{M}). To do so, we need to compute the different matrices appearing in the expression (\ref{Ktilde}). It is easily seen that
  
% \vspace*{5mm}
 %
% \hspace*{13mm}$ \stackrel{d}{\longleftrightarrow}$   $\stackrel{d+1}{\longleftrightarrow}$     \hspace{10mm} $\stackrel{d+N}{\longleftrightarrow}$  \\[0.5cm]
$$D^{-1}=\left(
\begin{array}{ccccccc%llllllll
 }
\Id & W_{[0,1)} &   \ldots &  W_{[0,N )}\\
&&&\\
  &   & &   \vdots\\
&&&\\
0 & \Id &   &  \\
&&& \\
\vdots & \ddots &   &  W_{[N-1,N )} \\
&&& \\
0 & 0 & 0 &   \Id
\end{array}
\right) 
\mbox{with}~W_{[n,m)} :=W_{n,n+1}   \ldots   W_{m-1,m} \Id_{m>n}.$$

%\vspace*{-48mm}

%\hspace*{60mm}
%\begin{minipage}{60mm}
% $\updownarrow d$

% \vspace{2.8cm}
 
%$ \updownarrow d + N-1$
 
% \vspace{0.4cm}
 
% $\updownarrow d+N$ 
 
% \end{minipage}

%\vspace*{5mm}

  \noindent Coordinatewise, the ${\frak X}^{(n)}\times {\frak X}^{(m)}$-matrices $W_{[n,m)}$ are given by
% $$
 %W_{[n,m)} =
% \left\{
% \begin{array}{lllll}
% W_{[n,m)} :=W_{n,n+1}   \ldots   W_{m-1,m} \Id_{m>n}%& \mbox{for} \quad m >n
 %\\
 %0 & & & \mbox{for} \quad m \leq n
% \end{array}
% \right.
%$$
 %with 
$$\begin{aligned}
  W_{[n,m)}( x^{(n)},x^{(m)})&=  \varphi_{d+1+n  } (x^{(n)} ,x^{(n+1)} )  \ast \ldots \ast  \varphi_{d+m } (x^{( m-1)} , x^{ (m)} ) 
 \\
 &=   \varphi^{d+1+n,d+1+m} (x^{(n)} ,x^{(m)} )
.\end{aligned}$$
 Moreover,
 $$\footnotesize\mbox{$
 BD^{-1} = \left(\stackrel{\stackrel{{\frak X}^{(0)}}{\leftrightarrow}}{E_0},~~    E_0 W_{[0,1)}\stackrel{\stackrel{{\frak X}^{(1)}}{\longleftrightarrow}} 
 { + } E_1, \ldots ,\sum^{m-1 }_{k=0} E_{k } W_{[k ,m)}\stackrel{\stackrel{{\frak X}^{(m)}}{\longleftrightarrow}}{ +}E_{m },\ldots,  \sum^{N-1}_{k=0} E_{k } W_{[k ,N )}\stackrel{\stackrel{{\frak X}^{(N)}}{\longleftrightarrow}}{ +}E_N\right)~\updownarrow d+N$},
 $$
 
  \newpage
 
  \vspace*{-1cm}
  
  \noindent where
 % 
% Notice that for $k \geq 1$, the following matrix of size $(d+N)\times {\frak X}^{(m)}$ reads as
 %
 %$E_k  W_{k+1,m}$
 $$
 \begin{aligned}
% &
 E_k   W_{[k,m)} &=\left(\begin{array}{ccc}
 O\\
  \varphi_{d+k} (x^{(k-1)}_{d+k} , x^{(k )} )%, \ldots ,\tilde\varphi^{k,k+1} (x^{(k)}_{d+k} \ , \ x^{(k+1)}_{d+k})
 \\
 O
 \end{array}\right)  \leftarrow d+k
 \\
 &~~~~\times \left(\varphi^{d+1+k,d+1+m}(x^{(k )}  \ , \ x^{(m)}_j) \delta_{m>k }\right)%_{1\leq i \leq d+k \atop 1\leq j \leq d+m-1}
 \\ \\
 &={\footnotesize \left(\begin{array}{ccc}
 O\\
 (\varphi_{d+k} \ast \varphi^{d+1+k,d+1+m}) (x^{(k-1)}_{d+k} \ , \ x^{(m)}) \delta_{m>k }%,\dots,  (\tilde\varphi_k \ast \tilde\varphi^{k+1,m}) (x^{(k)}_{d+k} \ , \ x^{(m)}_{d+m-1}) \delta_{m>k }
 \\
 O
 \end{array}\right)\leftarrow d+k }.
 \end{aligned}
 $$

 %\newpage
 
 \noindent Hence 
 $BD^{-1}$ consists of $N+1$ column-blocks, the $m$th block being given below, for $0\leq m \leq N $, with $O_{N-m}$ a $(N-m)\times {\frak X}^{(m)}$ matrix of zeros,  
 \medbreak
 
 $ \sum^{m-1}_{k=0} E_{k } W_{[k,m)}+E_{m }=$%
 \begin{equation}
\left(\begin{array}{c}
( \chi_1 \ast  \varphi^{d+1,d+1+m})(x^{(m)}) \\
\vdots \\
( \chi_d \ast  \varphi^{d+1,d+1+m})(x^{m}) \\ \\
( \varphi_{d+1} \ast  \varphi^{d+2,d+1+m} )(x^{(0)}_{d+1} , x^{(m)}) \\
\vdots \\
( \varphi_{d+k} \ast  \varphi^{d+1+k,d+1+m }) (x^{(k-1)}_{d+k} , x^{(m)})  \\
\vdots \\
( \varphi_{d+m-1} \ast  \varphi^{d+m,d+1+m}) (x^{(m-2)}_{d+m-1} , x^{(m)})  \\ \\
 \varphi_{d+m}  (x^{(m-1)}_{d+m } , x^{(m)}) \\ \\
 O_{N-m }
\end{array}\right)_{ ~~~~~~~~~~}
\label{BD}
\end{equation}

\vspace*{-55mm}

\hspace*{95mm}
\begin{minipage}{60mm}
$ \leftarrow d+1$

$\vdots$

$\leftarrow d+k$

$\vdots$

\vspace*{3mm}

$\leftarrow d+m-1$

\vspace*{4mm}

$\leftarrow d+m $ .
 \end{minipage}

  \vspace*{1.2cm}
  \bigbreak

\noindent Notice that in the matrix above, we have that the arguments $x^{(k-1)}_{d+k}=\mbox{\em virt}$. Moreover 
\be
 D^{-1}C =
 \left(\begin{array}{c}
 W_{[0,N)} \Psi^{d+N} \\
 \vdots\\
 W_{[k,N)} \Psi^{d+N}\\
 \vdots\\
 W_{[N-1,N )} \Psi^{d+N}\\ \\
 \Psi^{d+N}
 \end{array}
 \right) 
=  \left(\begin{array}{l}
 %\Psi^{(d)}_{d-j}
   \psi_j^{(1)}(x^{(0)})\\
 \vdots\\
% \Psi^{(d+k )}_{d+k -j} 
 \psi_j^{(k+1)}(x^{(k)})
 \\
 \vdots
 \\
 %\Psi^{(d+N-1)}_{d+N-1-j} 
  \psi_j^{(N)}(x^{(N-1)})
 \\ \\
 %\Psi^{(d+N)}_{d+N-j} 
  \psi_j^{(N+1)}(x^{(N )})
 \end{array}
 \right)_{1\leq j \leq d + N} \!\!\!\!\!\!\!\!\!\!\!\!,\mbox{   with  }\psi_j^{(N+1)}=\psi_j ,%~~~~~~~~~~~~~~~~~ ,~~~~~~~~~~~~~~~~~~~~~~~~~~~~
  \label{DC}\ee
 %
%\vspace*{-46mm}
%
%\hspace*{100mm}
%\begin{minipage}{60mm}
%$ \updownarrow d$
%
%\vspace*{6mm}
%
%$\updownarrow d+k-1$
 %\vspace*{2mm}
%$\vdots$
%
%$\updownarrow d+N-1$
%
%\vspace*{4mm}
%
%$\updownarrow d+N$
 %\end{minipage}
 %
  %\vspace*{4mm}
  \noindent because, for $0\leq k\leq N-1$ and $1\leq j\leq d+N$, we have, using the notation of  (\ref{Psi}),
  $$ \begin{aligned}
    ( W_{[k,N )}  \Psi^{d+N})(x^{(k)},j)&= W_{[k,N )} (x^{(k)},x^{(N)}) \Psi^{d+N} (x^{(N)},j)
     \\
     &=   \varphi^{d+1+k ,d+1+N }(x ^{(k)}, \mbox{\tiny$\circ$})\ast \psi_j(\mbox{\tiny$\circ$}) 
    %\\&
    = \psi^{(k+1) }_{j} (x^{(k)} )
  , \end{aligned}$$
 % %
and $\Psi^{d+N}(x^{(N)},j)=\psi_j^{(N+1)} (x^{(N)})=\psi_j (x^{(N)}) $. Hence the $(m+1,n+1)^{th}$ block of $(D^{-1}C)M^{-1}(BD^{-1})$ reads for\footnote{In the second summation below, we also have $\varphi^{d+1+n ,d+1+n  } =1$.} $0\leq m,n\leq N$,
 $$
 \begin{aligned}
 & \sum_{1\leq k \leq d+N \atop 1\leq \ell \leq d} %\Psi^{(d+n )}_{d+n -j} 
  \psi_k^{(m+1)}(x^{(m )})  (M^{-1})_{k\ell} (\chi_\ell \ast  \varphi^{d+1,d+1+n }) (x^{(n )})
 \\
 &+ \sum_{1\leq k \leq d+N \atop d+1\leq \ell \leq d+n}   
  \psi_k^{(m+1)}(x^{(m )}) (M^{-1})_{k\ell}  (\varphi_{\ell } \ast  \varphi^{\ell+1 ,d+1+n  }) (x^{(\ell-d-1 )}_\ell , x^{(n )}),
 \end{aligned}
 $$
with $x^{(\ell-d-1 )}_\ell=\mbox{\em virt}$ and the $(m+1,n+1)^{th}$ block of $(\Id_{\frak Y}-D^{-1})$ reads 
$$-W_{m ,n }=\varphi^{d+1+m,d+1+n}(x^{(m )},x^{(n )}).$$ 
 Then upon setting $x=x^{(m )} $ and $y=x^{(n )} $ as running variables, and combining the two last equations, we conclude that for $0\leq m,n\leq N$,
$$
  \begin{aligned}\tilde {\mathbb K}(m,x;n,y)&=(\Id_{\frak Y}-D^{-1} + D^{-1} C M^{-1} B D^{-1})_{m+1,n+1}
 % \\
%&  = -  \varphi^{d+n+1,d+m+1} (x,y)
%\\
%& + \sum_{1\leq k \leq d+N \atop  1\leq \ell \leq d} \Psi^{(d+n )}_{d+n -k} (x) (M^{-1})_{k\ell} (\chi_\ell \ast  \varphi^{d+1,d+m+1})(y)
%\\
%& + \sum_{1\leq k \leq d+N \atop d+1\leq \ell \leq d+N} 
 %\Psi^{d+n }_{d+n -k} (x) (M^{-1})_{k\ell} (\tilde\varphi_{\ell } \ast \tilde\varphi^{\ell +1,d+m+1}) (\mbox{\it virt}%x^{(\ell-d-1)}_\ell
 %  , y).
  \end{aligned}
$$
gives indeed formula (\ref{Kernel0}), using the notation (\ref{phipsi}) for $\varphi^{n,m}$.
 Finally, we have
 $M = (M_{ij})_{1\leq i,j \leq d+N}=BD^{-1} C = (BD^{-1})C$,
 whose entries, upon using (\ref{BD}), are given by
  $$
 M_{i,j} =
 \left\{
 \begin{array}{lll}
 (\chi_i \ast  \varphi^{d+1,d+1+N} ) (\mbox{\tiny$\bullet$},\mbox{\tiny$\circ$}) \ast \psi _j (\mbox{\tiny$\circ$}), &\mbox{for} \  1\leq i \leq d,\\ \\
 ( \varphi_{i } \ast  \varphi^{i +1,d+1+N }) (\mbox{\em virt},\mbox{\tiny$\circ$}) \ast \psi _j (\mbox{\tiny$\circ$}) , &\mbox{for} \  d+1\leq i \leq d +N -1,
 \\ \\
  \varphi_{d+N}   (\mbox{\em virt},\mbox{\tiny$\circ$}) \ast\psi _j (\mbox{\tiny$\circ$}) & \mbox{for} \ i=d+N.
 \end{array}
 \right. 
$$
  This ends the proof of Proposition \ref{Prop:Kern0}.\qed 
   
   \remark For each each $0\leq m\leq N$, consider the linear space $V_{m}$, 
   $$
   \begin{aligned}V_{m}:=\mbox{span}\{\mbox{functions in brackets of (\ref{BD})}\}=\mbox{span}\{\phi_j^{(m)},~1\leq j\leq d+m\},
   \end{aligned}
   $$
  where $\phi_j^{(m)},~1\leq j\leq d+m$ is a new basis of $V_{m}$, defined by the conditions $\phi_i^{(m)}\ast \psi_i^{(m)}=\Id_{i=j}$ for $1\leq i, j\leq d+m$. Assuming this new basis is such that $\varphi_{d+m}(\mbox{virt},x)=c_m \phi_{d+m}^{(m)}(x),~c_m\neq 0,~0\leq m\leq N$ (Assumption A in \cite{BR,BFP}), then it is shown that the kernel (\ref{Ktilde}) has the following simple form:
  $${\mathbb K}_q (m ,x;n,y)=-\varphi^{ d+1+m, d+1+n} (x,y)+\sum_1^{d+n}\psi_j^{(m)}(x)\phi_j^{(n)}(y). 
  $$
   Assumption A can in our case indeed be achieved for some appropriate choice of basis, which at the end of the day leads to a representation of the kernel in terms of $(2,3)$-mops on the circle.

% \newpage

\section{Transforming $M$ into $\tilde M$ and computing its inverse}

The purpose of this section is to transform the matrix $M$, which appears in the kernel (\ref{Kernel0}) for the multi-cut case, into a new matrix $\tilde M$, which is easily invertible. To do so, one needs to define $D_n:=\mbox{diag}(e_{n-1},\ldots, e_0)$, $\Dt^{(q)}_n$ a Vandermonde and $ \Pi_n$ a permutation matrix, (recall $ \PR^{ }_{N-1}(z)=\sum_{i=1}^Ne_{i-1}z^i$)
$$    \!\Dt^{(q)}_n:=\left(\!\!\begin{array}{cccccccccc}
1&     1&   \ldots & 1\\
(q^{-1})^{n-1}&(q^{-1})^{n-2}&\ldots&1
\\
(q^{-2})^{n-1}&(q^{-2})^{n-2}&\ldots&1
\\
\vdots&\vdots&  &   \vdots\\
(q^{-(n-1)})^{n-1}&(q^{-(n-1)})^{n-2}&\ldots&1
\end{array}\!\!\right)~,~
%~\mbox{and}~~
 \Pi_n:=\left(\!\!\begin{array}{cccccccc}
&&&&&1\\
O&&&&1& \\
 &&&1&&  \\
&&1&&&O \\
&1& &&& \\
\end{array}\!\!\right)  .$$

\begin{proposition}\label{Minverse} The inverse of the matrix $\widetilde M:=MT^{\top -1}$ reads as follows:
\be\begin{aligned}
 \widetilde M ^{-1}&=
 \left(\begin{array}{cccc}
\Id_d &| & O\\
\hline
O&|&\left(\begin{array}{ccccccc}
 {\dt_{2d+N-i+1,j}}{(q;q)_{i-d-1}}
\end{array}\right)_{{d+1\leq i,j\leq d+N} }
\end{array}\right)
%\\ \\
%&=\left(\begin{array}{cccc}\Id_d &| & O\\  \hline
%&\Bigr|\\
%O&\Bigr|& \begin{array}{ccccccc}
%&&&&(q;q)_0\\
%&&&.& \\
%O&&.& &  \\
%&.& & & O\\(q;q)_{N-1}&&&&
%\end{array} 
%\end{array}\right)
\end{aligned}\label{Minv}\ee
where $T$ is a $x$ and $y$ independent square matrix of size $d+N$,
$$T=\left(\begin{array}{ccccc}\Id_d&|&O\\
\hline 
O&|&T_N\end{array}\right),
$$
with \be\begin{aligned}
T_N&:=
\diag (q^{(N-j)d} )_{1\leq j\leq  N} \Dt_N ^{(q)-1}\Pi_ND_N^{-1}
\diag\left(q^{-(d+N )(d+N-j)}\right)_{1\leq j\leq  N}.
\end{aligned}\label{T}\ee

\end{proposition}

Before giving the proof of Proposition \ref{Minverse}, we need Lemma
\ref{lemma HT} below. 
 It is useful to modify the matrix $H$ of size $d+N$ (as in (\ref{Hdef})) to $\widetilde H$, by an $x$ and $y$ independent  transformation $T$ of block form, so as to be close to the Schur case rather than the skew-Schur case; that is, so that the $(d+N,N)$-block of the matrix $\widetilde H$ are pure powers of $q$.  

\begin{lemma}\label{lemma HT} Multiplying the block matrix $H$ with the matrix $T$ above (\ref{T}), yields ($O_{n,m}$ denotes a $n\times m$-matrix of zeros):%Given the notation above,
\be\begin{aligned}
\widetilde H:=HT&=:\left(\begin{array}{cccc}A&|&B\end{array}\right)
\left(\begin{array}{ccccc}\Id_d&|&O\\
\hline 
O&|&T_N\end{array}\right)
\\&=\left(\bigl[h_{x_i-y_j}(q^d,\ldots,q^{d+N-1})\bigr]_{{1\leq i\leq d+N}\atop{1\leq j\leq d}} \Bigr| ~~[q^{ x_i(d+N-j)}
 \bigr] _{{1\leq i\leq d+N}\atop{1\leq j\leq N}}\right)
%\\&=@\left((q^{d(x_i+j-m_1)}P_{N-1}(q^{x_i+j-m_1}))_{{1\leq i\leq d+N}\atop{1\leq j\leq d}} \Bigr| ~~(q^{ x_i(d+N-j)}
%))_{{1\leq i\leq d+N}\atop{1\leq j\leq N}}\right)
\\&=\left({{\bigl[q^{d( x_i-y_j)}  \PR^{ }_{N-1}(q^{x _i-y_j})  \bigr]_{{1\leq i\leq c}\atop{1\leq j\leq d}}
}\atop{O_{d+b,d}}}
 ~\Bigr| ~~ \bigl[q^{  dx_i }q^{x_i(N-j)}
\bigr]_{{1\leq i\leq d+N}\atop{1\leq j\leq N}}\right).
\end{aligned}\label{Htilde}\ee
%with $T_N$ given above in (\ref{T}).
%where (see Figure page 2 for $m_1$)
%$$ \tilde \PR^{(j)}_{N-1}(q _i ):=q^{d(j-m_1) }\PR_{N-1}(q^{x_i+j-m_1}),~\mbox{with}~
%\PR_{N-1}(z)=\prod_{i=1}^{N-1}\frac { 1-zq ^{ i}  }{ 1-q ^{i } }$$
%$T_N$ is given by

\end{lemma}

\proof At first, we have the following identity  (Remember: $\PR_{n-1}(z)=\sum_{i=1}^n e_{i-1} z^{i-1} 
$; see formula (\ref{calP}))
\be\begin{aligned}
(z^{n-1},\ldots,z^0)&=( \PR_{n-1}(z(q^{-1})^{n-1}),\ldots, \PR_{n-1}(z))  \widetilde T,~~~\mbox{with}~~ \widetilde T_n =   {\Dt^{(q)}_n} ^{-1}\Pi_nD_n^{-1}~.
%\\
%&=(q^d P_{n-1}(z(q^{-1})^{n-1}),\ldots,q^{dn}P_{n-1}(z))%\mbox{diag}(q^{ d},\ldots,q^{ dn} )^{-1}
 %T
\end{aligned}\label{Ttilde}\ee
Indeed, $\widetilde T$ is such that:
%$$\begin{aligned}
% z^{n-\ell} &=\sum_{k=1}^n \PR_{n-1}(z(q^{-1})^{n-k})  \widetilde  T_{k\ell}
% \\
% &=\sum_{i=1}^n z^{i-1}e_{i-1}\sum_{k=1}^n   (q^{-1})^{(n-k)(i-1)}  \widetilde T_{k\ell}
%\end{aligned}$$
%or what is the same
$$\begin{aligned}
 z^{ \ell-1} &=\sum_{k=1}^n \PR_{n-1}(z(q^{-1})^{n-k}) 
  \widetilde  T_{k,n-\ell+1}
 \\&
  =\sum_{i=1}^n z^{i-1}e_{i-1}\sum_{k=1}^n   (q^{-(i-1)})^{(n-k) }  \widetilde T_{k,n-\ell+1},
\end{aligned}$$
  implying, for all $1\leq \ell,i\leq n$, that 
$
e_{i-1}^{-1}\dt_{i,\ell}=\sum_{k=1}^n (q^{-(i-1)})^{(n-k)} 
\widetilde  T_{k,n-\ell+1}   
$, which is statement (\ref{Ttilde}) in matrix notation.
 
Apply this to the matrix $H$ and $n=N$, 
$$
H=\left(\begin{array}{cccc}\overbrace{A}^d&|&\overbrace{B}^{ N}\end{array}\right),
$$
where the matrix $A$ of size $(d+N,d)$ consists of the $d$ columns corresponding to the points in the lower-cut $y_1,\ldots,y_d$ and the matrix $B$ of size $(d+N,N)$ to the points $ y_{d+1},\ldots,y_{d+N} $ to the left of $\bf P\cap \{y=0\}$. One now uses formula (\ref{hP}) and notices that $N-1+x_i-y_{d+j}\geq 0$ for $i,j\geq 1$; this enables us to omit the indicator in the formula for the matrix $B=(B_{ij})_{{1\leq i\leq d+N}\atop{ 1\leq j\leq  N}}$ :
$$\begin{aligned}
B_{ij} &=\left(h_{x_i-y_{d+j}}(q^d,\ldots,q^{d+N-1}\right)\\&= 
 q^{d(x_i+d+j)}\PR_{N-1}(q^{x_i+d+j})\Id_{N-1+x_i-y_{d+j}\geq 0}\\
&=q^{d(x_i+d+N)}  \left. \PR_{N-1}\bigl(z(q^{-1})^{N-j}\bigr)\right|_{z=q^{x_i+d+N}} (q^{-1})^{ (N-j)d },
\end{aligned}$$
and thus, in matrix notation,
$$\begin{aligned}
\mbox{diag}(q^{-d(x_i+d+N)}) B
~\mbox{diag}(q^{(N-j)d} ) 
&=
\left(  \PR_{N-1}\bigl(z(q^{-1})^{N-j }) \Bigr |_{z=q^{x_i+d+N}} \right)
\Bigr|_{{1\leq i\leq d+N}\atop{ 1\leq j\leq  N}}.
%\\
%&=\left( P_{N-1}(q^{x_i+d+j})\right)
 \end{aligned}$$
From (\ref{Ttilde}), it follows that for $ \widetilde T_N=T$, as in (\ref{T}),
$$\begin{aligned}
\mbox{diag}(q^{-d(x_i+d+N)}) B
~\mbox{diag}(q^{(N-j)d} ) \widetilde T_N%&=( (q^{(x_i+d+N)})^{N-1} ,\ldots, (q^{(x_i+d+N)})^{0})_{1\leq i\leq d+N} \\
&=\left(( z^{N-j})  \Bigr|_{z= q^{ x_i+d+N } }\right)_{{1\leq i\leq d+N}\atop{1\leq j\leq  N}}, 
%&=\left(q^{(d+N+x_i)(}\right). 
\end{aligned}$$
%Therefore we have that
and so
$$\begin{aligned}
 B
~\mbox{diag}(q^{(N-j)d} ) \widetilde T_N%&%=(  q^{( d+N+x_i)(d+N-j)}  ,\ldots,  q^{(d+N+x_i)(d+N-j)}  )_{1\leq i\leq d+N}
%\\
&= \left(q^{(d+N+x_i)(d+N-j)}\right)_{{1\leq i\leq d+N}\atop{1\leq j\leq  N}}
\\
&= \left(q^{  x_i (d+N-j)}\right)_{{1\leq i\leq d+N}\atop{1\leq j\leq  N}}.
\diag\left(q^{(d+N )(d+N-j)}\right)_{1\leq j\leq  N},\end{aligned}$$ 
which we rewrite as 
%$$\begin{aligned}
%B~\mbox{diag}(q^{(N-j)d} ) \widetilde T_N
%\diag\left(q^{-(d+N )(d+N-j)}\right)_{1\leq j\leq  N}&=\left(q^{  x_i (d+N-j)}\right)_{{1\leq i\leq d+N}\atop{1\leq j\leq  N}}.
 %\end{aligned}$$
%Hence 
$$
BT_N=
\left(q^{  x_i (d+N-j)}\right)_{{1\leq i\leq d+N}\atop{1\leq j\leq  N}}
$$
for $T_N$ given in terms of $\widetilde T_N$, as defined in (\ref{Ttilde}),
$$\begin{aligned}
T_N&:= \mbox{diag}(q^{(N-j)d} ) \widetilde T_N
\diag\left(q^{-(d+N )(d+N-j)}\right)_{1\leq j\leq  N}
%\\
%&=
%\mbox{diag}(q^{(N-j)d} ) \Dt_n ^{-1}\Pi_nD_n^{-1}
%\diag\left(q^{-(d+N )(d+N-j)}\right)_{1\leq j\leq  N}
\end{aligned}$$
This corresponds to the matrix $T_N$ in (\ref{T}). The multiplication by $T$ leaves the first $d$ columns unchanged. 
Also the last equality in (\ref{Htilde}) is an immediate consequence of formula (\ref{hP}) which includes the indicator function, since by hypothesis $N-1+x_c-y_1\geq 0$. This ends the proof of Lemma \ref{lemma HT}. \qed

%\newpage

\bigbreak 

\noindent{\em Proof of Proposition \ref{Minverse}}: Inserting (\ref{Psi}) in the matrix $M$, as defined in (\ref{M}), one sees that $M$ can be written, for appropriate choices of $ \Phi_i$'s, as:
\be M_{ij}= \Phi_i(\cdot,\mbox{\tiny$\circ$})\ast \psi_j(\mbox{\tiny$\circ$})
=  \Phi_i(\cdot,\mbox{\tiny$\circ$})\ast \sum_{\al=1}^{d+N} H^{-1}_{ j\alpha}\Id_{z=x_\alpha}(\mbox{\tiny$\circ$}),~~~~1\leq i, j\leq N+d.\label{M1}\ee
 Since $\widetilde H=HT$ in Lemma \ref{lemma HT}, this suggests transforming the $\psi_k$ into a new expression $\tilde \psi_\al$, defined by    
$$
\widetilde\psi_j=\sum_{\al=1}^{d+N} T^{-1}_{j\al}   \psi_\al=
   \sum_{\al=1}^{d+N}\widetilde H^{-1}_{j\al}\Id_{z=x_\al},
 $$
 and consequently $ \psi_k^{(m+1)}$ into a new $\widetilde \psi_k^{(m+1)}$ as defined in (\ref{Psi}) and (\ref{psikm}), 
 \be\begin{aligned}
\widetilde \psi_k^{(m+1)}(x):=
&\sum_{\ell=1}^{d+N}(\widetilde H^{-1})_{k\ell} h_{x_\ell-x}
 (q^{d+m  },\ldots,q^{d+N-1}).%\Id_{x\leq y}\Id_{n_1<n_2}
\end{aligned}\label{psi-new}\ee
From (\ref{Kernel0}), (\ref{M}) and (\ref{Psi}), this in turn induces a transformation on $M$, preserving the kernel (\ref{Kernel0}): 
   $$M_{ij}=(\Phi_i\ast \sum_\al T_{j\al}\widetilde \psi_\al)=\sum_\al T_{j\al}(\Phi_i\ast \widetilde \psi_\al)= \sum_\al(\Phi_i\ast \widetilde \psi_\al)_{ }T^\top_{\al j}=(\widetilde M T^\top)_{i j},
 $$
where $\widetilde M=MT^{\top -1}$ is the matrix $M$ as in (\ref{M1}), but with $ \psi_\al$ replaced by $\tilde \psi_\al$. In view of the form of the kernel (\ref{Kernel0}), we check, using the previous expression, that the kernel is indeed preserved. It suffices by (\ref{Kernel0}) and (\ref{Psi}) to check the identity
$$\begin{aligned}
 \sum_k\psi_k M^{-1}_{k\ell}&=\sum_k \sum_j H^{-1}_{kj}\Id_{z=x_j}\sum_\al T^{\top -1}_{k\al}\widetilde M^{-1}_{\al,\ell}
 \\&=\sum_{\al,j}\Id_{z=x_j}\widetilde M^{-1}_{\al,\ell}\sum_k H^{-1}_{kj}T^{\top -1}_{k\al}  %\\&@=
 %\sum_{\al,j}\Id_{z=x_j}\tilde M^{-1}_{\al,\ell}\sum_k T^{  -1}_{\al k }H^{-1}_{kj} 
 \\&=\sum_{\al,j}(HT)^{-1}_{\al,j}\Id_{z=x_j}\widetilde M^{-1}_{\al,\ell} 
 \\
 &= \sum_{\al,j}(\tilde H )^{-1}_{\al,j}\Id_{z=x_j}\widetilde M^{-1}_{\al \ell} =\sum_{\al } \tilde \psi ^{ }_{\al } \widetilde M^{-1}_{\al \ell},
 \end{aligned}$$
inducing a similar relation $\sum_k\psi^{(m+1)}_k M^{-1}_{k\ell}=\sum_{\al } \tilde \psi^{(m+1)}_{\al } \widetilde M^{-1}_{\al \ell}$. %
%{\bf For matrices $H$ and $\tilde H$:}  %Setting $n=d$ and $m=d+N$ in (\ref{phipsi}) yields the first part of 

\noindent We now compute explicitly $\widetilde M$:
\newline {\bf For   $1\leq i\leq d$ and $1\leq j\leq d+N $ }, setting $n=d$ and $m=d+N$ in (\ref{phipsi})
$$\begin{array}{llllllll}
M_{ij}=\chi_i\ast \varphi_{d+1}\ast\ldots\ast \varphi_{d+N}\ast \psi_{ j} 
&=(\varphi_{d+1}\ldots \varphi_{d+N})(y_i,\cdot)\ast \psi_{ j}(\cdot)
%\\
%&=\sum_{z\in \BZ}h_{z-y_i}(q^d,\ldots,q^{d+N-1})\sum_{k=1}^{N+d}(A^{-1})_{ j,k}
%\Id_{z=x_k}
\\  \\
&\dis =\sum_{k=1}^{N+d}(H^{-1})_{ j,k}h_{x_k-y_i}(q^d,\ldots,q^{N+d-1})
\\  
&=\dis\sum_{k=1}^{N+d} (H^{-1})_{ j,k}H_{k,i}={\mathbb I}_{ij}
\\   \widetilde M_{ij}= (MT^{\top -1})_{ij}={\mathbb I}_{ij},
\end{array}$$
  since the left  $(d+N)\times d$-block of  $T$ is $\left({\Id_d}\atop{O}\right)$. From this discussion it follows that the $  \widetilde \psi_j,~  \widetilde H,~  \widetilde M$ are related to each other in the same way as the $\psi_j,~H,~M$, via the transformation $\psi_j\mapsto \widetilde \psi_j,~H\mapsto \widetilde H,~M\mapsto \widetilde M$ .
  
  %
%\newpage

%\vspace*{-1.8cm}

\noindent {\bf For $d+1\leq i\leq d+N$ and $1\leq j\leq d+N$}, one has, using the same identity for $\varphi_{i } (\mbox{virt},\cdot)\ast  (\varphi_{i +1} \ast\ldots\ast \varphi_{d+N})(\cdot,\mbox{\tiny$\circ$}) \ast \psi_{j}(\mbox{\tiny$\circ$})$ as in (\ref{phipsi}), but for $\psi_{j}$ replaced by $\widetilde \psi_{j}$, one checks that 
%
%{\bf For the matrix $H$ }, the second part of $M$ is given by ( for )
%$$\begin{aligned}%d+N-
%\varphi_{i } (\mbox{virt},\cdot)\ast  (\varphi_{i +1}&\ast\ldots\ast \varphi_{d+N})(\cdot,\mbox{\tiny$\circ$}) \ast \psi_{j}(\mbox{\tiny$\circ$})\\
%&=\sum_{z\in \BZ}\frac{q^{(i- 1)z} }{(1-q)\ldots(1-q^{N-i+d})}  \sum_{\ell=1}^{
%N+d}
%(H^{-1})_{j\ell}\Id_{z=x_\ell}
%\\
%&=\frac{1}{(1-q)\ldots(1-q^{N-i+d})} \sum_{\ell=1}^{
%N+d} 
%(H^{-1})_{ j\ell}~ q^{(i -1)x_\ell} ,
%\end{aligned}$$
%
%whereas for the {\bf  matrix $\widetilde H$}, the second part of $M$ is  given by (also for $d+1\leq i$)
$$\begin{aligned}%d+N-
\widetilde M_{ij}&=\varphi_{i } (\mbox{virt},\cdot)\ast  (\varphi_{i +1} \ast\ldots\ast \varphi_{d+N})(\cdot,\mbox{\tiny$\circ$}) \ast \widetilde \psi_{j}(\mbox{\tiny$\circ$})
\\
%&=\sum_{z\in \BZ}\frac{q^{(i- 1)z} }{(1-q)\ldots(1-q^{N-i+d})}  \sum_{\ell=1}^{
%N+d}
%(\widetilde H^{-1})_{j\ell}\Id_{z=x_\ell}
%\\
&=\frac{1}{(1-q)\ldots(1-q^{N-i+d})} \sum_{\ell=1}^{
N+d} 
(\widetilde H^{-1})_{ j\ell}~ q^{(i -1)x_\ell} 
\\
&=\frac{1}{(1-q)\ldots(1-q^{N-i+d})} \sum_{\ell=1}^{
N+d} 
(\widetilde H^{-1})_{ j\ell}~\widetilde H_{\ell,2d+N-i+1} 
\\
&=\frac{1}{(1-q)\ldots(1-q^{N-i+d})} 
 \dt _{ j,2d+N-i+1}   =\frac{ \dt _{ j,2d+N-i+1} }{(q;q)_{d+N-i}}.
\end{aligned}$$
So, the matrix $
\widetilde M$ reads:
$$
\widetilde M=\left(\begin{array}{cccc}
\Id_d &| & O\\
\hline
O&|&\left(\begin{array}{ccccccc}
\frac{\dt_{2d+N-i+1,j}}{(q;q)_{d+N-i}}
\end{array}\right)_{{d+1\leq i,j\leq d+N} }
\end{array}\right)
%=\left(\begin{array}{cccc}
%\Id_d &| & O\\
%\hline
%&\Bigr|\\
%O&\Bigr|& \begin{array}{ccccccc}
%&&&&\frac 1{(q;q)_{N-1}}\\
%&&&.& \\
%O&&.& &  \\
%&.& & & O\\\frac 1{(q;q)_{0}}&&&&
%\end{array} 
%\end{array}\right)
 $$
and so
\be\begin{aligned}
 \widetilde M ^{-1}
 %&=
 %\left(\begin{array}{cccc}
%\Id_d &| & O\\
%\hline
%
%O&|&\left(\begin{array}{ccccccc}
% {\dt_{2d+N-i+1,j}}{(q;q)_{i-d-1}}
%\end{array}\right)_{{d+1\leq i,j\leq d+N} }
%\end{array}\right)
%\\ \\
&=\left(\begin{array}{cccc}
\Id_d &| & O\\
\hline
&\Bigr|\\
O&\Bigr|& \begin{array}{ccccccc}
&&&&(q;q)_0\\
&&&.& \\
O&&.& &  \\
&.& & & O\\
(q;q)_{N-1}&&&&
\end{array} 
\end{array}\right)
\end{aligned},\label{Minv}\ee
establishing Proposition \ref{Minverse}.\qed

%\newpage

\section{The integral representation of the ${\mathbb K}_q$-kernel}
%All statements in this section refer to the multi-cut case. 
 The kernel will be given in terms of multiple contour integrals, one of them being a $d+2$-fold integral. But at first, we express the kernel in terms of the basic functions $h_r$ and $\widetilde \psi_k^{(m+1)}(x)$, as in (\ref{psi-new}).

 \begin{proposition}\label{Prop5.1} For $0\leq m,n\leq N,$ the $q$-kernel reads as follows:
%The kernel is given by
\be\begin{aligned}
 {\mathbb K}_q(&m,x;n,y) \\
 { =}&-  h_{y-x}(q^{d+m },\ldots,q^{d+n-1})\Id_{n>m}   ~~~~~~~~~~~~~~~~~~~~~~~ (\mbox{\bf  i}) %\varphi_{d+m}\ast\ldots\ast \varphi_{d+n-1} (x,y)
\\
&+q^{dy}\sum_{k=1}^{n }  \widetilde \psi_{d+N-k+1}^{(m+1)}(x) q^{ (k-1) y}
\prod_{i= n+1  }^{N  }(1-q^{i-k})~~~~~~~ (\mbox{\bf ii})
\\
&+ \sum_{k=1}^{d}  \widetilde \psi_k^{(m+1)}(x) h_{y-y_k }(q^{d},\ldots,q^{d+n-1}).~~~~~~~~~ ~~~~~~ (\mbox{\bf iii})
%\frac{(q;q)_{ N-k+1}}{(q;q)_{n-k-2}} 
 %
\end{aligned}\label{Kern}\ee

%where (see Figure page 2 for $m_1$)
%$$ \tilde \PR_{N-1}^{j}(z):=q^{d(j-m_1) }\PR_{N-1}(zq^{ j-m_1}).$$

\end{proposition}

%\proof %From  , we have the following:
%
%$$\begin{aligned}
%(\varphi_{n +1}\ast \ldots\ast\varphi_{m })(x,\cdot) \ast \psi_j(\cdot)
%&=\sum_{\ell=1}^{d+N}(\widetilde H^{-1})_{j\ell} h_{x_\ell-x}(q^{ n },\ldots,q^{m-1})%
%\end{aligned}$$

%$$\begin{aligned}
%\psi_k^{(m)}(x):=(\varphi_{d+m}\ast \ldots\ast\varphi_{d+N })(x,\cdot) \ast \psi_k(\cdot)
%&=\sum_{\ell=1}^{d+N}(\widetilde H^{-1})_{k\ell} h_{x_\ell-x}(q^{d+m-1 },\ldots,q^{d+N-1})%\Id_{x\leq y}\Id_{n_1<n_2}
%\end{aligned}$$
%$$\begin{aligned}
%(\chi_{\ell}\ast\varphi_{d+1 }\ast \ldots\ast\varphi_{d+n-1})( y)
%&=h_{y-y_\ell }(q^{d},\ldots,q^{d+n-2}) \Id_{2\leq n }%\Id_{x \leq y_\ell}
%\\
%\varphi_{d+\ell} (\mbox{virt},\cdot)\ast (\varphi_{d+\ell+1}\ast\dots\ast \varphi_{d+n-1})(\cdot,y)&=
%\Id_{\ell< n-1}\frac{q^{(d+\ell-1)y}}{(1-q)\ldots(1-q^{ n-\ell-1 })}
%\end{aligned}$$
%See Lemma \ref{lemma:det}.

  %\vspace*{-.9cm}
\proof We use the kernel in Proposition \ref{Prop:Kern0}.  Equality $\stackrel{\ast}{ =}$ follows from the inverse (\ref{Minv}) of the matrix $\widetilde M$. Using Lemma \ref{lemma:det}, we then have:
$$\begin{aligned}
&{\mathbb K}_q(m,x;n,y)\\
=&-  \varphi_{d+m+1}\ast\ldots\ast \varphi_{d+n } (x,y)\\
&+\sum_{{1\leq k\leq d+N}\atop{1\leq \ell\leq d}}  \widetilde \psi_{k}^{(m+1)}(x)(\widetilde M^{-1})_{k\ell}(\chi_{\ell}\ast\varphi_{d+1 }\ast 
 \ldots\ast\varphi_{d+n })(y)
\\
&+\sum_{{1\leq k\leq d+N}\atop{d+1\leq \ell\leq d+n }} 
  \widetilde \psi_{k}^{(m+1)}(x)(\widetilde M^{-1})_{k\ell}(\varphi_\ell (\mbox{virt},\cdot)\ast (\varphi_{\ell+1}\ast\dots\ast \varphi_{d+n })(\cdot,y))
\\
\stackrel{\ast}{ =}&-  \varphi_{d+m+1}\ast\ldots\ast 
 \varphi_{d+n } (x,y)\\
&+\sum_{{1\leq k\leq d } }  \widetilde \psi_{k}^{(m+1)}(x) (\chi_{k}\ast\varphi_{d+1 }\ast \ldots\ast\varphi_{d+n })(y)
\\
&+\sum_{{d+1\leq k\leq d+N}\atop{d+1\leq \ell\leq d+n }}  \widetilde \psi_{k}^{(m+1)}(x)\dt_{2d+N-k+1,\ell}(q;q)_{k-d-1}(\varphi_\ell (\mbox{virt},\cdot)\ast (\varphi_{\ell+1}\ast\dots\ast \varphi_{d+n })(\cdot,y))
\end{aligned}$$
$$\begin{aligned}
 = &-  h_{y-x}(q^{d+m },\ldots,q^{d+n-1})\Id_{n>m} 
+\sum_{{1\leq k\leq d }  }   \widetilde \psi_k^{(m+1)}(x) h_{y-y_k }(q^{d},\ldots,q^{d+n-1})
\\
&+\sum_{{d+N-(n-1)\leq k\leq d+N} }  \widetilde  \psi_{k}^{(m+1)}(x) q^{(2d+N-k)y}\frac{(q;q)_{ k-d-1}}{(q;q)_{k-d-(N-n   +1)}}.
\end{aligned}$$
Then relabeling indices, we find:
$$\begin{aligned}
 {\mathbb K}_q(&m,x;n,y) \\
 { =}&-  h_{y-x}(q^{d+m },\ldots,q^{d+n-1})\Id_{n>m}%\varphi_{d+m}\ast\ldots\ast \varphi_{d+n-1} (x,y)
  + \sum_{{1\leq k\leq d }  }  \widetilde \psi_k^{(m+1)}(x) h_{y-y_k }(q^{d},
   \ldots,q^{d+n-1})
\\
&+\sum_{0\leq k\leq n-1}   \widetilde  \psi_{d+N-k}^{(m+1)}(x) q^{( d+k)y}
 \frac{(q;q)_{ N-k-1}}{(q;q)_{n-k-1}} 
\\
 { =}&-  h_{y-x}(q^{d+m },\ldots,q^{d+n-1})\Id_{n>m}  %~~~~~~~~~~~~~~~~~~~% (\mbox{\bf term (i)}) %\varphi_{d+m}\ast\ldots\ast \varphi_{d+n-1} (x,y)
\\
&+q^{dy}\sum_{k=1}^{n }   \widetilde  \psi_{d+N-k+1}^{(m+1)}(x) q^{ (k-1) y}
\prod_{ n+1  }^{N  }(1-q^{i-k})%~~~ ~~ %(\mbox{\bf term (ii)})
\\
&+ \sum_{k=1}^{d}  \widetilde  \psi_k^{(m+1)}(x) h_{y-y_k }(q^{d},\ldots,q^{d+n-1})%~~~~ ~~~ ~~~~~~~ %(\mbox{\bf term (iii)})
%\frac{(q;q)_{ N-k+1}}{(q;q)_{n-k-2}} 
 %
,\end{aligned}%\label{Kern}
 $$
%Note that since $1\leq n\leq N+1$, we have that  $d+1\leq d+N-k+1\leq d+N.$
establishing (\ref{Kern}).\qed

Theorem \ref{Kern2} below gives an integral representation of the  kernel for the ${\mathbb K}_q$-process. To do so, recall from (\ref{calP}), (\ref{hP}), the notation $q_i=q^{x_i}$ and the identities and definitions:
\be \label{hr} h_r(q^d,\ldots,q^{d+n })= q^{rd}\PR_{n }(q^r) \Id_{n+r\geq 0}
~~~\mbox{and}~~~\tilde \PR_{n-1}^{y}(z):=q^{-dy  }\PR_{n-1}(zq^{ -y } ).
\ee
and also that for $1\leq r\leq d$, $0\leq m\leq N-1$ and $ n\geq 1$, 
 \be\begin{aligned}
h_{x_k-x}(q^{d+m },\ldots,q^{d+N-1})&=
 q^{(x_k-x)(d+m )}\PR_{N-m-1}(q^{x_k-x})\Id_{x_k\geq x}
 %\\&=
%q^{-x(d+m )}v^{ d+m  }\PR_{N-m-1}(vq^{-x})\Bigr|_{v=q^{x_k}}\Id_{x_k\geq x}.
\\
 h_{y-y_r }(\mbox{\tiny$\bullet$}) =h_{y-y_r }(q^{d},\ldots,q^{d+n-1}) %=q^{(y-y_r)d} \PR_{n-1}(q^{y-y_r})=
 %=q^{(y-y_r)d}{\cal P}_{n-1}(q^{y-y_r})\Id_{y-y_r+n-1\geq 0}
 %
  &=\frac {1}{2\pi \I}\oint_{\Gamma_0}\frac{du}{u^{y-y_r+1}}\prod_{\ell=1}^{ n }\frac 1 {1-uq^{d-1+\ell}} \mbox{}%q^{yd}\tilde \PR ^{y_r}_{n-1}(q^y).
 .\end{aligned}\label{hP'}\ee
 %with the latter expression vanishing for $y\leq y_r$. %
The reader is also reminded of the expression, considered in Lemma \ref{qlim}, 
\be
\Phi_q(z^{-1}):=(q^{N-1};q^{-1})_{N-n }~{}_{2}\phi_1
(q^{-1},q^{n-1};q^{N-1}\Bigr|q^{-1},z^{-1}).
\ee
\noindent In the next Theorem, one will need $q$-analogues of the three determinants (\ref{Deltacut}), depending on the points $ {\bf y}_{\mbox{\tiny cut}} =(y_1,\dots, y_d)$ in the lower-cuts. The second will involve the $q$-analogue $ h_{y-y_r}(\mbox{\tiny$\bullet$}):=h_{y-y_r}(\mbox{\footnotesize$q^{d},\ldots,q^{d+n-1}$})$ of the function $h_{y-y_r}(1^n)$ which has the integral representation (\ref{hP'}). The definitions are:
$$\begin{aligned}
\Dt_{q,d}^{({\bf y}_{\mbox{\tiny cut}})}(u_1,\dots,u_d)&:=\det (\tilde \PR_{N-1}^{y_\beta} (u_\alpha))_{1\leq \alpha,\beta \leq d}
\end{aligned}
$$
\be\begin{aligned}
 \widetilde\Dt_{q,d}^{({\bf y}_{\mbox{\tiny cut}})}( w;u_2,\dots,u_d )&=\det\left(\begin{array}{cccccccc}
w^{ y_1}&\dots&w^{y_d}\\
 \tilde \PR^{y_1}_{N-1}(u_{2})&\dots&\tilde \PR^{{y_d}}_{N-1}(u_{2})\\
 \vdots&&\vdots\\
 \tilde \PR^{y_1}_{N-1}(u_{d })&\dots&\tilde \PR^{{y_d}}_{N-1}(u_{d })\\
 \end{array}
\right)
\\
\widetilde\Dt_{q,d,n}^{({\bf y}_{\mbox{\tiny cut}})}( y;u_2,\dots,u_d )
 &  :=
 \oint_{\Gamma_0} \frac {dw}{2\pi \I w^{  y  +1}  {\prod_{k=1}^{n}(1-wq^{d+k-1})}} 
 \widetilde\Dt_{q,d}^{({\bf y}_{\mbox{\tiny cut}})}( w,u_2,\dots,u_d )
\\&=
\det\left(\!\begin{array}{cccccccc}
h_{y-y_1}(\mbox{\tiny$\bullet$}) &\dots&h_{y-y_d}(\mbox{\tiny$\bullet$})\\
 \tilde \PR^{y_1}_{N-1}(u_{2})&\dots&\tilde \PR^{{y_d}}_{N-1}(u_{2})\\
 \vdots&&\vdots\\
 \tilde \PR^{y_1}_{N-1}(u_{d })&\dots&\tilde \PR^{{y_d}}_{N-1}(u_{d })\\
 \end{array}
\!\right)
.
\end{aligned}\label{vdmy}\ee
The last expression is obtained by performing the $w$-integration on the first row of the matrix in $\widetilde\Dt_{q,d}^{({\bf y}_{\mbox{\tiny cut}})}$. The expression $\widetilde\Dt_{q,d,n}^{({\bf y}_{\mbox{\tiny cut}})}(y;u_2,\dots,u_d)$ clearly vanishes, when $y\leq y_d-1$, or what is the same, when $y$ is strictly to the left of the lower-cuts.

\begin{theorem}\label{Kern2} For $0\leq m,n \leq N $, the kernel for the ${\mathbb K}_q$-process reads as follows:
\be\label{Kern2'}\begin{aligned}
&q^{(d+m )(x-y)}  {\mathbb K}_q( m,x;n,y) \\
& { =}  
 \frac{(zq;q)_{n-m-1}}{( q;q)_{n-m-1}} \Bigr|_{z=q^{y-x}}
\Id_{n>m}\Id_{y\geq x}  %
\\
&+    %(q^{N-1};q^{-1})_{N-n+1}~    
 \oint_{ {\Gamma(q^{x-y+{\mathbb N}} })}\frac { v^{m }dv}{2\pi \I  }
  \oint_{\Gamma_\infty}\frac {\Phi_q(  z^{-1})dz}{2\pi \I  z^{   }}~
  %\\&\times
   \frac{\PR_{N-m-1}(vq^{y-x})}{ z-v  }\frac{Q_q(zq^y )}{Q_q(vq^y) } ~ %\\&\times 
   \frac{\Om_q (vq^y,zq^y)}{\Om_q(0,0)}~ 
\\
%&\\
  &+
   {     d \over    q^{ (d-1)   y   }    }  
%\left(
 \oint_{\Gamma_{\Gamma(q^{x-y+{\mathbb N}})  } }\frac{v^{m }dv  }{   2\pi i  }\frac{ \PR_{N-m-1}(vq^{y-x})}{Q_q(vq^y)}%\prod _1^{d+N}(q^yv-q^{x_i })
    % \prod^{d-1}_1   (v-u_i) 
   %\\
%&\times  
 \frac{ \widetilde{\Om}_q(vq^y,y)}{\Om_q(0,0)}  
\\&=:({\mathbb K}^{(0)}_q+{\mathbb K}^{(1)}_q+{\mathbb K}^{(2)}_q)( m,x;n,y),
\end{aligned}\ee
%
%\end{aligned}$$
where\footnote{$\Om_q=1$ for $d=0$. $F_r$ and $\widetilde \Om_q=1$ for d=1.} % (see (\ref{hP'}) for notation $\mbox{\tiny $\bullet$}$)
$$%\label{Omq}
 \begin{aligned}
\Om_q (v,z) &:=%\frac 1{{\tilde D}_q}
 \left(\prod^d_{j=1} \oint_{\Gamma (q_1,\ldots,q_c)} { du_j \over  2\pi \I Q_q(u_j)}\right)  \prod_1^d \frac{v -u_i }{z  -u_i }
   % OMIT\det (\tilde \PR_{N-1}^{y_\beta} (u_\alpha))_{1\leq \alpha,\beta \leq d}
    \\
   &~~~~~~~~~~~~~~~~~~~~~~~ \times\Dt_{d}(u_1,u_2,\ldots,u_d)
   \Dt^{({\bf y}_{\mbox{\tiny cut}})}_{q,d}(u_1,u_2,\ldots,u_d),
\end{aligned} 
$$  
and

\newpage

\vspace*{-2cm}

\be
\label{Omq}\begin{aligned}
    \widetilde{\Om}_q(v ,y)  &:= 
\left(  
 \prod^{d }_{j=2} \oint_{\Gamma	(q_1,\ldots,q_c)}   {du_j \over 2\pi \I Q_q(u_j)}
\right)
\\  
& ~~~~~~~~~~~~~~~~~~~~~~~\times 
\Dt_{d }   (v ,u_2,\ldots,u_{d })      \widetilde\Dt_{q,d,n}^{({\bf y}_{\mbox{\tiny cut}})}(q^y,u_2,\dots,u_d)
\\&= \sum_{{1\leq r\leq d}\atop{y_r\leq y}}  (-1)^{r-1} h_{y-y_r}(\mbox{\tiny$\bullet$})%\Id_{y-y_r \geq 0}
  ~F_r(v),
 \end{aligned} \ee
  with $\mbox{\tiny $\bullet$}=(q^{d},\ldots,q^{d+n-1})$ as in (\ref{hP'})  and
\be\begin{aligned} F_r(v): =\left(  
 \prod^{d -1}_{j=1} \oint_{\Gamma	(q_1,\ldots,q_c)}   {du_j \over 2\pi \I Q_q(u_j)}\right)&  \det (\widetilde \PR_{N-1}^{y_\beta} 
(u_\alpha))_{1 \leq \alpha \leq d -1 \atop 1\leq \beta \leq d, \beta \neq r}  
%\right)
%
%\\&\times 
  \Dt_{d}   (v ,u_1,\ldots,u_{d-1 })  .
\end{aligned}\label{Fr}\ee
Finally, ${\mathbb K}^{(2)}_q(m,x;n,y)=0$, when $y\leq y_d-1$; when $y$ is to the left of the lower-cuts.
\end{theorem}

%\remark Note that $ \widetilde{\Om}_q(v ,y)$ can also be written as 
%\be\label{Omq1}\begin{aligned}   \widetilde{\Om}_q(v ,y) & =
%\left(\oint_{\Gamma_0} \frac {du_1}{2\pi \I u_1^{  y- y_d  +1}  {\prod_{k=1}^{n}(1-u_1q^{d+k-1})}} 
% \prod^{d }_{j=2} \oint_{\Gamma	(q_1,\ldots,q_\ell)}   {du_j \over 2\pi \I Q(u_j)}
%\right)
%\\  
%&~~~~\times 
%\Dt_{d }   (v ,u_2,\ldots,u_{d })     
%\det\left(\begin{array}{cccccccc}
%u_1^{d-1}&\dots&u_1^0\\
 %\tilde \PR^{y_1}_{N-1}(u_{2})&\dots&\tilde \PR^{{y_d}}_{N-1}(u_{2})\\ \vdots&&\vdots\\ \tilde \PR^{y_1}_{N-1}(u_{d })&\dots&\tilde \PR^{{y_d}}_{N-1}(u_{d })\\
% \end{array}
%\right)
%\end{aligned}\ee

%The latter representation implies $ \widetilde{\Om}_q(v ,y) =0$ when $y\leq y_d-1$, i.e., when $y$ is to the left of the lower indentation. 

%\proof To be given here.

\bigbreak
\noindent {\em Proof of Theorem \ref{Kern2}}:    Expressed as sum of residues, the multiple contour integral $\Om_q(v,z)$, $\widetilde\Om_q(v,z)$ and $F_r (v )$  are  as follows:
\be\begin{aligned}
\Om_q (v,z)&:=   \sum_{1\leq i_1,\dots,i_d\leq c}\frac{ \Dt_d(q^{x_{i_1}},\ldots,q^{x_{i_d}}) \Dt_{q,d}^{({\bf y}_{\mbox{\tiny cut}})}(q^{x_{i_1}},\dots,q^{x_{i_d}})%@@\det (\tilde \PR_{N-1}^{y_\beta} ( q^{x_{i_\al}}))_{1\leq \alpha,\beta \leq d}
   }{\prod^d_{\al=1}Q_q'(q^{x_{i_\al}} ) }\prod_{\al=1}^d \frac{v -q^{x_{i_\al}}}{z  -q^{x_{i_\al}}}
 \\
 \widetilde\Om_q (v,z)&:=   \sum_{1\leq i_1,\dots,i_{d-1}\leq c}\frac{ \Dt_d(v,q^{x_{i_1}},\ldots,q^{x_{i_{d-1}}}) \widetilde\Dt_{q,d,n}^{({\bf y}_{\mbox{\tiny cut}})}(q^y,q^{x_{i_1}},\dots,q^{x_{i_{d-1}}})%@@\det (\tilde \PR_{N-1}^{y_\beta} ( q^{x_{i_\al}}))_{1\leq \alpha,\beta \leq d}
   }{\prod^{d-1}_{\al=1}Q_q'(q^{x_{i_\al}} ) }%\prod_{\al=1}^d \frac{v -q^{x_{i_\al}}}{z  -q^{x_{i_\al}}}
 \\
F_r (v )&:=%\frac 1{{\tilde D}_q}
 %\left(\prod^d_{j=1} \oint_{\Gamma (q_1,\ldots,q_\ell)} { du_j \over  2\pi \I Q_q(u_j)}\right)\prod_1^d \frac{v -u_i }{z  -u_i }
 % \det (\tilde \PR_{N-1}^\beta (u_\alpha))_{1\leq \alpha,\beta \leq d} \Dt_d(u_1,\ldots,u_d)\\
 % &
   \sum_{1\leq i_1,\dots,i_{d-1}\leq c}\frac{ \det (\tilde \PR_{N-1}^{y_\beta} ( q^{x_{i_\al}}))_{{{1\leq \alpha \leq d-1 }\atop{1\leq \beta\leq d,~\beta\neq r}} }
   }{\prod^{d-1}_{\al=1}Q_q'(q^{x_{i_\al}} ) }\prod_{\al=1}^{d-1}  
 \Dt_d(v,q^{x_{i_1}},\ldots,q^{x_{i_{d-1}}}).
 \end{aligned} \label{OmqRes}\ee
 %@@@@
%
%
%
{\bf Expressing the determinant of $\widetilde H$ of size $(d\!+\!N)$ as a multiple contour integral}, using the Laplace expansion for the determinant of a block matrix and using (\ref{Htilde}) for $\widetilde H$ and Lemma \ref{Vanderm1} in the third equality\footnote{Given the set $(1,\dots,M)$, and a subset $(i_1<\dots<i_d)$, define
\be
\sigma_{M}(i_1,\dots,i_d):=\#\left\{\begin{array}{ccc}\mbox{transpositions needed to map }\\
\mbox{ $(1<\dots <i_1<\dots<i_d<\dots<M)$}\\
 \mbox{into ~$(i_1,\dots, i_d,1,2,\dots , \hat i_1,\dots ,\hat i_d,\dots ,M)$}\end{array}\right\}.
\label{sigma}\ee}, (remember $q_i:=q^{x_i} $)
$$
\begin{aligned}
 %&\hspace*{-.5cm}
 &
\frac{      \det \widetilde H }{\Dt_{d+N}(q_1,\dots,q_{d+N})}
 \\ =&\prod^{d+N}_{{i=1}\atop{ }}  q^{ d x_i} \frac{\det
\left(\dis {{(\tilde \PR_{N-1}^{y_j}(q^{x_i}))_{1\leq i \leq c \atop 1 \leq j \leq d}}\atop{O_{d+b,d}}} ~\Bigl | ~(q^{x_i(N-j)})_{1\leq i \leq d + N \atop 1 \leq j \leq N}\right)}{\Dt_{d+N}(q_1,\dots,q_{d+N})}
\end{aligned}
$$

\newpage

\vspace*{-2cm}

\be \begin{aligned}%\\
  =&   (-1)^{d(d-1)/2}\prod^{d+N}_{{i=1}\atop{ }}  q^{ d %\displaystyle 
 x_i} \sum_{1\leq i_1 < \ldots < i_d \leq c} (-1)^{\sigma(i_1,\ldots,i_d)} \det
\left(\begin{array}{ccc}
\tilde \PR_{N-1}^{y_j}(q_{i_1}) \\
\vdots \\
\tilde \PR_{N-1}^{y_j}(q_{i_d})
\end{array}\right)_{1\leq j \leq d}
\\
&~ \times {\Dt_N(q_1,\ldots,\widehat{q_{i_1}},\ldots, \widehat{q_{i_d}},\ldots, q_{d+N})   \over  \Dt_{d+N}(q_1,\ldots,q_{d+N})}
\\
  =& {(-1)^{d(d-1)/2}   \over  d!} \prod^{d+N}_{{i=1}\atop{ }}  q^{ d %\displaystyle 
 x_i}\sum_{1\leq i_1,\ldots,i_d \leq c} %@@\det\left(\begin{array}{ccc}
%\tilde \PR_{N-1}^{y_j}(q_{i_1}) \\
%\vdots \\
%\tilde \PR_{N-1}^{y_j}(q_{i_d})
%\end{array}\right)_{1\leq j \leq d}
%\\
%& \times
 { \Dt_{q,d}^{({\bf y}_{\mbox{\tiny cut}})}(q_{i_1},\ldots,q_{i_d})\Dt_d(q_{i_1},\ldots,q_{i_d})   \over \displaystyle  \prod^d_{j-1} Q_q' (q_{i_j})  }
\\
%  =@& {(-1)^{d(d-1)/2} \over  d!}  \prod^{d+N}_{{i=1}\atop{ }}  q^{ d %\displaystyle 
% x_i}  \left(\prod^d_{j=1} \oint_{\Gamma (q_1,\ldots,q_\ell)} { du_j \over  2\pi \I Q_d(u_j)}\right) \det (\tilde \PR_{N-1}^{y_\beta} (u_\alpha))_{1\leq \alpha,\beta \leq d} \Dt_d(u_1,\ldots,u_d)\\
  %:=&  \prod^{d+N}_{{i=1}\atop{ }}  q^{ d %\displaystyle 
 % x_i} D_q
  =& {(-1)^{d(d-1)/2} \over  d!}  \Bigl(\prod^{d+N}_{{i=1}\atop{ }}  q^{ d %\displaystyle 
  x_i}\Bigr) \Om_q(0,0),
\end{aligned}
\label{detH}\ee
upon using the residue formula (\ref{OmqRes}) for $\Om_q(0,0)$ in the last equality. 
%$$\begin{aligned}
% \tilde D_q:=(-1)^{\frac{d(d-1)}2}d!D_q&=       \left(\prod^d_{j=1} \oint_{\Gamma (q_1,\ldots,q_\ell)} { du_j \over  2\pi \I Q_d(u_j)}\right) \det (\tilde \PR_{N-1}^\beta (u_\alpha))_{1\leq \alpha,\beta \leq d} \Dt_d(u_1,\ldots,u_d)
% \\
% &=\sum_{1\leq i_1,\ldots,i_d \leq \ell} 
% {\Dt_d(q_{i_1},\ldots,q_{i_d})   \over \displaystyle  \prod^d_{j=1} Q_d' (q_{i_j})  }
 % \det
%\left( 
%\tilde \PR_{N-1}^\beta(q_{i_\al})
 %\right)_{1\leq \al,\beta \leq d}
 %\end{aligned}$$

\medbreak

\noindent{\bf Expressing the entries of inverse ${\widetilde H}^{-1}$ as integrals}. Computing its entries ${\widetilde H_{ r,k} } ^{-1}$ will  depend on whether  $1\leq r\leq d$  or $d+1\leq r\leq d+N$. To do so, the determinant of the adjoint matrix will be expanded as a sum of the determinants of blocks taken from the first part of the matrix times the determinant of a complementary block of the second part. Then using in $\stackrel{*}{=}$ the second relation of Lemma \ref{Vanderm1}  one finds for $1\leq r\leq d$ and $1\leq k\leq d+N$:
\be \label{Hinv}
\begin{aligned}
{\widetilde H_{ r,k} }^{-1}
=&\frac{   {\prod^{d+N}_{_{{i=1}\atop{i\neq k}} }
 q^{ d  x_i} 
\mbox{adj} \left(\dis {{(\tilde \PR_{N-1}^{y_j}(q^{x_i}))_{1\leq i \leq c \atop 1 \leq j \leq d}}\atop{O_{d+b,d}}} ~\Bigl | ~(q^{x_i(N-j)})_{1\leq i \leq d + N \atop 1 \leq j \leq N}\right)_{ r,k}}}{   \det \widetilde H        }
%\\=&\frac{   {\prod^{d+N}_{_{{i=1}\atop{i\neq k}} }
% q^{ d  x_i} 
%\mbox{adj} \left((\tilde \PR_{N-1}^j(q_i))_{1\leq i \leq d +N \atop 1 \leq j \leq d} \Bigl | (q^{ x_i(N-j)})_{1\leq i \leq d + N \atop 1 \leq j \leq N}\right)_{ r,k}}}{  \prod^{d+N}_{_{i=1} }  q^{ d  x_i}
%D_q\Dt_{d+N}(q_1,\dots,q_{d+N})          }
\\
=& \frac{d!(-1)^{r+k+d(d-1)/2}}{\Om_q(0,0) q^{dx_k} }\sum_{1\leq i_1<\dots<i_{d-1}\leq c} (-1)^\sigma
\\&\times \det \left(
\begin{array}{cccccccc}
\tilde \PR_{N-1}^ {y_1}(q_{i_1})&\dots& \widehat{\tilde \PR_{N-1}^{y_r}(q_{i_1})}&\dots& \tilde \PR_{N-1}^{y_d}(q_{i_1})\\
\vdots &  &  \vdots &&\vdots\\
\tilde \PR_{N-1}^{y_1}(q_{i_{d-1}})&\dots& \widehat{\tilde \PR_{N-1}^{y_r}(q_{i_{d-1}})}&\dots& \tilde \PR_{N-1}^{y_d}(q_{i_{d-1}})
\end{array}
\right)
\\
&\times \frac{\Dt_N(q_1,\dots,\widehat{q_{k}},\widehat {q_{i_1}},\dots,\widehat q_{i_{d-1} },\dots,q_{d+N})}{(-1)^{k-1}\Dt_{d+N}(q_k,q_1,\dots,\widehat{q}_{k},\dots, q_{d+N})}
\\
\stackrel{*}{=}&  \frac{d!(-1)^{r+k+d(d-1)/2}}{\Om_q(0,0)q^{dx_k}}\sum_{1\leq i_1<\dots<i_{d-1}\leq c} (-1)^\sigma  \frac
{ \det (\widetilde \PR_{N-1}^{y_\beta} 
(q^{x_{i_\al}} ))_{1 \leq \alpha \leq d-1  \atop 1\leq \beta \leq d, \beta \neq r}}{\prod_{\al=1}^{d-1}Q_q'(q^{x_{i_\al}})}   
 \\
&\times  \frac{(-1)^{\sigma  +d(d-1)/2} \Dt_d(q_k, {q_{i_1}},\dots,  q_{i_{d-1} })}{(-1)^{k-1}Q_q'(q_k)}
 \\ =&{ d!(-1)^{ r-1}  \over {(d-1)!  }}    %\left(\prod^{d }_{j=2} \oint_{\Gamma (q_1,\ldots,q_\ell)} { du_j \over  2\pi \I Q_q(u_j)}\right)
\frac{q^{-dx_k}F_r(q^{x_k})}{Q'_q(q^{x_k})\Om_q(0,0)}
 %\oint_{\Gamma_{q_k}}\frac {v^{-d} dv}{2\pi \I Q_q(v)}
% \\
%&   
%  \times\det (\tilde \PR_{N-1}^{y_\beta} (u_\alpha))_{{{2\leq \alpha \leq d }\atop{1\leq \beta\leq d,~\beta\neq r}}\atop{}} \Dt_{d-1}(u_2,\ldots,u_{d}) \prod_2^{d }  (v-u_i)
 %
= 
 (-1)^{r-1}d\oint_{\Ga(q^{x_k})}\frac{v^{-d}dv}{2\pi \I Q_q(v)}\frac{F_r(v)}{\Om_q(0,0)}.
\end{aligned}
\ee

\newpage

\vspace*{-1cm}

\noindent
Also the last equality follows from replacing $\sum_{1\leq i_1<\dots<i_{d-1}\leq c} $ by $\sum_{1\leq i_1,\dots,i_{d-1}\leq c} $, which brings in the denominator $(d-1)!$. Remember $F_r$ is the $d-1$-fold integral (\ref{Fr}).%Expresssing $D_q$ in terms of $\tilde D_q$ brings in $d!$ in the numerator.

%\newpage

Using (\ref{detH}) above, we now show that for $1\leq r\leq N$ the inverse ${{\widetilde H}_{d+r,k} }^{-1}$ has a multiple integral  representation, as in the previous computation. Namely one uses in $\stackrel{\ast}{=}$ the first identity (\ref{Vanderm2}) (Lemma \ref{Vanderm1}) and the definition (\ref{vdmy}) of 
$\Dt_{q,d}^{({\bf y}_{\mbox{\tiny cut}})}$. Also notice that the restriction $i_\ell \neq k$ in the sum appearing in $\stackrel{\ast}{=}$ can be removed, since $ \Dt^{\widehat {N-r}} _{N-1} =0$ for $i_\ell = k$. In $\stackrel{\ast\ast }{=}$, one uses (\ref{symmf}) (Lemma \ref{Vanderm1}) and the terms containing $q^{x_k}$ can be rewritten as a contour integral about $q^{x_k}$, because of the presence of $Q'_{q}( q^{x_k})$ in the denominator. Finally, in $\stackrel{\ast\ast\ast }{=}$, one uses the definition (\ref{OmqRes}) of $\Om_q(v,z)$, thus yielding:
 \be
\begin{aligned}
 {{\widetilde H}_{d+r,k} }^{-1} \!%= & 
 %\frac{   {\prod^{d+N}_{{i=1}\atop{i\neq k}} 
% q^{ d  x_i} 
%\mbox{adj} \left(\dis {{(\widetilde \PR_{N-1}^{y_j}(q^{x_i}))_{1\leq i \leq c \atop 1 \leq j \leq d}}\atop{O_{d+b,d }}} ~\Bigl | ~(q^{x_i(N-j)})_{1\leq i \leq d + N \atop 1 \leq j \leq N}\right)_{d+r,k}}}{ \det \widetilde H% \prod^{d+N}_{_{i=1} }  q^{ d  x_i}
%D_q\Dt_{d+N}(q_1,\dots,q_{d+N})        
  % }
%\\ 
 =&
  \frac{d!(-1)^{d(d-1)/2}  }{\Om_q(0,0)q^{dx_k}} ~\frac{\mbox{adj} \left(\dis{ {(\widetilde \PR_{N-1}^{y_j}(q^{x_i}))_{1\leq i \leq d +N \atop 1 \leq j \leq d} }\atop{O_{d+b,d }}}\Bigl | (q^{x_i(N-j)} )_{1\leq i \leq d + N \atop 1 \leq j \leq N}\right)_{d+r,k}}{\Dt_{d+N}(q_1,\dots,q_{d+N})}
  % \frac{    \det \widetilde H }{V_{d+N}(q_1,\dots,q_{d+N})}
% {\widetilde H_{d+r,k} }^{-1}=Dq^{dx_k}{\widetilde H_{d+r,k} }^{-1}
%
\\
 =&\frac{d!(-1)^{d(d-1)/2}  }{\Om_q(0,0)q^{dx_k}} ~
 (-1)^{d+r+k}\sum_{{1\leq i_1<\dots<i_{d}\leq c}\atop{\mbox{\tiny all}~i_\ell\neq k} }(-1)^\sigma
 \det \left(\widetilde \PR_{N-1}^{y_\beta}(q^{x_{i_\al}}))\right)_{1\leq \al,\beta\leq d}
\\
 &\qquad \qquad\qquad
\times \frac{\Dt^{\widehat {N-r}}_{N-1}(q_1,\dots,\widehat{q_{k}},\widehat {q_{i_1}},\dots,\widehat q_{i_{d} },\dots,q_{d+N})}{\Dt_{d+N}(q_1,\dots,q_{d+N})} %
\\
\stackrel{\ast}{=}&\frac{d!(-1)^{d(d+1)/2+r+k}  }{\Om_q(0,0)q^{dx_k}} ~
 \sum_{{1\leq i_1<\dots<i_{d}\leq c}  }(-1)^\sigma
\Dt_{q,d}^{({\bf y}_{\mbox{\tiny cut}})}(q^{x_{i_1}},\dots,q^{x_{i_d}})% \det @@\left(\widetilde \PR_{N-1}^{y_\beta}(q^{x_{i_\al}}))\right)_{1\leq \al,\beta\leq d}
\\ &\times \! (-1)^{\frac{ d(d+1)}2+\!\sg\!+\!k\!-\!1}\frac{\Dt_{d+1}(q_k, {q_{i_1}},\dots,  q_{i_{d } })e_{r\!-\!1}(q_1,\dots,\widehat{q_{k}},\widehat {q_{i_1}},\dots,\widehat q_{i_{d} },\dots,q_{d+N})}{\prod_{\al=1}^{d }Q_q'(q^{x_{i_\al}})Q_q'(q_k)}
\\
 \stackrel{\ast\ast}{=}&\frac{1}{\Om_q(0,0)q^{dx_k}} ~
 \sum_{{1\leq i_1<\dots<i_{d}\leq c}\atop{  } }
\Dt_{q,d}^{({\bf y}_{\mbox{\tiny cut}})}(q^{x_{i_1}},\dots,q^{x_{i_d}})
\Dt_{d }(   q^{x_{i_1}} ,\dots,  q^{x_{i_d}})
\\ &\times   \frac{1 }{Q_q'(q^{x_k}) }
\prod_{\al=1}^{d }\frac{(q^{x_{k}}-q^{x_{i_\al}})}{Q_q'(q^{x_{i_\al}})}
\oint_{\Gamma_\infty}\frac {dz}{2\pi \I z^{N-r+1}}\frac{Q_q(z)}{(z-q^{x_k})\prod_{\ell=1}^d (z-q^{x_{i_\ell}})}
\\
  \stackrel{\ast\ast\ast}{=} &    
 \oint_{\Gamma_{q^{x_k}}}\frac {dv}{2\pi \I v^{d} }
  \oint_{\Gamma_\infty}\frac {dz}{2\pi \I z^{N-r+1}} 
\frac{1}{(z-v) }\frac{Q_q(z)}{Q_q(v) }\frac{\Om_q(v,z)}{\Om_q(0,0)} . \end{aligned}
\label{Hmin}\ee
%\end{document}

%\newpage

\noindent {\bf The computation of the $\widetilde \psi^{(m+1)}_k$-function}, as defined in (\ref{psi-new}) of Proposition \ref{Prop5.1}, comes in two parts.
 For $1\leq r\leq N$, we have, using directly (\ref{Hmin}) and (\ref{hP'}),
$$\begin{aligned}
 \widetilde \psi_{d+r}^{(m+1)} (x)&=\sum_{k=1}^{d+N}(\widetilde H^{-1})_{d+r,k} h_{x_k-x}(q^{d+m  },\ldots,q^{d+N-1})\Id_{x_k\geq x} 
\end{aligned}$$
\be\begin{aligned} \label{psi(d+r)}&=q^{-x(d+m )}\sum_{k=1}^{d+N} {\widetilde H^{-1} }_{d+r,k}
v^{ d+m  }\PR_{N-m-1}(vq^{-x})\Bigr|_{v=q_k}\Id_{x_k\geq x}
\\&= q^{-x(d+m )} \oint_{\Gamma({q^x,q^{x+1},\ldots})}\frac {v^{m}dv}{2\pi \I  }
\oint_{\Gamma_\infty}\frac {dz}{2\pi \I z^{ N-r+1 }}
\frac{\PR_{N-m-1}(vq^{-x})}{(z-v) }\frac{Q_q(z)}{Q_q(v) }\frac{\Om_q(v,z)}{\Om_q(0,0)}
%\\
%\end{aligned}$$
%$$\begin{aligned}
% &= (omit?) \frac{q^{-x(d+m )}}{\Om_q(0,0)}    \left(\prod^d_{j=1} \oint_{\Gamma (q_1,\ldots,q_\ell)} { du_j \over  2\pi \I Q_q(u_j)}\right) \oint_{\Gamma({q^x,q^{x+1},\ldots})}\frac {v^{m}dv}{2\pi \I  }\oint_{\Gamma_\infty}\frac {dz}{2\pi \I z^{ k }}
 %\\
%&\times 
%\frac{\PR_{N-m-1}(vq^{-x})}{(z-v) }\frac{Q_q(z)}{Q_q(v) }\prod_1^d \frac{v-u_i}{z-u_i}
 % \det (\tilde \PR_{N-1}^{y_\beta} (u_\alpha))_{1\leq \alpha,\beta \leq d} \Dt_d(u_1,\ldots,u_d)
.
\end{aligned}
\ee
  Similarly for $1 \leq r \leq d$, using (\ref{Hinv}), (\ref{psi-new}) and (\ref{Fr}),
\be \label{psi(r)}
\begin{aligned}
\widetilde \psi^{(m+1)}_r (x) =& \sum^{d+N}_{k=1} 
(\widetilde H^{-1})_{rk} h_{x_k-x} (q^{d+m },\ldots,q^{d+N-1})
\Id_{x_k\geq x}
\\
&= \frac{d(-1)^{r-1}}{q^{x(d+m)}}  
  \sum_{{1\leq k\leq d+N} \atop{x_k\geq x}}  
   q^{x_k(d+m)}{\cal P}_{N-m-1}(q^{x_k-x})\oint_{\Ga_{q^{x_k}}}\frac{v^{-d}dv}{2\pi \I Q_q(v)}\frac{F_r(v)}{\Om_q(0,0)}
\\
&=\frac{d(-1)^{r-1}}{q^{x(d+m)}}  
 % \sum_{{1\leq k\leq d+N} \atop{x_k\geq x}}  
    \oint_{\Ga({q^{x+{\mathbb N}})}}\frac{v^{m}{\cal P}_{N-m-1}(vq^{ -x})dv}{2\pi \I Q_q(v)} 
    \frac{F_r(v)}{\Om_q(0,0)}.
 %\\
% =OMIT:& {d(-1)^{ r-1}      \over  q^{ x(d+m )} \Om_q(0,0)}  \left(
%\prod^{d }_{j=2} \oint_{\Gamma	(q_1,\ldots,q_\ell)}   {du_j \over 2\pi \I Q_q(u_j)} \right)\oint_{\Gamma({q^{x+{\mathbb N}}  })}{v^{m } dv   \over 2\pi i Q_q(v)}  
%\\
%&\det (\widetilde \PR_{N-1}^{y_\beta} (u_\alpha))_{2 \leq \alpha \leq d  \atop 1\leq \beta \leq d, \beta \neq r}   \Dt_{d-1}   (u_2,\ldots,u_{d })
%  \prod^{d }_2   (v-u_i)      \PR_{N-m-1}(vq^{-x})
%    \\
% =OMIT:& {d(-1)^{ r-1}      \over  q^{ x(d+m )} \Om_q(0,0)}  
%\left(
%\prod^{d }_{j=2} \oint_{\Gamma	(q_1,\ldots,q_\ell)}   {du_j \over 2\pi \I Q_q(u_j)}
%\right)\oint_{\Gamma({q^{x+{\mathbb N}} })}{v^{m } dv   \over 2\pi i Q_q(v)}  
%\\
%&\det (\widetilde \PR_{N-1}^{y_\beta} (u_\alpha))_{2 \leq \alpha \leq d  \atop 1\leq \beta \leq d, \beta \neq r}   \Dt_{d }   (v,u_2,\ldots,u_{d })     
%   \PR_{N-m-1}(vq^{-x})
\end{aligned}
\ee

\noindent{\bf Computing each of the terms ${\mathbb K}^{(i)}_q$ in the kernel (\ref{Kern2'}) for $i=0,1,2$ and for $0\leq m,n\leq N$}. Indeed, using (\ref{hP'}), one finds for $(i)$ of (\ref{Kern}):
%
%\newline{\bf (i)}. 
$$\begin{aligned}
  {\mathbb K}^{(0)}_q( m,x;n,y)&= h_{y-x}(q^{d+m },\ldots,q^{d+n-1})\Id_{n>m}\Id_{y\geq x}
  \\&= 
z^{d+m }\PR_{n-m-1}(z)\Bigr|_{z=q^{y-x}}\Id_{n>m}\Id_{y\geq x}
\\&= 
z^{d+m }\prod_{i=0}^{n-m-2}\frac { 1-zq ^{ i+1}  }{ 1-q ^{i+1 } }\Bigr|_{z=q^{y-x}}
\Id_{n>m}\Id_{y\geq x}
\\&= 
q^{(d+m )(y-x)}\frac{(zq;q)_{n-m-1}}{( q;q)_{n-m-1}} \Bigr|_{z=q^{y-x}}
\Id_{n>m}\Id_{y\geq x}.
\end{aligned}$$
For $1\leq k\leq N,~m\geq 0$, 
using formula (\ref{psi(d+r)}) for $r=N-k+1$ and using Petrov's expression (\ref{Phiq}), one checks for $(ii)$ of (\ref{Kern}):
$$\begin{aligned}
 {\mathbb K} ^{(1 )}_q( m,x;n,y) 
 &=q^{dy}\sum_{k=1}^{n }   \psi_{d+N-k+1}^{(m+1)}(x) q^{ (k-1) y}
\prod_{ r=n+1 }^{N  }(1-q^{r-k})
\\
&=q^{d(y-x)-xm} \oint_{ {\Gamma(q^{x+{\mathbb N}})}}\frac {v^{m }dv}{2\pi \I  }
  \oint_{\Gamma_\infty}\frac {dz}{2\pi \I z^{   }}\frac{\PR_{N-m-1}(vq^{-x})}{(z-v) }\frac{Q_q(z)}{Q_q(v) }\frac{\Om_q(v,z)}{\Om_q(0,0)}
  \\&~~~~\times ~z\sum_{k=1}^{n }z^{-k}q^{ (k-1) y}
\!\prod_{ r=n+1 }^{N  }\!(1\!-\!q^{r-k}) 
\end{aligned}
$$
$$\begin{aligned}
%
% 
%\\&=:\Phi_q(\frac {q^y }z)) 
 %@@@@
\\&= q^{d(y-x)-xm} \oint_{ {\Gamma(q^{x+{\mathbb N}})}}\frac {v^{m }dv}{2\pi \I  }
  \oint_{\Gamma_\infty}\frac {\Phi_q( q^yz^{-1})dz}{2\pi \I z^{   }}\frac{\PR_{N-m-1}(vq^{-x})}{(z-v) }\frac{Q_q(z)}{Q_q(v) }\frac{\Om_q(v,z)}{\Om_q(0,0)}
\\
&\stackrel{*}{=}   {q^{ (y-x)(d+m )}} %(q^{N-1};q^{-1})_{N-n+1}~    
 \oint_{ {\Gamma(q^{x-y+{\mathbb N}} })}\frac { v^{m }dv}{2\pi \I  }
  \oint_{\Gamma_\infty}\frac {\Phi_q(  z^{-1})dz}{2\pi \I  z^{   }}~
  %\\&\times
   \frac{\PR_{N-m-1 }(vq^{y-x})}{(z-v) }\frac{Q_q(zq^y )}{Q_q(vq^y) } ~ %\\&\times 
   \frac{\Om_q (vq^y,zq^y)}{\Om_q(0,0)}
.\end{aligned}$$
This last equality $\stackrel{*}{=}$ is obtained by the substitution $v\mapsto vq^y$ and $z\mapsto zq^y$. Substituting (\ref{psi(r)}) for $\widetilde \psi_r^{(m+1)}$ and (\ref{hP'}) for $h_{y-y_r }$, and using (\ref{Omq}), one finds for $(iii)$ 
%
%
%\newpage
%\vspace*{-1.5cm}
%\noindent 
of (\ref{Kern}):
\be\begin{aligned}
 {\mathbb K}^{(2 )}_q( m,x;n,y)
&=\sum_{{1\leq r\leq d } }   \widetilde  \psi_r^{(m+1)}(x) h_{y-y_r }(q^{d},\ldots,q^{d+n-1})
\\
 &= {d      \over  q^{ x(d+m )} }  
 \oint_{ {\Gamma(q^{x+{\mathbb N}}})}{v^{m } \PR_{N-m-1}(vq^{-x})dv   \over 2\pi\I Q_q(v)}
 \\
 &~~\times\sum_{1\leq r\leq d}(-1)^{r-1}
h_{y-y_r}(\mbox{\tiny$\bullet$})
\frac{F_r(v)}{\Om_q(0,0)}
\\ &= {d      \over  q^{ x(d+m )} }  
 \oint_{ {\Gamma(q^{x+{\mathbb N}}})}{v^{m } \PR_{N-m-1}(vq^{-x})dv   \over 2\pi\I Q_q(v)}\frac{\widetilde\Om_q(v,y)}{\Om_q(0,0)}.
\end{aligned}
\ee
Expression ${\mathbb K}^{(2 )}_q$ as in the kernel (\ref{Kern2'}) follows by the substitution $v\mapsto vq^y$. This ends the proof of Theorem \ref{Kern2}.\qed

%\medbreak

% \newpage

\section{The ${\mathbb K}^{\mbox{\tiny red}}$-kernel as a limit of the ${\mathbb K}_q$-kernel, for $q\to 1$%$d+2$-fold integral  ($d=$ total size of lower-cuts).
 }

Before stating Proposition \ref{Kernlim}, we consider the limits for $q\to 1$ of the expressions (\ref{vdmy}), yielding by Lemma \ref{qlim} the corresponding expressions (\ref{Deltacut}):
\be \begin{aligned}
 \lim_{q\to 1}  \Dt_{q,d}^{({\bf y}_{\mbox{\tiny cut}})}(q^{u_1},\dots,q^{u_d})&=\Dt_{ d}^{({\bf y}_{\mbox{\tiny cut}})}( u_1,\dots, u_d)
\\
  \lim_{q\to 1}~   \widetilde\Dt_{q,d }^{({\bf y}_{\mbox{\tiny cut}})}(  w;u_2,\dots,u_d )&=\widetilde\Dt_{ d }^{({\bf y}_{\mbox{\tiny cut}})}(  w;u_2,\dots,u_d )
  \\
   \lim_{q\to 1}~   \widetilde\Dt_{q,d,n}^{({\bf y}_{\mbox{\tiny cut}})}(   y;u_2,\dots,u_d )&=\widetilde\Dt_{ d,n}^{({\bf y}_{\mbox{\tiny cut}})}(  y;u_2,\dots,u_d ).
\end{aligned}\label{limDel}\ee

%\noindent This section establishes the following Theorem for the hexagon $\bf P$.

%\remark The version of Theorem \ref{Kernlimtot} for the multi-cut model  will be discussed in Section 10.%can be extended to the multi-cut case by inserting the symmetric functions $E(u_1,\dots,u_d)$ and $E(z,u_2,\dots,u_d)$ in the integrands of the expressions $\Om_\RR(v,z)$ and $\widetilde\Om^{(1)}_\RR(v,z)$ respectively. For the symmetric function $E$, see the remark after Corollary \ref{Dt23}.

\bigbreak

\noindent %Before proving Theorem \ref{Kernlimtot}, we first need the proposition below.
 The limiting kernel  $ {\mathbb K}^{\mbox{\tiny red}} $ in Proposition \ref{Kernlim} is expressed as a $d+2$-fold integral, where -remember- $d$ is the sum of the sizes of the cuts below. Incidentally, the form (\ref{Kernlim''}) of the kernel in this proposition is the most convenient one to show that the kernel $ {\mathbb K}^{\mbox{\tiny red}} $ is the inverse Kasteleyn matrix, up to some trivial conjugation; see \cite{AJvM2}.

\begin{proposition} \label{Kernlim}The limiting kernel for $q\to 1$ has the following form for $d\geq 0$, and $(m,x)$, $(n,y)\in {\bf P}$:
\be \begin{aligned}
   {\mathbb K}^{\mbox{\tiny red}} &(m,x;n,y)  =:({\mathbb K}_0+ {\mathbb K}_1+{\mathbb K}_2)(m,x;n,y)
\\  =& -\frac{(y-x+1)_{n-m-1}}{(n-m-1)!}\Id_{n>m}\Id_{y\geq x}
+  
\oint_{ {\Gamma( {x +{\mathbb N}}) }}  \frac{dv}{2\pi \I}
 \left(\frac{  (v-x+1)_{N-m-1}}{ (N-m-1)! Q(v)}\right)
\\
&\times\left(\oint_{\Gamma_{\infty}} \frac{dz }{2\pi \I(z\!-\!v)}\left(\frac{(N\!-\!n\! )!Q(z)}{(z-y)_{N-n+1}}\right) 
\frac{\Om_\RR(v,z)
 }{\Om_\RR(0,0)}\!\! \right.
 \\& \left.~~~~~~~~~~~~~~~~~~~~~~~~~~~~~~+d\oint_{\Ga_0}\frac{dw}{2\pi \I w^{y+1}(1-w)^n } 
 \frac{ {\widetilde{\Om}^{(1)}}_\RR(v,w)}{\Om_\RR(0,0)}\right), 
\end{aligned}\label{Kernlim''}\ee
%
%@@@@@@@@@@
where, using the expressions $\Dt_{ d}^{({\bf y}_{\mbox{\tiny cut}})}$ and $\widetilde \Dt^{({\bf y}_{\mbox{\tiny cut}})}_{d }$ in (\ref{Deltacut}), (for later use ${\widetilde \Om}^{}_\RR(v,w)$ is defined below with $P(z)$ as in (\ref{P}) and for brevity, set $u=(u_1,\dots,u_d)$)
\be\begin{aligned}
\Om_\RR(v,z):= & \left(\prod_{\al=1}^d  \oint_{\Gamma(\RR)}
\frac{du_\al  }{2\pi \I    Q(u_\al )}
   \frac{v-u_\al}{z-u_\al}
    \right)
  \Dt_d (u ) \Dt_{ d}^{({\bf y}_{\mbox{\tiny cut}})}(u) 
 % \\
 %= &  C_{N,d} \left(\prod_{\al=1}^d  \oint_{\Gamma(\RR)}
%\frac{du_\al P(u_\al)}{2\pi \I ~  Q(u_\al )}
 %  \frac{v-u_\al}{z-u_\al}
 %   \right)
 % \Dt_d (u)^2 %,\mbox{   for }n+y\geq y_1+1
%
%$$
\\ {\widetilde\Om}^{(1)}_\RR(v,z):=&  P(z){\widetilde \Om}^{}_\RR(v,z)   :=\left( 
\prod^{d }_{\al=2} \oint_{\Gamma(	\RR)}   {du_\al \over 2\pi \I Q(u_\al)}
\right)\Dt_{d }   (  v, {u_2 },\ldots, {u_d} )
\\
&\hspace*{5cm}\times \widetilde \Dt^{({\bf y}_{\mbox{\tiny cut}})}_{d }(z;u_2,\dots, u_d)
%\end{aligned}\ee
%
%\be\begin{aligned} 
%\prod_{\al=2}^d(v-u_\al)(u_1-w_\al) 
,\end{aligned}\label{OmR'}\ee
 %
% \newpage
 %
 %\vspace*{-1.5cm}
 %
% \noindent 
 where $ \Om_\RR(v,z)=1$ for $d=0$, as before, and $\widetilde\Om^{(1)}_\RR(v,w)=w^{y_1}$ for $d=1$. %Note the two expressions (\ref{OmR}) and (\ref{OmR'}) for $\Om_\RR(v,z)$ are equivalent in the ????one-cu-cut model.

\end{proposition}

\begin{corollary} \label{CorPetrov}
Given an hexagon ${\bf P}$, with $\ell-1$ cuts on top and none at the bottom, the kernel $  {\mathbb K}^{\mbox{\tiny red}}$ reads
\be \begin{aligned}
   {\mathbb K}^{\mbox{\tiny red}} & (m,x;n,y)  
   =  -\frac{(y-x+1)_{n-m-1}}{(n-m-1)!}\Id_{n>m}\Id_{y\geq x}
\\&+  
\frac{(N-n)!}{ (N-m-1)!}\oint_{ {\Gamma( {x +{\mathbb N}}) }}  \frac{dv}{2\pi \I}
\oint_{\Gamma_{\infty}} \frac{dz }{2\pi \I(z-v)}
 \left(\frac{  (v-x+1)_{N-m-1}Q(z)}{ (z-y)_{N-n+1} Q(v)}\right),
%\\
%&\times\left(\left(\frac{Q(z)}{}\right) 
%\frac{\Om_\RR(v,z)
% }{\Om_\RR(0,0)}\!\! \right.
% \\& \left.~~~~~~~~~~~~~~~~~~~~~~~~~~~~~~+d\oint_{\Ga_0}\frac{dw}{2\pi \I w^{y+1}(1-w)^n } 
% \frac{ {\widetilde{\Om}^{(1)}}_\RR(v,w)}{\Om_\RR(0,0)}\right), 
\end{aligned}\label{KernlimPetrov}\ee
where (setting $a:=n_1+n_2$)
\newline {\bf (i)} ~$Q(z)$ as in (\ref{P}) for $\ell>1$.  (Petrov \cite{Petrov})
\newline {\bf (ii)} $Q(z)=(z-a+1)_c(z+c+1)_b$, for $\ell=1$ (hexagon $(a,b,c)$).)(Johansson \cite{Jo05b})

\end{corollary}

%\newpage

\noindent{\em Proof of Proposition \ref{Kernlim}}: Referring to the kernel (\ref{Kern2'}) in Theorem \ref{Kern2}, we first have $\lim_{q\to 1} {\mathbb K} ^{(0 )}_q = {\mathbb K} _{0}$. The other terms require some argumentation.
\newline {\bf Next we prove $\lim_{q\to 1} {\mathbb K} ^{(1 )}_q = {\mathbb K} _{1}$}. The %one makes the change $n\to n+1$ and $m\to m+1$. So, 
term $ {\mathbb K} ^{(1 )}_q $ can be expressed as follows, taking into account the roots $v=q^{x_j-y}$  of $Q_q(vq^y)$ and the integration contour $ {\Ga(q^{x-y+{\mathbb N}})}$,
\be\begin{aligned}
&{\mathbb K} ^{(1 )}_q(m,x;n,y)= \oint_{\Gamma_{\infty}}\frac{dz}{2\pi \I z}
\sum_{{1\leq j\leq d+N}\atop{x_j\geq x}} {{\cal R}_j}(z),
\end{aligned}\label{termii}\ee
where
\be\begin{aligned}
  {\cal R}_j(z)
   :=&\mbox{Res}_{v=q^{x_j-y}} 
%&\mbox{Res}_{v=q^{x_j-y}}
    v^{m }     \frac{\Phi_q(z^{-1})\PR_{N-m-1}(vq^{y-x}) Q_q(zq^y)}{(z-v)Q_q(vq^y)}\frac{\Om_q(vq^y,zq^y)}{\Om_q(0,0)} 
  \\
 %
 %
% :=\mbox{Res}_{v=q^{x_j-y}}\\
%&\mbox{Res}_{v=q^{x_j-y}}
%&\frac {q^{ (y-x)(d+m-1)}v^{m-1}}{{\widetilde D}_q}\left( \frac{P_{N-m}(vq^{y-x}) Q_q(zq^y)\Phi_q(z^{-1})}{(z-v)Q_q(vq^y)}\sum_{1\leq i_1,\dots,i_d\leq \ell}\frac{1}{\prod^d_{\al=1}Q_q'(q^{x_{i_\al}} ) }\right.\\
%&\left. ~~~~~~~~~~~~\times\prod_{\al=1}^d \frac{vq^y-q^{x_{i_\al}}}{zq^y -q^{x_{i_\al}}}
%\det (\tilde P_{N-1}^\beta ( q^{x_{i_\al}}))_{1\leq \alpha,\beta \leq d} V_d(q^{x_{i_1}},\ldots,q^{x_{i_d}})\right)  \\
 =&   {q^{m(x_j-y) }}  \Phi_q(z^{-1}) \PR_{N-m-1}(q^{x_j-x})%\sum_{1\leq i_1,\dots,i_d\leq \ell} %q^y%
 %\frac{\prod_{{1\leq r\leq d+N}\atop{r\neq j}}(zq^y-q^{x_r})}{Q(q^{x_i})}
\prod_{{1\leq r\leq d+N}\atop{r\neq j}}
 \frac{z-q^{x_r-y}}{q^{x_j-y}-q^{x_r-y}}  \frac{\Om_q(  q^{x_j},zq^y)}{\Om_q(0,0)} 
%\prod_{{1\leq r\leq d+N}\atop{r\neq j}}
\\
%&\prod_{\al=1}^d \frac{ q^{x_j-y}-q^{x_{i_\al}-y}}{z  -q^{x_{i_\al}-y}}
% ~ \frac{\det (\tilde P_{N-1}^\beta ( q^{x_{i_\al}}))_{1\leq \alpha,\beta \leq d}}{\prod^d_{\al=1}Q_q'(q^{x_{i_\al}} ) } V_d(q^{x_{i_1}},\ldots,q^{x_{i_d}})
%\\
\stackrel{*}{=}&   \sum_{{1\leq i_1,\dots,i_d\leq c}\atop{j \notin(i_1,\dots,i_d) }}\frac{q^{m(x_j-y) }\PR_{N-m-1}(q^{x_j-x})}{\prod_{{1\leq r\leq d+N}\atop{r\neq j}}(q^{x_j-y}-q^{x_r-y})}%q^y%\frac{\prod_{{1\leq r\leq d+N}\atop{r\neq j}}(zq^y-q^{x_r})}{Q(q^{x_i})}
~~~(\star)
\\
&\times 
 {\Phi_q(z^{-1})\prod_{{1\leq r\leq d+N}\atop{r\neq j,i_1,\dots,i_{d}}}(z-q^{x_r-y})} 
~~~(\star\star )\\
&\times \frac{ \Dt_{q,d}^{({\bf y}_{\mbox{\tiny cut}})}(q^{x_{i_1}},\dots,q^{x_{i_d}})   \Dt_d(q^{x_{i_1}},\ldots,q^{x_{i_d}})}{{\Om}_q(0,0)}\prod_{\al=1}^d \frac{ q^{x_j-y}-q^{x_{i_\al}-y}}{Q_q'(q^{x_{i_\al}} ) }
  \\&~~~~~~~~~~~~~~~~~~~~~~~~~~~~~~~~~~~~~~~~~~~~~~~~~~~~~~~~~~~~~~~~~~~  ~~~~(\star\star\star )
\\ =:&
\sum_{{1\leq i_1,\dots,i_d\leq c}\atop{j \notin(i_1,\dots,i_d) }}((\star)\times(\star\star)\times(\star\star\star)).\end{aligned}\label{Rj}\ee
Equality $\stackrel{*}{=}$ is obtained by inserting the expression (\ref{OmqRes}) for $\Om_q(v,z)$ into the formula preceding $\stackrel{*}{=}$. One also notices that the  term  $\prod_{\al=1}^d (z-q^{x_{i_\al}-y})$ in the denominator of $\Om_q(  q^{x_j},zq^y)$ cancels $d$ terms in the product $\prod_{{1\leq r\leq d+N}\atop{r\neq j}}
  ({z-q^{x_r-y}})$, yielding equality ($\stackrel{\ast}{=}$). Also $j$ can be taken different from $i_1,\dots,i_d$, because if $j$ would figure in that set the term would vanish. 
  
  Now we let $q\to 1$. Since the terms ($\star$) and ($\star\star \star$) do not depend on $z$, the integration in (\ref{termii}) will act on term $(\star\star)$ only. So, at first for given $x_j$, we have, using Lemma \ref{qlim}, the estimate
 \be\begin{aligned}
 (\star)\simeq &(q-1)^{1-N-d}\frac{ {(x_j-x+1)_{N-m-1}}}{{(N-m-1)!}\prod_{{1\leq r\leq d+N}\atop{r\neq j}} (x_j-x_r)} 
 \\
 =&(q-1)^{1-N-d}\frac{ {(x_j-x+1)_{N-m-1}}}{{(N-m-1)!}Q'(x_j)}.
\label{star1}\end{aligned}\ee
Next, again using Lemma \ref{qlim}, we have for $q\sim 1$, 
 \be\begin{aligned}
 &\oint_{\Gamma_{\infty}} \frac{dz}{2\pi \I z}( \star\star)
 \\&= \oint_{\Gamma_{\infty}} \frac{dz}{2\pi \I z}%\sum_{{1\leq i_1,\dots,i_d\leq \ell}\atop{j\notin (i_1,\dots,i_d)  }}
 {\Phi_q(z^{-1})\prod_{{1\leq r\leq d+N}\atop{r\neq j,i_1,\dots,i_{d}}}(z-q^{x_r-y})}
 \\&\simeq (q-1)^{N-1}(N-n )!
 %\sum_{{1\leq i_1,\dots,i_d\leq \ell}\atop{j\notin (i_1,\dots,i_d)  }}
 \oint_{\Gamma_\infty}\frac{dz}{2\pi \I (z-y)_{N-n+1}}\prod_{{1\leq r\leq d+N}\atop{r\neq j,i_1,\dots,i_{d}}}  {(z-x_r)}{}
\\&\simeq (q-1)^{N-1}(N-n )!\oint_{\Gamma_\infty}\frac{dz Q(z)}{2\pi \I (z-y)_{N-n+1}(z-x_j)
 }%\sum_{{1\leq i_1,\dots,i_d\leq \ell}\atop{j\notin (i_1,\dots,i_d)  }}
 \frac 1{\prod_{\al=1} ^d {(z-x_{i_\al})}}.
 \end{aligned}\label{star2}\ee 
Next we need to estimate ($\star\star\star$).
 From (\ref{limDel}), combined with some of the limits in Lemma \ref{qlim}, one finds the estimate
 \be\begin{aligned}
&  {\Dt_d(q^{x_{i_1}},\ldots,q^{x_{i_d}})\Dt_{q,d}^{({\bf y}_{\mbox{\tiny cut}})}(q^{x_{i_1}},\dots,q^{x_{i_d}})} \prod^d_{\al=1}\frac{q^{x_j-y}-q^{x_{i_\al}-y}}{ \displaystyle  Q_q' (q^{x_{i_\al}})  }
  \\&
  \simeq \! (q\!-\!1)^{-d(N+\frac{d-1}2-1) } 
    {\Dt _d(x_{i_1},\dots,x_{i_d})    }
    \Dt_{ d}^{({\bf y}_{\mbox{\tiny cut}})}( {x_{i_1}},\dots, {x_{i_d}})
   \prod^d_{\al=1}  \frac{ x_j-x_{i_\al} }{\displaystyle  Q' (x_{ i_{\al}  })  } .
  \end{aligned}\ee
 and thus, when $q\to 1$, the residue version of expression 
 $\Om_\RR(0,0)$ defined in (\ref{OmqRes}) tends to the residue version of (\ref{OmR'}), modulo a power of $(q-1)$:
 \be\begin{aligned}
 %\tilde D_q&=\Om_q(0,0) \\ 
 \Om_q(0,0)&=\sum_{1\leq i_1,\ldots,i_d \leq c} 
 \frac{\Dt_d(q^{x_{i_1}},\ldots,q^{x_{i_d}})\Dt_{q,d}^{({\bf y}_{\mbox{\tiny cut}})}(q^{x_{i_1}},\dots,q^{x_{i_d}})}{ \displaystyle  \prod^d_{\al=1} Q_q' (q^{x_{i_\al}})  }
 \\&
  \simeq% (q-1)^{d(d-1)/2 -d(N+d-1)} \sum_{1\leq i_1,\ldots,i_d \leq \ell}  {V_d(x_{i_1},\ldots,x_{i_d}) ^2  \over \displaystyle  \prod^d_{j=1} Q' (x_{i_j})  }\prod_{\al=1}^d(x_\al+d)^{(N-d)}
 %\\
 %&=
   (q-1)^{-d(N+\frac{d-1}2)}  \sum_{1\leq i_1,\ldots,i_d \leq c} 
    \frac {\Dt_d(x_{i_1},\dots,x_{i_d})  \Dt_d^{({\bf y}_{\mbox{\tiny cut}})}( {x_{i_1}},\dots, {x_{i_d}})    }{\displaystyle  \prod^d_{\al=1} Q' (x_{ i_{\al}  })  } 
 %   @@@
% \left(\prod^d_{j=1} \oint_{\Gamma (x_1,\ldots,x_c)} { du_j P(u_j)\over  2\pi \I Q  (u_j)}%\oint_{\Gamma_0} \frac {dw_j }{2\pi \I w_j^{u_j +d- m_1  +1}  { (1-w_j )^{N }}}
%  \right) \Dt_d (u_1,\ldots,u_d)^2  
 %\\
% &~~~~\times %\prod_{\al=1}^d
 %{\cal P}_{N-d}(u_\al) 
\\
&=    (q-1)^{-d(N+\frac{d-1}2)}  \Om_\RR(0,0).
 \label{OmegaLim}\end{aligned}\ee
Upon using (\ref{limDel}), it follows that the last line ($\star\star\star$) of formula (\ref{Rj}) is estimated by 
\be
 \star\star\star \simeq \frac{(q-1)^{ d }}{\Om_\RR(0,0)}
    {  }
 \left(\prod^{d }_{\al=1}  \frac {  (x_j-x_{i_\al}) }{ Q' (x_{ i_\al}) }\right)  { \Dt_d(x_{i_ 1},\ldots,x_{i_ d}) \Dt_d^{({\bf y}_{\mbox{\tiny cut}})}( {x_{i_1}},\dots, {x_{i_d}})          }.
\label{star3}\ee
%
%\vspace*{7cm}
%
Multiply  the three contributions (\ref{star1}), (\ref{star2}) and (\ref{star3}) together and do  the summation, in which the requirement $j\notin (i_1,\dots,i_d)$ can be removed; indeed whenever $j\in (i_1,\dots,i_d)$, the sum below would automatically vanish. So we find for each $x_j\geq x$,
$$\begin{aligned}
\lim_{q\to 1}&\oint_{\Gamma_\infty} \frac{dz}{2\pi \I z}
{\cal R}_j(z)\\
 =& \frac{(N-n )!}{(N-m-1)!}   \frac{(x_j-x+1)_{N-m-1}}{Q'(x_j)}
%\sum_{{1\leq i_1,\dots,i_d\leq \ell}\atop{(i_1,\dots,i_d)\ni \!\!\!\!   \backslash~  j}} \frac{ \prod_{\al=1}^d  (  x_j- x_{ \al}  )}{{\prod_{{1\leq r\leq d+N}\atop{r\neq j}}( x_j-x_r })}  
 \oint_{\Gamma_\infty}\frac{dz Q(z)}{2\pi \I (z-y)_{N-n+1}(z-x_j)
%\prod_{\al=1} ^d {(z-x_{i_\al})} 
}
 \\&\times %\frac{1}{2\pi \I }% \sum_{{1\leq i_1,\dots,i_d\leq \ell}\atop{j \notin(i_1,\dots,i_d) }}
 \sum_{{1\leq i_1,\dots,i_d\leq c}\atop{  }}
  \left(\prod^{d }_{\al=1}  \frac { (x_j-x_{i_\al}) }
 {  (z-x_{i_\al})Q' (x_{i_\al}) }\right)   \frac { \Dt_d(x_{ i_1},\ldots,x_{i_ d}) \Dt_d^{({\bf y}_{\mbox{\tiny cut}})}( {x_{i_1}},\dots, {x_{i_d}})       }{\Om_\RR(0,0)}%@@ %
% \left(\prod^d_{\al=1}  {   {\cal P}_{N-d} (u_\al)\over  2\pi \I Q'(u_\al)}\right)   
% \oint_{\Gamma_\infty}\frac{dz}{2\pi \I}\prod_1^{N-1} \frac{(z-x_r)}{(z-y)_{N-n+2}}
\\
=&\frac{(N-n )!} {(N-m-1)!}   \frac{(x_j-x+1)_{N-m-1}}{Q'(x_j)}
 \oint_{\Gamma_\infty}\frac{dz Q(z)}{2\pi \I (z-y)_{N-n+1}(z-x_j)}
 \frac{\Om_\RR(x_j,z)}{\Om_\RR(0,0)}.
\end{aligned}$$
%where
%$${\cal P}_{N-d} (u ):=\prod_{\ell=1}^{N-d}( u+d-m_1+\ell)=(u+d-m_1+1)_{_{ N-d }}
%$$
%
So we conclude, upon summing the $v$-residues,
$$\begin{aligned}\lim_{q\to 1} {\mathbb K} ^{(1 )}_q(m,x;n,y)= \frac{(N-n )!}{(N\!-\!m\!-\!1)!}
\oint_{ {\Gamma( {x +{\mathbb N}}) }}
& \frac{dv}{2\pi \I}
  \oint_{\Gamma_{\infty}}  \frac{dz}{2\pi \I}
\frac{(v-x+1)_{N-m-1}}{(z-v)(z-y)_{N-n+1}}
\\&\times\frac{Q(z)}{Q(v)}\frac{\Om_\RR(v,z)}{\Om_\RR(0,0)}={\mathbb K}_1(m,x;n,y).
\end{aligned}$$
 
 %

%\newpage

%\vspace*{-1.5cm}

%\mbox{\bf Third term}
\noindent {\bf Consider now the ${\mathbb K} ^{(2 )}_q$ }part of the kernel (\ref{Kern2'}), given in residue form as \be\begin{aligned}
{\mathbb K} ^{(2 )}_q(m,x;n,y)=&
  \frac{      d  }{  q^{ y(2d+N-1) }  }  
%\left(
 \oint_{\Gamma_{\Gamma(q^{x-y+{\mathbb N}})  } }{dv~v^{m } \PR_{N-m-1}(vq^{y-x})    \over 2\pi i  \prod _1^{d+N}(v-q^{x_i-y})} % \prod^{d-1}_1   (v-u_i) 
   %\\
%&\times  
 \frac{\widetilde\Om_q(vq^y,y)}{\Om_q(0,0)}
 \\
  =&\frac{      d  }{  q^{ yd }  } \sum_{x_i\geq x}\frac{q^{m(x_i-y)}\PR_{N-m-1}(q^{x_i-x})}{ Q'_q(q^{x_i}) }\frac{\widetilde\Om_q(q^{x_i},y)}{\Om_q(0,0)},
\end{aligned}\label{Kq3}\ee
with $\widetilde\Om_q( q^{x_i},y)$ given in (\ref{OmqRes}), from which  it follows that, upon using (\ref{Deltacut}), (\ref{vdmy}), (\ref{hP'}), (\ref{psi(d+r)}), Lemma \ref{qlim} and (\ref{limDel}),
\be\begin{aligned}  \label{Omqtilde}
\widetilde\Om_q(& q^{x_j},y) 
 \\& =
\sum_{1\leq i_2,\dots,i_d\leq c}   
  \frac{   \Dt_{d }   ( q^{x_j},q^{x_{i_2}},\ldots,q^{x_{i_d}}) 
  \widetilde\Dt_{q,d,n}^{({\bf y}_{\mbox{\tiny cut}})}( q^y,q^{x_{i_2}},\ldots,q^{x_{i_d}} )}
   {   \prod_{\al=2}^d Q_q'(q^{x_{i_\al}}) }.
%\\  
%&\times 
%\end{aligned}\ee
%So, inserting this limit into (\ref{Omqtilde}), we find ({\bf Mark: here one interchanges the integrals!OK?)}
%$$\begin{aligned}
%\widetilde\Om_q(  q^{x_j}&, y) 
\\ &= \frac{1}{(q-1)^{(N -1+d/2)(d-1)}}
\left(\oint_{\Gamma_0} \frac {dw}{2\pi \I w^{  y  +1}  { (1-w )^n}} 
\right)
\\&~~\times \left( 
\prod^{d }_{\al=2} \oint_{\Gamma	(x_1,\ldots,x_c)}   {du_\al \over 2\pi \I Q(u_\al)}
\right)\Dt_{d }   (  {x_j}, {u_2 },\ldots, {u_d} )
\widetilde \Dt^{({\bf y}_{\mbox{\tiny cut}})}_{d }(w;u_2,\dots, u_d)
\\
&= \frac{1}{(q-1)^{(N -1+d/2)(d-1)}}
 \oint_{\Gamma_0} \frac {dw~ {\widetilde\Om}^{(1)}_\RR( {x_j},w)}{2\pi \I w^{  y  +1}  { (1-w )^n}} ,
%
%
%&@\times\oint_{\Gamma( y,\dots , y-N+n)}\frac{(N-n)!dz}{2\pi \I(z-y)_{N-n+1}}
 %\Dt_d(z,u_2,\dots,u_d) P(z)\prod_{\al=2}^dP(u_{\al})\\=& \frac{C_{N,d}}{(q-1)^{(N -1+d/2)(d-1)}}\oint_{\Gamma( y,\dots , y-N+n)}\frac{(N-n)!P(z)dz}{2\pi \I(z-y)_{N-n+1}}\widetilde\Om_\RR(v,z)
\end{aligned}\ee
and thus, combined with the estimate (\ref{OmegaLim}), this gives
$$\begin{aligned}
\frac{\widetilde\Om_q(  q^ {x_j},y)}{ \Om_q(0,0)}
%&\simeq
%(q-1)^{d+N-1} 
%\frac{\Xi_1 (  {x_j},y)}{ C_{N,d}\Om_\RR (0,0)}\\
&=
(q-1)^{d+N-1} \oint_{\Gamma_0} \frac {dw~}{2\pi \I w^{  y  +1}  { (1-w )^n}} \frac{ {\widetilde\Om}^{(1)}_\RR( {x_j},w)}{\Om_\RR(0,0)} .\end{aligned}$$
Inserting this formula into (\ref{Kq3}), and using the limits (\ref{limPQ}), we  find 
$$\begin{aligned}
\lim_{q\to 1}~ {\mathbb K} ^{(2 )}_q(m,x;n,y)
%\\&=
%  \frac{d}{(N-m-1)!}\oint_{\Gamma( {x +{\mathbb N}})} \frac{dv}{2\pi \I Q(v)}  (v-x+1)_{N-m-1}\frac{\Xi_1(v,y)}{\Om_\RR(0,0)}
%\\&=
 =&\frac{d}{(N-m-1)!}\oint_{\Gamma( {x +{\mathbb N}})} \frac{dv}{2\pi \I Q(v)}  (v-x+1)_{N-m-1}
\\
&~~~ ~\times\oint_{\Gamma_0} \frac {dw~}{2\pi \I w^{  y  +1}  { (1-w )^n}}\frac{ {\widetilde \Om}^{(1)}_\RR(v,z)}{\Om_\RR(0,0)}={\mathbb K}_2(m,x;n,y),
\end{aligned}$$
 yielding the third part ${\mathbb K}_2$ of the kernel (\ref{Kernlim''}), %. Using the formulas in the beginning of the section then 
thus proving Proposition \ref{Kernlim}.   \qed

\bigbreak

 \noindent{\em Proof of Corollary \ref{CorPetrov}}: Setting $d=0$ in the kernel (\ref{Kernlim''}) leads to formulas (\ref{KernlimPetrov}). Expanding $1/(z-v)=\sum v^j/z^{j+1}$ formula (\ref{KernlimPetrov}) can be written as a finite sum $\sum p_j(x)q_j(y)$, with $p,~q$ polynomials. It is unclear in case ({\bf ii}) how this relates to the extended Hahn kernel in \cite{Jo05b}.\qed

% \newpage

\section{The ${\mathbb K}^{\mbox{\tiny red}}$-kernel as a $r+3$-fold integral, with $r=|\LR|-d$% number of particles on the oblique lines of the strip $\{\rho\}$
 } 

In this section it will be shown that ${\mathbb K}^{\mbox{\tiny red}}$-kernel can be reduced from a $d+1$-fold integral (\ref{Kernlim''}) to a $r+3$-fold integral (\ref{Kernlim4}), where (remember!) $d$ is the total sizes of the lower-cuts; this will establish  Theorem \ref{Kr}.

%will prove Theorem \ref{Kr}, which holds {\bf for the two-cut model}. It shows that the  

This reduction will be instrumental in performing the asymptotic analysis on ${\mathbb L}^{\mbox{\tiny blue}}$ for the two-cut model in \cite{AJvM2}, when all the sides and cuts of the polygon $\bf P$ tend to $\infty$, while $r$ and $\rho$ remain finite. Remember that in the two-cut model $r:=|\LR|-d$ happens to be the number of blue dots on the oblique lines $\{\eta=m_1+k\}$ within the strip $\rho$.

Proposition \ref{Kernlim} expresses the kernel in terms of two multiple integrals $\Om_\RR $ and $\widetilde\Om_\RR$ as in (\ref{OmR'}), which become after substituting (\ref{Dt2}) and (\ref{Dt3}) for $  \Dt^{({\bf y}_{\mbox{\tiny cut}})}_{d }$ and $\widetilde \Dt^{({\bf y}_{\mbox{\tiny cut}})}_{d }$ in these expressions,
\be\begin{aligned}
\Om_\RR(v,z) = & C_{N,d} \left(\prod_{\al=1}^d  \oint_{\Gamma(\RR)}
\frac{du_\al P(u_\al)}{2\pi \I ~  Q(u_\al )}
   \frac{v-u_\al}{z-u_\al}
    \right)
  E^{({\bf y}_{\mbox{\tiny cut}})}_g(u )\Dt_d (u)^2  %,\mbox{   for }n+y\geq y_1+1
%
%$$
\\
 \widetilde\Om_\RR(v,z):=%&
&  
 C_{N,d}  \left(\prod^{d }_{\al=2} \oint_{\Gamma	(\RR)} \!\!  {du_\al P(u_\al)  \over 2\pi \I ~ Q(u_\al)}\frac{ v- {u_\al}  }{  z- {u_\al}   }
 \right)
  \\& 
 \times~E^{({\bf y}_{\mbox{\tiny cut}})}_g(z,u_2 ,\dots,u_d ) \Dt_{d  }   (   z, {u_2 },\ldots, {u_d} )^2%\prod_{\al=2}^d(v-u_\al)(u_1-w_\al) 
.\end{aligned}\label{OmR}\ee
 Define the same expressions, but with the contours $\Ga(\RR)$ replaced by $\Ga(\LR)$:
  \be   \Om_\LR(v,z)\mbox{    and   }   \widetilde\Om_\LR(v,z),
 \label{OmL}\ee
 with $ \Om_\RR(v,z)=1$ for $d=0$ and $\widetilde\Om_\RR(v,z)=1 $ for $d=1$ and similarly for $ \Om_\LR(v,z),~\widetilde\Om_\LR(v,z)$.
  We now state :%Proposition \ref{KernFinal} we give two expressions for the kernel, a first one involving (\ref{OmR}) and a second one involving (\ref{OmL}). %These expressions contain two multiple integrals, on the one hand the expression $\Om_\RR $, the same as the one in (\ref{OmR'}), but with (\ref{Dt2}) substituted for $\Dt^{({\bf y}_{\mbox{\tiny cut}})}_d$; and on the other hand, a new expression $\widetilde\Om_\RR$, to wit,%\footnote{In view of the ratios, the precise value of the constant $C_{N,d}$ and  will actually never appear.},
%
%aims at simplifying the kernel ${\mathbb K}^{\mbox{\tiny red}}$ in (\ref{Kernlim''}) by working out the $\Gamma_0$-integration in (\ref{Kernlim''}), leading to a new multiple integral \footnote{In view of the ratios, the precise value of the constant $C_{N,d}$ will actually never appear.} $\widetilde\Om_\RR$. The first expression $\Om_\RR $ is the same as the one in (\ref{OmR'}), but with (\ref{Dt2}) substituted for $\Dt^{({\bf y}_{\mbox{\tiny cut}})}_d$. So, 

%Proposition \ref{KernFinal} rewrites the $\Om$-expressions in the ${\mathbb K}^{\mbox{\tiny red}}$-kernel (\ref{Kernlim'''}) in terms of a contour $\Gamma({\mathcal L})$ about ${\mathcal L}$, instead of $\RR$. So, replacing the contour $\Ga(\RR)$ by $\Ga(\LR)$ in (\ref{OmR}) yields two new multiple integrals

%the kernel ${\mathbb K}^{\mbox{\tiny red}}$ in terms of rewrite the .%for in Proposition \ref{Kernlimtot}

%, which will occur in Paper II, for the kernels ${\mathbb K}$ and , when all the sides and cuts of the polygon $\bf P$ tend to $\infty$, while $r$ and $\rho$ remain finite. 

% {\bf Remember the polynomials $P$ %=P_{ \rho } Q_{ \CR } P_{ \sg }$,~  
 %and $Q$ %=Q_{ \LR } Q_{ \CR } Q_{ \RR }$ 
 % from (\ref{PQdecomp}), the rational function $h $ from (\ref{h}) and the $\Om^\pm_r$-kernel from (\ref{Omr}). }%set
 % \be h(u):=\frac{  Q _{ \RR }(u)}{  P_\rho(u  )P_\sg(u  )Q_\LR(u )}.
%\label{h}\ee

%Consider the following  We also define the same expressions as in (\ref{OmR}), but with  

\begin{proposition}\label{KernFinal} For $(m,x),~(n,y)\in {\bf P}$, the kernel for the ${\mathbb K}^{\mbox{\tiny red}}$-process takes on two different forms, a first one involving (\ref{OmR}) and a second one involving (\ref{OmL}),%. integration along contours about $\LR$ and another one involving contours about $\RR$:%the following form:%The kernel ${\mathbb K}^{\mbox{\tiny red}}$ %for the red dots in the {\bf lower one-cut model} 
 %has the following form for $d\geq 0$, and $(m,x)$, $(n,y)\in {\bf P}$:
%
\be \begin{aligned}
     {\mathbb K}^{\mbox{\tiny red}} &(m,x;n,y) 
 \\   
 =& {\mathbb K}_0(m,x;n,y)%-\frac{(y-x+1)_{n-m-1}}{(n-m-1)!}\Id_{n>m}\Id_{y\geq x}
+  
\frac{(N-n)!}{(N-m-1)! } \oint_{\Ga(x+{\mathbb N})}\frac{ dv~ (v-x+1)_{N-m-1}}{ 2\pi \I Q(v)}  
\\
&\times  \left(\oint_{\Gamma_{\infty}} \!\!\frac{dz ~Q(z)  }{2\pi \I(z\!-\!v)(z-y)_{N-n+1}} 
\frac{\Om_\RR(v,z)+d(z-v)\frac{P(z)}{Q(z)}\widetilde\Om_\RR(v,z)}{\Om_\RR(0,0)} \right.
\\&~~~~\hspace*{5cm}~~~  -   d %\Id_{\tau<0}
%\oint_{ {\Gamma( {x +{\mathbb N}}) }}  \frac{dv}{2\pi \I}\! \frac{  (v-x+1)_{N-m-1}}{ (N-m-1)! Q(v)} 
%\\
%&\times
\left. \oint_{\Gamma_{\tau}} \frac{dz P(z) }{2\pi \I (z-y)_{N-n+1}}%\left(\frac{ }{}\right) 
\frac{ \widetilde\Om_\RR(v,z)}{\Om_\RR(0,0)} \right)
%,
%  \end{aligned}\label{Kernlim'''}\ee
% \be \begin{aligned}
 %  {\mathbb K}^{\mbox{\tiny red}}& (m,x;n,y) 
\\  = & {\mathbb K}_0(m,x;n,y)%-\frac{(y-x+1)_{n-m-1}}{(n-m-1)!}\Id_{n>m}\Id_{y\geq x}
 +  
\frac{(N-n)!}{(N-m-1)!}\oint_{ {\Gamma( {x +{\mathbb N}}) }}  %\frac{}{}
\!\!\!  \frac{ dv (v-x+1)_{N-m-1}}{2\pi \I Q(v)}
 \\
 & 
%\\&
\times \left(\oint_{\Gamma_{\infty}} \frac{dzQ(z) }{2\pi \I(z\!-\!v)(z-y)_{N-n+1}}  
 %
%\\
%&
 %\times 
 \frac{\Om_{\LR} (v,z)}{\Om_\LR (0,0)} 
%\right.
%\\& \left. \hspace*{5.5cm}   %\Id_{y+n<m_1 }d
%
%\\
%&\times
 %\left.
 + d\oint_{ \Gamma_{\tau}} \frac{dz P(z) }{2\pi \I (z-y)_{N-n+1}}%\left(\frac{ }{}\right) 
\frac{ \widetilde\Om_\LR(v,z)}{\Om_\LR(0,0)} \right),
  \end{aligned}\label{Kernlim3}\ee
%  where $\tau=(y+n)-(y_1+1)$ and the contour $ \Ga_\tau $ were  defined in (\ref{cont0}).% and (\ref{gatau}). %thus containing $m_1-(y+n)$ points. 
  % 
  with the $\Om$'s and $ \widetilde\Om$'s defined in (\ref{OmL}) and the contour $\Ga_\tau$  defined in (\ref{cont0}). Contour $\Gamma_\infty=\Gamma(v;y,\dots , y-N+n)$ is a large circle about all the poles of the integrand. 

\end{proposition}

To prove Proposition \ref{KernFinal}, we need the following Lemma:

\begin{lemma} \label{LemOmtOm}
Given a rational function $R(u)$ with possibly poles within a contour $\Gamma$ and a point $z$ not within $\Gamma$, not a pole of $R(u)$ and given a symmetric polynomial  $S({\bf u})$, with ${\bf u}=(u_1,\dots,u_\ell)$. Then for $0\leq k-1\leq \ell$, we have (the notation $\Gamma\cup z$ refers to the contour $\Gamma$, deformed so as to contain $z\in \BC$), setting ${\bf u}=(u_1,\dots,u_\ell)$,
\be\begin{aligned}
\Bigl(\prod_{\al=1}^\ell \oint_{\Ga \cup z}&\frac{du_\al R(u_\al)}{2\pi \I( u_\al-z)}\Bigr)\Dt_\ell ^2({\bf u})S({\bf u})
%-%@@+
 %\prod_{\al=1}^{k-1}   \oint_{\Ga} \frac{du_\al R(u_\al)}{2\pi \I( u_\al-z)}
%\prod_{\al=k }^{\ell } \oint_{\Ga\cup z}\frac{du_\al R(u_\al)}{2\pi \I( u_\al-z)}\Dt_\ell^2({\bf u})S({\bf u})
 \\
=& (k-1)R(z)\left(\prod_{\al=1}^{\ell-1}\oint_{\Ga} \frac{du_\al}{2\pi \I} R(u_\al) 
(u_\al-z)\right) \Dt_{\ell-1}^2(u_1,\dots,u_{\ell-1})S(z,u_1,\dots,u_{\ell-1})
\\
&+\prod_{\al=1}^{k-1}   \oint_{\Ga} \frac{du_\al R(u_\al)}{2\pi \I( u_\al-z)}
\prod_{\al=k }^{\ell } \oint_{\Ga\cup z}\frac{du_\al R(u_\al)}{2\pi \I( u_\al-z)}\Dt_\ell^2({\bf u})S({\bf u}).
\end{aligned}\label{OmtOm}\ee
In particular for $k=\ell+1$,
\be\begin{aligned}
\Bigl(\prod_{\al=1}^{\ell} &\oint_{\Ga \cup z} \frac{du_\al R(u_\al)}{2\pi \I( u_\al-z)}\Bigr) \Dt_\ell ^2({\bf u}) S({\bf u})
-\Bigl(\prod_{\al=1}^{\ell }   \oint_{\Ga} \frac{du_\al R(u_\al)}{2\pi \I( u_\al-z)}\Bigr)\Dt_\ell^2({\bf u})  S({\bf u})
\\
=& \ell R(z)\Bigl(\prod_{\al=1}^{\ell-1}\oint_{\Ga} \frac{du_\al}{2\pi \I}R(u_\al) 
(u_\al-z)\Bigr)\Dt_{\ell-1}^2(u_1,\dots,u_{\ell-1})S(z,u_1,\dots,u_{\ell-1}).
\end{aligned}\label{OmtOm'}\ee
%So we have the following residue result:

%$$\begin{aligned}
%\mbox{Residue}_z
%\end{aligned}$$

\end{lemma}

\noindent{\em Proof of Lemma \ref{LemOmtOm}}: The identity (\ref{OmtOm}) is obviously true for $k=1 $. We now proceed by induction: let it be true for fixed $k\geq 1$; then we show its truth for $k+1$. In the second expression on the right hand side of equation (\ref{OmtOm}) one computes the residue at $u_k=z$, yielding%
$$\begin{aligned}
\prod_{\al=1}^{k-1} &  \oint_{\Ga} \frac{du_\al R(u_\al)}{2\pi \I( u_\al-z)}
\prod_{\al=k }^{\ell } \oint_{\Ga\cup z}\frac{du_\al R(u_\al)}{2\pi \I( u_\al-z)}\Dt_\ell^2({\bf u})S({\bf u})
\\=&
\prod_{\al=1}^{k-1}    \oint_{\Ga} \frac{du_\al R(u_\al)}{2\pi \I( u_\al-z)}
\oint_{\Ga\cup z}\frac{du_{k} R(u_k)}{2\pi \I( u_k-z)}
\prod_{\al=k+1 }^{\ell } \oint_{\Ga\cup z}\frac{du_\al R(u_\al)}{2\pi \I( u_\al-z)}\Dt_\ell^2({\bf u})S({\bf u})
\\=&
\prod_{\al=1}^{k }    \oint_{\Ga} \frac{du_\al R(u_\al)}{2\pi \I( u_\al-z)}
\prod_{\al=k+1 }^{\ell } \oint_{\Ga\cup z}\frac{du_\al R(u_\al)}{2\pi \I( u_\al-z)}\Dt_\ell^2({\bf u})S({\bf u})
\\&
+\prod_{\al=1}^{k-1}    \oint_{\Ga} \frac{du_\al R(u_\al)}{2\pi \I( u_\al-z)}\left(R(z)
\prod_{\al=k+1 }^{\ell } \oint_{\Ga\cup z}\frac{du_\al R(u_\al)}{2\pi \I( u_\al-z)}(\Dt_{\ell }^2S)(u_1,\dots,\stackrel{k}{\stackrel{ \downarrow}{z}} ,\dots,u_\ell)
\right)
\end{aligned}$$
$$\begin{aligned}
\\= &
\prod_{\al=1}^{k }    \oint_{\Ga} \frac{du_\al R(u_\al)}{2\pi \I( u_\al-z)}
\prod_{\al=k+1 }^{\ell } \oint_{\Ga\cup z}\frac{du_\al R(u_\al)}{2\pi \I( u_\al-z)}\Dt_\ell^2({\bf u})S({\bf u})
\\&
+R(z)\prod_{\al=1}^{k-1}    \oint_{\Ga} \frac{du_\al  R(u_\al)}{2\pi \I }\left( 
\prod_{\al=k+1 }^{\ell } \oint_{\Ga\cup z}\frac{du_\al R(u_\al) }{2\pi \I }\right)
\\
&\hspace*{1.7cm}\times \Dt_{\ell-1}^2(u_1,\dots,\hat u_k ,\dots,u_\ell)
S(u_1,\dots,\stackrel{k}{\stackrel{ \downarrow}{z}} ,\dots,u_\ell)\prod_{{j=1}\atop{j\neq k}}^{\ell}(u_j-z)
\\=&
\prod_{\al=1}^{k }    \oint_{\Ga} \frac{du_\al R(u_\al)}{2\pi \I( u_\al-z)}
\prod_{\al=k+1 }^{\ell } \oint_{\Ga\cup z}\frac{du_\al R(u_\al)}{2\pi \I( u_\al-z)}\Dt_\ell^2({\bf u})S({\bf u})
\\&
+R(z)\prod_{\al=1}^{\ell-1}\oint_{\Ga}  \frac{du_\al}{2\pi \I} R(u_\al) 
(u_\al-z)\Dt_{\ell-1}^2(u_1,\dots,u_{\ell-1})S(z, u_1,\dots,u_{\ell-1})  .\end{aligned}$$
The latter is obtained by renaming the variables $(u_1,\dots,u_{k-1},u_{k+1},\dots,u_\ell)\to (u_1,\dots,u_{\ell-1})$ and upon noticing that in the second expression the integration over $\Ga\cup z$ can be replaced by $\Ga$, since the integrand has no residue at $z$. One also uses the symmetry of the function $S$. This ends the proof of Lemma \ref{LemOmtOm}.\qed

\noindent{\em Proof of Proposition \ref{KernFinal}}:
%\newline
 %\proof %\noindent{\em Proof of proposition \ref{Kernlimtot}}:  
{\bf First expression in (\ref{Kernlim3}).}  The starting point is expression (\ref{Kernlim''}) for the kernel ${\mathbb K}^{\mbox{\tiny red}}  $, and in particular its ${\mathbb K}_2$-piece. Use the expression (\ref{Deltacut}) for ${\widetilde{\Om}^{(1)}}_\RR(v,w) $ in (\ref{OmR'}), combined with expression
 (\ref{Dt3}) in Corollary \ref{Dt23} and (\ref{Deltacut}), leading to:
$$ \begin{aligned}
d\oint_{\Ga_0}&\frac{dw~~{\widetilde{\Om}^{(1)}}_\RR(v,w) }{2\pi \I w^{y+1}(1-w)^n } 
\\&=d
\left( 
\prod^{d }_{\al=2} \oint_{\Gamma(	\RR)}   {du_\al \over 2\pi \I Q(u_\al)}
\right)
 \Dt_{d  }   (  v, {u_2 },\ldots, {u_d} )
\widetilde \Dt^{({\bf y}_{\mbox{\tiny cut}})}_{d,n }(y,u_2,\dots, u_d)
\\&=d
 \left(\oint_{\Ga(y,y-1,\dots,y+n-N)}-\oint_{\Gamma_\tau}\right)\frac{(N-n)! P(z) dz}{2\pi \I(z-y)_{N-n+1}}
~     \widetilde\Om_\RR(v,z),
%
%,
%\left( \prod^{d }_{\al=2} \oint_{\Gamma(	\RR)}   {du_\al P(u_\al )  \over 2\pi \I Q(u_\al)}\right)\Dt_{d  }   (  v, {u_2 },\ldots, {u_d} )\Dt_d(z,u_2 ,\dots,u_d )  
 \end{aligned}$$
in terms of $ \widetilde\Om_\RR(v,z)$, as in (\ref{OmR}). %The  final identity uses formula (\ref{IntId1}) in Lemma \ref{LemmaInt}, with the contour $\gamma_\tau$ defined in (\ref{gatau}), thus leading to formula (\ref{Kernlim'''}) for the kernel. 
 This establishes the first formula (\ref{Kernlim3}).% and thus ends the proof of Proposition \ref{Kernlimtot}. %\qed 

\noindent {\bf Second expression in (\ref{Kernlim3}).} The following identities holds:
 \be\label{reduce3}
\begin{aligned}  \frac{\Om_\RR(v,z)+d(z-v)\frac{P(z)}{Q(z)}\widetilde\Om_\RR(v,z)}{\Om_\RR(0,0)}
 &=\frac{\Om_\LR(v,z)}{\Om_\LR(0,0)}\mbox{    and   }
  \frac{ \widetilde\Om_\RR(v,z)}{\Om_\RR(0,0)}
 &=-\frac{\widetilde\Om_\LR(v,z)}{\Om_\LR(0,0)}.
\end{aligned}\ee
It suffices to show that 
\be \begin{aligned}
(-1)^{d}\Om_\LR(v,z) &= \Om_\RR( v,z)+d(z-v) \frac{P(z )}{Q(z)}\widetilde\Om_\RR(v,z)
 %\mbox{   and   }
 \\
  (-1)^{d-1}\widetilde\Om_\LR(v,z)&=\widetilde\Om_\RR(v,z).
 \end{aligned}\label{reduce2}\ee
%Indeed, remembering from formula (\ref{PQ}), that \be\frac{P }{Q }=\frac{P_{\GR}  }{Q_{\LR} Q_{\RR} },\label{PQ1}\ee  
Remembering the degrees of the polynomials $P$ and $Q$ in (\ref{PQ}), we have that each integrand in the $d$-fold integral $\Om_\LR(v,z)$ and in the $d-1$-fold integral $\widetilde\Om_\LR(v,z)$ (as in Proposition \ref{KernFinal}) have degree (in a fixed $u_\al$) equal to 
$$\bl
&\stackrel{{P}\atop{\downarrow}}{(N- y_1+y_d-1)}-\stackrel{{Q}\atop{\downarrow}}{(N+d)}+\stackrel{{\Dt^2_d(z,u_2,\dots,u_d)}\atop{\downarrow}}{2(d-1)}+\stackrel{{E^{({\bf y}_{\mbox{\tiny cut}})}_g(z,u_2 ,\dots,u_d )}\atop{\downarrow}}{(y_1-y_d-d+1)}=-2.
\el$$
%using the expressions (\ref{rho}) and (\ref{sg})  for $\rho,~\sg$. 
 So the integrands have no pole at $\infty$. %, but its only poles  are, in view of formula (\ref{PQ1}), at the points of $\RR$ and $\LR$. 
 Thus the contour $-\Ga(\LR)$ in $\Om_\LR$ (as in (\ref{reduce2})) can be deformed to a contour $\Ga(\RR \cup z)$; so we have ($C_{N,d}$ is the constant in Proposition \ref{detV'}) 
\be (-1)^d\frac{\Om_\LR(v,z)}{C_{N,d} }=\left(\prod_{\al=1}^d  \oint_{\Gamma(\RR\cup z)}
\frac{du_\al P(u_\al)}{2\pi \I ~  Q(u_\al )}
   \frac{ u_\al-v}{ u_\al-z}
    \right)
   E^{({\bf y}_{\mbox{\tiny cut}})}_g(u )\Dt_d (u )^2 ,
\label{reduce3}\ee
which has the form of Lemma \ref{LemOmtOm}, with $\ell=d$ and $R(u)=\frac{P(u )}{Q(u)}(u-v )$,  $S=E^{({\bf y}_{\mbox{\tiny cut}})}_g$, modulo a sign of $(-1)^d$. Then the right hand side of (\ref{reduce3}) equals, by formula (\ref{OmtOm'}) and (\ref{OmR}), 
$$ \begin{aligned}
 \frac{\Om_\RR(v,z)}{C_{N,d} }&+d( z-v)\frac{P(z )}{Q(z)}\prod_2^{d }\oint_{\Ga(\RR)} 
\frac{du_\al P(u_\al)}{2\pi \I ~  Q(u_\al )}
   (v-u_\al)( z-u_\al )
   \\
   &~~~~~~~~~~~\times \Dt^2_{d-1} (u_2,\dots,u_d )   E^{({\bf y}_{\mbox{\tiny cut}})}_g(z,u_2,\dots,u_d )
\\
&=\frac{1}{C_{N,d} }  \left(\Om_\RR(v,z)+d( z-v)\frac{P(z )}{Q(z)}\widetilde \Om_\RR(v,z)\right),%
\end{aligned}$$
establishing the first identity (\ref{reduce3}). The second identity (\ref{reduce2}) is simpler, since the integrands have no pole about $u_\al=z$. Therefore the contour about $\LR$ can be deformed to a contour about $\RR$ in each of the $d-1$ integrals. This shows both identities (\ref{reduce2}), upon using $\Om_\LR(0,0)=(-1)^d \Om_\RR(0,0)$, 
ending the proof of Proposition \ref{KernFinal}. \qed

\bigbreak

%\newpage

\noindent\noindent{\em Proof of Theorem \ref{Kr}}: It suffices to prove: %the second identity in (\ref{reduce1}), with $\Om_r$ defined in (\ref{Omr}). 
\be\begin{aligned}
 \frac{\Om_\LR (v,z)}{\Om_\LR (0,0)} =\frac{Q_{ \LR }(v)}{Q_{ \LR }(z)}
  \frac{ \Om^+_r (v,z)}{ \Om^+_r (0,0)},~~~~~
   d\frac{\widetilde\Om_\LR (v,z)}{\Om_\LR (0,0)} =\tfrac{Q_{ \LR }(v) Q_{ \LR }(z)}{r+1} 
   %Q_{ \LR }(v) Q_{ \LR }(z)
  \frac{ \Om^-_{r+1} (v,z)}{ \Om^+_r (0,0)}
, \end{aligned}\label{d-to-r'}\ee
 $\Om_\LR$, $\widetilde\Om_\LR$ were defined in (\ref{OmL}), and $\Om^+_r$ and $ \Om^-_{r+1}$ in (\ref{Omr}).

  The proof will depend on the two identities below (\ref{complement}) and (\ref{vandermk}). Set $J:=\{i_1<\dots<i_k,\mbox{  with  } x_{i_\al}\in \LR\}$, viewed as particles and the 
corresponding set of holes $j_1<\dots<j_\ell$, also with $x_{j_\al}\in \LR$ and $|\LR |=k+\ell$.   Let $\varphi(u)$ be a rational function  with no roots or poles along $\LR$.  For the ease of notation set momentarily $\Dt (\LR):=\{$vandermonde determinant of variables $x_i$ in $\LR\}$. Then%\Dt_b(x_{d+c+1},\dots,x_{d+c+b})$. Then% , so that, for instance, 
  \be\prod_{\al=1}^k \frac{\varphi(x_{i_\al})}{Q_\LR'(x_{i_\al})}=\prod_{x_i\in \LR}
\frac{ \varphi(x_{i })}{Q_\LR'(x_{i })}
\prod_{\al=1}^\ell \frac{Q_\LR'(x_{j_\al})}{\varphi(x_{j_\al})}.\label{complement}\ee

 In formula (\ref{Vanderm2}) of Lemma \ref{Vanderm1},
 set $n=k$, $m=\ell$ and $n+m=k+\ell=|\LR|$,  and ${\bf x}=(x_{i_1},\dots, x_{i_k})$ and ${\bf x}^c=(x_{j_1},\dots, x_{j_\ell})$ 
, and $ Q  :=Q_{\LR}(z)$, leading to:
%
%$$\begin{aligned}
%\Dt_d(x_{i_1},\dots, x_{i_d})=\pm %(-1)^{r(r-1)/2}
%\frac{ \Dt_{d+r}(x_{ d+c+1},\dots,x_{d+N})   \Dt_{ r}({ x_{j_1},\dots, x_{j_r}}) }
%{\prod_{\al=1}^r {Q'_{ \LR }} (x_{j_\al})}.
% \end{aligned}$$
%
%
\be \label{vandermk}\begin{aligned}
\Dt_k({\bf x})=\pm %(-1)^{r(r-1)/2}
\frac{ \Dt_{ }(\LR)   \Dt_{\ell}({ {\bf x}^c}) }
{\prod_{\al=1}^\ell {Q'_{ \LR }} (x_{j_\al})}.
 \end{aligned}\ee
  %  
 %  \newpage
%  \vspace*{-1.5cm}
 % (\ref{symcompl})
  
\noindent Then, combining the three facts (\ref{complement}), (\ref{vandermk}) and (\ref{symcompl}) above, we have for $k,\ell\geq 0$ such that $k+\ell=|\LR|$, with ${\bf x}$ and ${\bf x}^c$ as above:
  \be\label{complement'}
  \begin{aligned}
  \Bigl(&\prod_{\al=1}^k   \oint_{\Gamma(\LR)}
\frac{du_\al \varphi(u_\al)}{2\pi \I ~  Q_\LR(u_\al )}
       \Bigr)
 (S \Dt_k ^2 )(u)
 %\\& 
 =
    \sum_{{i_1<\dots<i_k}\atop{x_{i_\al}\in \LR}}
k! (S   \Dt_k^2)({\bf x} )  
 \left( \prod_{\al=1}^k\frac{\varphi(x_{i_\al})}{Q_\LR'(x_{i_\al})}
      \right)
 \\
 &=\frac{k!}{\ell!}% \sum_{{i_1<\dots<i_d}\atop{x_{i_\al}\in \LR}}
 {\Dt^2_{ }(\LR) }   \prod_{x_i\in \LR} 
  \frac{\varphi(x_{i })}{Q_\LR'(x_{i })} 
     %@@\frac{Q_{ \LR }(v)}{Q_{ \LR }(z)}
  %\\&\qquad\times 
   \ell!   \!\! \sum_{{j_1<\dots<j_\ell}\atop{x_{j_\al}\in \LR}} \left( \prod_{\al=1}^\ell\frac{Q_\LR'(x_{j_\al})}{\varphi(x_{j_\al})(Q'_{ \LR }  (x_{j_\al}))^2}
   \right)
    (\widetilde S\Dt^2_{ \ell})( {\bf x}^c)
 \\ 
 &%\stackrel{\ast\ast}
  {=} \frac{k!}{\ell!} 
  {\Dt^2_{ }(\LR) }  \left( \prod_{x_i\in \LR} 
  \frac{\varphi(x_{i })}{Q_\LR'(x_{i })}
  \right)  
 % \\&@@@
  \left(\prod_{\al=1}^\ell  \oint_{\Gamma(\LR)}
\frac{du_\al  }{2\pi \I ~ \varphi(u_\al )Q_\LR(u_\al )}
     \right)
( \widetilde S \Dt^2_\ell) (u).
  \end{aligned}\ee
%
 % $ Q'(x)=Q _{ \CR }(x)Q _{ \RR }(x)Q'_{ \LR }(x)$
 %
 We now apply the formula above to two different $k,~\ell$ and $\varphi(u)$ (themselves depending on $v,z$):
  \be\label{substphi}\begin{aligned}\varphi(u)&:=\frac{P(u)(v-u)}{Q_{\RR}(u)Q_{\CR}(u)(z-u)}\\
   S(u_1,\dots&,u_d)=E^{({\bf y}_{\mbox{\tiny cut}})}_g(u_1,\dots,u_d )
  \\
  \widetilde S(u_1,\dots&,u_r)= \widetilde E^{({\bf y}_{\mbox{\tiny cut}})}_g(u_1,\dots,u_r )
  \\
  &k =d,~~\ell=r
  \end{aligned}
  \mbox{   ~~ or~~~   } \begin{aligned}\varphi(u)&:=\frac{P(u)(v-u)(z-u)}{Q_{\RR}(u)Q_{\CR}(u)}
 \\ S(z,u_2,\dots,&u_{d  }) =E^{({\bf y}_{\mbox{\tiny cut}})}_g(z;u_2,\dots,u_{d } )
 \\
  \widetilde S(z, u_1,\dots,&u_{r+1 }) =\widetilde E^{({\bf y}_{\mbox{\tiny cut}})}_g(z;u_1,\dots,u_{r+1} )
 \\
 & k =d-1,~~\ell=r+1.
  \end{aligned} \ee
 Te map $S\mapsto \widetilde S$ above refers to the operation on symmetric functions explained in (\ref{symcompl}). Using the expression (\ref{h}) for $h(u)$, and remembering the definitions\footnote{as in Proposition \ref{KernFinal}.} (\ref{OmR}), (\ref{OmL}) and (\ref{Omr}) of $\Om_\LR$ and $\Om_r$, the first substitution (\ref{substphi}) in (\ref{complement'}) leads to the following expression, where $C_{N,d} $ is the constant in (\ref{detV}) and where 
  $C_{\LR}$ is a constant\footnote{To be precise, $C_\LR=\Dt(  \LR)^2\prod_{x_i\in \LR}\frac{P(x_i)}{Q'_\LR(x_i)Q_\RR(x_i)Q_\CR(x_i)} =\Dt(  \LR)^2\prod_{x_i\in \LR}\frac{P(x_i)}{Q'(x_i) }$.}  depending on the geometry of $\bf P$ only:    
 $$\begin{aligned}
 \frac{{\Om_\LR(v,z)}}{C_{N,d}C'_{N,d}} &= \left(\prod_{\al=1}^d  \oint_{\Gamma(\LR)}
\frac{du_\al P(u_\al)}{2\pi \I ~  Q(u_\al )}
   \frac{v-u_\al}{z-u_\al}
    \right)
  (E^{({\bf y}_{\mbox{\tiny cut}})}_g\Dt^2_d) (u )  
  \\
  & =\frac{d!}{r!} C_{\LR} \frac{Q_{ \LR }(v)}{Q_{ \LR }(z)}\left(\prod_{\al=1}^r  \oint_{\Gamma(\LR)}
\!\!\!\frac{du_\al h(u_\al) }{2\pi \I  }
   \frac{z\!-\!u_\al}{v\!-\!u_\al}
    \right)
  (\widetilde E^{({\bf y}_{\mbox{\tiny cut}})}_g\Dt^2_r) (u_1,\dots,u_r)
   \\
   &
  =\frac{d!}{r!} C_{\LR} \frac{Q_{ \LR }(v)}{Q_{ \LR }(z)}\Om^+_r(v,z),
  \end{aligned}
  $$
  whereas the second substitution (\ref{substphi}) in (\ref{complement'}) yields, remembering the definitions (\ref{OmR}), (\ref{OmL}) and (\ref{Omr}) of 
  $\widetilde\Om_\LR$ and $\widetilde\Om_{r+1}$, 
   $$\begin{aligned}
\frac{ \widetilde\Om_\LR   (v,z)}{C_{N,d} }  %\\
  &= \left(\prod_{\al=2}^{d}  \oint_{\Gamma(\LR)}
\frac{du_\al P(u_\al)}{2\pi \I ~  Q(u_\al )}
     (v-u_\al) (z-u_\al) 
    \right)
 \\&~~~~~~~~~~~~~~~~~~~~~~~~~~~~~~~~~~~   E^{({\bf y}_{\mbox{\tiny cut}})}_g  (z;u_2,\dots,u_{d} ) 
   \Dt_{d-1}^2( u_2,\dots,u_{d} )
  \\
  & =\frac{(d-1)!}{(r+1)!} C_{\LR}  {Q_{ \LR }(v)}{Q_{ \LR }(z)}\left(\prod_{\al=1}^{r+1}  \oint_{\Gamma(\LR)}
\frac{du_\al h(u_\al) }{2\pi \I (v-u_\al)(z-u_\al)}
    \right)
\\
&~~~~~~~~~~~~~~~~~~~~~~~~~~~~~~~~~~
 \widetilde E^{({\bf y}_{\mbox{\tiny cut}})}_g  (z;u_1,\dots,u_{r+1} ) 
   \Dt_{r+1}^2( u_1,\dots,u_{r+1} )
  \\&=\frac{(d-1)!}{(r+1)!} C_{\LR}  {Q_{ \LR }(v)}{Q_{ \LR }(z)} \Om^-_{r+1}(v,z).
  \end{aligned}
  $$
  These two identities lead at once to the ratios (\ref{d-to-r'}). Substituting these ratios in the kernel (\ref{Kernlim3}) establishes Theorem \ref{Kr}.\qed
  
  \begin{corollary} \label{bisd} For $|\LR|=d$, we have that
\be\begin{aligned}
\frac{\Om_\LR(v,z)}{\Om_\LR(0,0)}= \frac{Q_{\LR}(v)}{Q_{\LR}(z)}.\end{aligned}\label{b=d}\ee
\end{corollary}

\noindent{\em Proof of Corollary \ref{bisd}:} Here $r=0$ and thus $ \Om_r (v,z)=1$ and the corollary follows from (\ref{d-to-r'}).\qed 

%@@@@@@@@@

%\newpage

%\vspace*{-3cm}


\begin{thebibliography}{10}

 
  
\bibitem{ACJvM} Mark Adler,  Sunil Chhita, Kurt Johansson and
Pierre van Moerbeke: {\em Tacnode GUE-minor processes and double Aztec diamonds}, Probab. Theory Related Fields {\bf 162}, no. 1-2, 275-325 (2015)

\bibitem{AJvM}  Mark Adler, Kurt Johansson and Pierre van Moerbeke:  {\em Double Aztec diamonds and the tacnode process}. Adv. Math. 252 (2014), 518-571.

\bibitem{AJvM2}  Mark Adler, Kurt Johansson and Pierre van Moerbeke:  {\em  Lozenge tilings of hexagons with cuts and asymptotic fluctuations}. To appear.

\bibitem{AvM}  Mark Adler and Pierre van Moerbeke: {\em Coupled GUE-minor Processes}. Intern Math Research Notices, {\bf 21} (2015) (arXiv:1312.3859)
 
 \bibitem{BBNV} Dan Betea, J. Bouttier, P. Nejjar and M. Vuletic: {\em The free boundary Schur process and applications}. (arXiv:1704.05809)


\bibitem{BuK} A. Bufetov and A. Knizel: {\em Asymptotics of random domino tilings of rectangular Aztec diamonds} (arXiv: 1604.01491)


 \bibitem{BFP}
Alexei Borodin, Patrik L. Ferrari, Michael Pr\"ahofer, Tomohiro Sasamoto: \emph {Fluctuation properties of the TASEP with periodic initial configuration} J. Stat. Phys. {\bf 129} (2007) (arXiv:math-ph/0608056)
 
\bibitem{BGR}
 A. Borodin, V. Gorin and E. M. Rains, {\em $q$-Distributions on boxed plane partitions}, Selecta Math. 16 (2010),
731-789.

 \bibitem{BR} Alexei Borodin, Eric M. Rains: {\em Eynard-Mehta theorem, Schur process, and their Pfaffian analogs} J. Stat. Phys. 121 (2005), no. 3-4, 291-317.   (arXiv:math-ph/0409059)
 
  \bibitem{B } Alexei Borodin: {\em Determinantal point processes}, The Oxford handbook of random matrix theory, 231Ð249, Oxford Univ. Press, Oxford, 2011.

\bibitem{Ciucu} Mihai Ciucu and Ilse Fischer {\em Lozenge tilings of hexagons with arbitrary dents}. Adv. in Appl. Math. 73, 1-22.  (2016)



\bibitem{Cohn} H. Cohn, M. Larsen and J. Propp. {\em The shape of a typical boxed plane partition}, The New York Journal of Mathematics. {\bf 4} 137-165 (1998) 

 \bibitem{Defos} M. Defosseux: {\em Orbit measures, random matrix theory and interlaced determinantal processes}, Ann. Inst. H. Poincar Probab. Statist. {\bf 46},  209-249. (2010)
 
 \bibitem{DJM}Erik Duse, Kurt Johansson, Anthony Metcalfe {\em The Cusp-Airy Process} (arXiv:1510.02057)

 \bibitem{Duse} Erik Duse and Anthony Metcalfe:  
{\em Asymptotic geometry of discrete interlaced patterns: Part I.}
Internat. J. Math. {\bf 26} (2015),  1550093. 

\bibitem{Duse1} Erik Duse and Anthony Metcalfe:  
{\em Asymptotic geometry of discrete interlaced patterns: Part II.} (arXiv:1507.00467)

% Asymptotic Geometry of Discrete Interlaced Patterns: Part II

\bibitem{Jo01b}
Kurt~Johansson, \emph{Discrete orthogonal polynomial ensembles and the Plancherel measure}, Ann. of Math. {\bf 153}, 259Ð296. (2001) 
 

 
\bibitem{Jo02b}
Kurt~Johansson, \emph{Non-intersecting paths, random tilings and random
  matrices}, Probab. Theory Related Fields \textbf{123} (2002), 225--280.

\bibitem{Jo03b}
Kurt~Johansson, \emph{Discrete polynuclear growth and determinantal processes},
  Comm. Math. Phys. \textbf{242} (2003), 277--329.


 \bibitem{Jo05b}Kurt~Johansson,  \emph{Non-intersecting, simple, symmetric random walks and the extended Hahn kernel}, Ann. Inst. Fourier (Grenoble) {\bf 55}, 2129-2145. (2005) 
 
 \bibitem{Jo05c}Kurt~Johansson,  \emph{The arctic circle boundary and the Airy process}, Ann. Probab. {\bf 33}, 1Ð30. (2005)
 
  \bibitem{Jo16}Kurt~Johansson:  \emph{Edge Fluctuations og Limit Shapes}, Harvard Lectures (Nov. 2016) (arXiv:1704.06035)

 \bibitem{JN} Kurt Johansson and Eric Nordenstam: {\em Eigenvalues of GUE minors}, Electron. J. Probab. {\bf 11} , 1342Ð1371 (2006).
 
  \bibitem{Gorin} Vadim E. Gorin: \emph{Nonintersecting paths and the Hahn orthogonal polynomial ensemble},  Funct. Anal. Appl. {\bf 42}, 180Ð197 (2008).
  
   \bibitem{Gorin1} Vadim E. Gorin: \emph{Bulk universality for random lozenge tilings near straight boundaries and for tensor products}, to appear in Communications in Mathematical Physics. (arXiv:1603.02707)


 \bibitem{Kast} Pieter W. Kasteleyn: {\em Graph theory and crystal physics}.  Graph Theory and Theoretical Physics pp. 43Ð110 Academic Press, London (1967).
 
 \bibitem{K} Richard Kenyon: {\em Lectures on dimers} Statistical mechanics, 191-230, IAS/Park City Math. Ser., 16, Amer. Math. Soc., Providence, RI, 2009. arXiv: 0910.3129
 
  \bibitem{KO} Richard Kenyon and Andrei Okounkov: {\em Limit shapes and the complex Burgers equation}, Acta Math. {\bf 199}, no. 2, 263-302 (2007)
  
 \bibitem{Krat} Christian Krattenthaler: {\em Advanced determinantal calculus}, S\'eminaire Lotharingien Combin. 42 (1999) (The Andrews Festschrift), paper B42q, 67 pp
 
\bibitem{Krat2} Christian Krattenthaler, {\em Descending plane partitions and rhombus tilings of a hexagon with a triangular hole},  European J. Combin. {\bf 27}  no. 7, 1138-1146 (2006)

\bibitem{McD} I. Macdonald. {\em Symmetric functions and Hall polynomials}. Oxford Mathematical Monographs. (1995).
 
\bibitem{McMa} P. A. MacMahon, {\em Memoir on the theory of the partition of numbersÑPart V. Partitions in two-dimensional
space}, Phil. Trans. R. S., 1911, A.
 


 \bibitem{Metc} Anthony Metcalfe: {\em Universality properties of GelfandÐTsetlin patterns}, Probab. Theory Related Fields {\bf 155}(1-2) 303-346 (2013).
 
 \bibitem{Nov} Jonathan Novak, {Lozenge tilings and Hurwitz numbers}, Journal of Stat. Phys., 161 , 509-517 (2015) (arXiv:math/0309074)
 
 \bibitem{OR1} Andrei Okounkov and Nicolai Reshetikhin: {\em Correlation function of Schur process with application to local geometry of a random 3-dimensional Young diagram} J. of the American Math. Society {\bf 16}, 581-603 (2003)
 
 \bibitem{OR} Andrei Okounkov and Nicolai Reshetikhin: {\em The birth of a random matrix} Mosc. Math. J. {\bf 6} , 553-566, 588.(2006)
 
  \bibitem{Petrov1} Leonid Petrov: {\em Asymptotics of uniformly random lozenge tilings of polygons. Gaussian free field}, Ann. Probab. {\bf 43} 1Ð43 (2015).
 
 
 \bibitem{Petrov} Leonid Petrov: {\em Asymptotics of random lozenge tilings via Gelfand-Tsetlin schemes}. Probability Theory
and Related Fields. 160, 3-4 (2014), 429-487.

\end{thebibliography}
 \end{document}